"The genetic architecture of adaptations to high altitude in Ethiopia"

Gorka Alkorta-Aranburu[1], Cynthia M. Beall[2]*, David B. Witonsky[1], Amha Gebremedhin[3], Jonathan K. Pritchard[1,4], Anna Di Rienzo[1]*

[1] Department of Human Genetics, University of Chicago, Chicago, Illinois, United States of America,

[2] Department of Anthropology, Case Western Research University, Cleveland, Ohio, United States of America,

[3] Department of Internal Medicine, Faculty of Medicine, Addis Ababa University, Addis Ababa, Ethiopia,

[4] Howard Hughes Medical Institute

* E-mail: dirienzo@bsd.uchicago.edu and cmb2@case.edu

Corresponding authors:
Anna Di Rienzo
Department of Human Genetics
University of Chicago
920 E. 58th Street
Chicago, IL 60637, USA.

Cynthia M. Beall
Anthropology Department
Case Western Reserve University
238 Mather Memorial Building
11220 Bellflower Road
Cleveland, OH  44106, USA.

ABSTRACT

Although hypoxia is a major stress on physiological processes, several human populations have survived for millennia at high altitudes, suggesting that they have adapted to hypoxic conditions. This hypothesis was recently corroborated by studies of Tibetan highlanders, which showed that polymorphisms in candidate genes show signatures of natural selection as well as well-replicated association signals for variation in hemoglobin levels. We extended genomic analysis to two Ethiopian ethnic groups: Amhara and Oromo. For each ethnic group, we sampled low and high altitude residents, thus allowing genetic and phenotypic comparisons across altitudes and across ethnic groups. Genome-wide SNP genotype data were collected in these samples by using Illumina arrays. We find that variants associated with hemoglobin variation among Tibetans or other variants at the same loci do not influence the trait in Ethiopians. However, in the Amhara, SNP rs10803083 is associated with hemoglobin levels at genome-wide levels of significance. No significant genotype association was observed for oxygen saturation levels in either ethnic group. Approaches based on allele frequency divergence did not detect outliers in candidate hypoxia genes, but the most differentiated variants between high- and lowlanders have a clear role in pathogen defense. Interestingly, a significant excess of allele frequency divergence was consistently detected for genes involved in cell cycle control, DNA damage and repair, thus pointing to new pathways for high altitude adaptations. Finally, a comparison of CpG methylation levels between high- and lowlanders found several significant signals at individual genes in the Oromo.

AUTHOR SUMMARY

Although hypoxia is a major stress on physiological processes, several human populations have survived for millennia at high altitudes, suggesting that they have adapted to hypoxic conditions. Consistent with this idea, previous studies have identified genetic variants in Tibetan highlanders associated with reduction in hemoglobin levels, an advantageous phenotype at high altitude. To compare the genetic bases of adaptations to high altitude, we collected genetic and epigenetic data in Ethiopians living at high and low altitude, respectively. We find that variants associated with hemoglobin variation among Tibetans or other variants at the same loci do not influence the trait in Ethiopians. However, we find a different variant that is significantly associated with hemoglobin levels in Ethiopians. Approaches based on the difference in allele frequency between high- and lowlanders detected strong signals in genes with a clear role in defense from pathogens, consistent with known differences in pathogens between altitudes. Finally, we found a few genome-wide significant epigenetic differences between altitudes. These results taken together imply that Ethiopian and Tibetan highlanders adapted to the same environmental stress through different variants and genetic loci.

**INTRODUCTION**

Hypoxia is a major stress on human physiological processes and a powerful homeostasis system has evolved in animals to cope with fluctuations in oxygen concentration [1]. High altitude (HA) hypoxia, such as that experienced at 2500m of altitude or greater, engages this system and elicits physiological acclimatization when lowlanders become exposed to hypoxia. In addition to lower oxygen levels, lower biodiversity and extreme day-to-night temperature oscillations challenge HA living. The classic response is an increase in hemoglobin (Hb) concentration that during acclimatization compensates for the unavoidable lowered percent of oxygen saturation ($O_2$ sat) of Hb due to ambient hypoxia. Acclimatization is not completely effective, however. For example, birth weights are lower than at low altitude (LA) as is physical exercise performance [2,3]. In addition, lowlanders residing at altitudes higher than 2500 meters (m) are at risk for chronic health problems arising in part from acclimatization processes. For example, long-term high Hb levels increase blood viscosity as well as the risk of thrombosis and stroke [4,5,6] and poorer pregnancy outcomes [7]. These results taken together suggest that the acclimatization response does not assure that fitness is unaltered at HA. Several distantly related human populations have survived for 5-50 thousand years (ky) [8,9,10] at altitudes above 2500 m. Indeed, in most cases, sufficient time has elapsed since HA settlement for natural selection to have changed the frequency of adaptive alleles. Interestingly, even though all HA residents are exposed to the same, constant, ambient hypoxia, indigenous highlander populations show distinctive physiological characteristics thought to offset HA stress: Andeans show some reduction in $O_2$ sat, but a marked increase in Hb levels [11], Tibetans present markedly low $O_2$ sat, but relatively little increase in Hb levels [12], and Amhara in Ethiopia present little reduction in $O_2$ sat or increase in Hb levels [13,14]. Whether these phenotypic contrasts reflect different genetic adaptations across populations remains an open question.

Substantial evidence in Tibetan highlanders suggests that variation in Hb levels and $O_2$ sat is adaptive. In the Tibetan population, a major gene effect on $O_2$ sat was detected, with the inferred genotypes associated with higher $O_2$ sat also associated with higher reproductive success [15]; though the locus underlying this effect has not yet been identified, its effect on phenotypes directly related to fitness points to the presence of adaptive variation. With regard to Hb levels, which are surprisingly similar between Tibetan highlanders and lowlanders at sea level [11], population genetics and genotype-phenotype association analyses have identified alleles at two loci (endothelial PAS domain protein 1 (*EPAS1*) and egl nine homolog 1 (*EGLN1*)) that are consistently associated with signatures of positive natural selection and with lower Hb levels, suggesting that natural selection in Tibet favored variants that counteract the deleterious effects of long-term acclimatization [16,17,18,19].

An analysis of genome-wide genotype data in Tibetan and Andean highlanders suggested that natural selection acted on largely distinct loci in the two populations [20]. In addition, a recent study comparing Ethiopian Amhara highlanders with other ethnic groups at LA identified yet another set of candidate targets of selection [21]. However, the Tibetans remain unique with regard to the strength of the evidence for natural selection and the marked genetic effects on the Hb level phenotype.

HA populations offer a rare opportunity to investigate the impact of natural selection on the genetic architecture of adaptation because independent realizations of the adaptive process can be examined in different parts of the world. Specifically, the Ethiopian highlands offer a unique opportunity to study HA adaptation because individuals from distinct, but closely related ethnic groups have communities at HA and LA, thus allowing more informative genetic and phenotypic comparisons. In this study, we extended genomic analysis to two Ethiopian ethnic groups, Amhara and Oromo, with the goal of determining whether Tibetans and Ethiopian highlanders share the same adaptations and of elucidating the genetic bases of adaptive HA phenotypes in Ethiopia. We also measured genome-wide methylation levels to explore the contribution of epigenetic modifications to HA adaptations.

## RESULTS
**The Ethiopian Amhara and Oromo differ in adaptive phenotypes**

We obtained phenotype data in two distinct, but closely related ethnic groups, the Amhara and the Oromo (Texts S1-S2; Figures S1-S5), that include communities of HA and LA residents. All individuals were born and raised at the same altitude where they were sampled. These samples allow comparing phenotypes across altitudes within ethnic groups as well as across ethnic groups. While historical records indicate that the Oromo have moved to HA only in the early 1500s [22,23], the Amhara have inhabited altitudes above 2500m for at least 5ky and altitudes around 2300-2400m for more than 70ky [24,25]. Therefore, sufficient time has elapsed for the Amhara to have evolved genetic adaptations to hypoxia.

As shown in Figure 1, the HA samples of both ethnic groups had higher Hb than the LA samples, however the Oromo had twice as much elevation in Hb as the Amhara. The elevation in Hb levels is particularly evident for the measurements in males, raising the possibility that other factors (e.g. menstrual cycle) in females affect the power to detect significant phenotypic differences between groups. With regard to $O_2$ sat, HA Amhara had a 5.6% lower $O_2$ sat compared to LA Amhara while HA Oromo had 10.5% lower $O_2$ sat than their LA counterparts. Therefore, we detected significant phenotypic differences not only between populations from the same ethnic group that live at different altitudes, but also across populations from closely related ethnic groups (Oromo and Amhara) that live at the same altitude. Given the low genetic divergence between these two ethnic groups at the genome-wide level (mean $F_{ST} = 0.0098$), the phenotypic differences between Amhara and Oromo highlanders are unlikely to be due to independent genetic adaptations in these ethnic groups; rather they are likely to reflect genetic adaptations that evolved in the Amhara, due to their longer residence at HA.

We also measured pulse and calculated arterial oxygen content, but these phenotypes did not show significant differences across ethnic groups or altitudes and were omitted from further analyses. For details on the phenotypic variation in Amhara and Oromo, see Text S3, Tables S1-S2 and Figure S6.

**Genome-wide association signals in Ethiopia**

To learn about the genetic bases of variation in Hb level and $O_2$ sat in Ethiopia, we tested for an association between SNP genotype in the 260 unrelated Ethiopian samples and Hb levels or $O_2$ sat. We considered the total Ethiopian sample (*i.e*. HA and LA Amhara and Oromo) as well as each ethnic group and each altitude separately (Figures S7-S15). No genome-wide significant signal was observed for either Hb levels or $O_2$ sat in the Oromo and in the total Ethiopian sample and for $O_2$ sat in the Amhara (Figures S7-S15 and Tables S3-S20). Likewise, no excess of low p-values was observed in these association analyses relative to null expectations obtained by permutations (Figures S7-S15). In contrast, the Amhara showed an excess of low association p-values in the analysis of Hb levels compared to expectations obtained by permutations (Figure 2A), indicating a genetic contribution to variation in Hb levels. In addition, one SNP (rs10803083) on chromosome 1 was associated with variation in Hb levels (p= $4.96 \times 10^{-8}$) in Amhara; this association is genome-wide significant after correction for the 985,385 tests performed (Figure 2B and Table S3). In addition, the next six most strongly associated SNPs are in strong LD with rs10803083 ($r^2 \geq 0.69$). There are no known genes within 600kb of these SNPs, and the closest genes, *i.e.* phospholipase D family member 5 (*PLD5)* and centrosomal protein 170kDa (*CEP170*), are not obvious candidate genes for variation in Hb levels. These SNPs do not reside within an ultra conserved sequence element. Notably, the effect size of SNP rs10803083 (~0.83 g/dL, Figure 2C) is half that of the *EGLN1* SNPs [16] but comparable to that of Hb associated *EPAS1* SNPs in Tibetans [17]. Although SNP rs10803083 reaches genome-wide significance levels, replication studies will be needed to further assess the evidence for an association with Hb levels.

Our sample size is small relative to traditional GWAS, thus liming the power of our analysis. In addition, due to the correlation between linked SNPs, the Bonferroni correction we applied is overly conservative. Therefore, many true Hb concentration variants may not reach genome-wide levels of significance. In this regard, it is interesting to note that the second strongest association signal (rs2899662) is located within the established hypoxia-candidate gene retinoid-related orphan receptor alpha (*RORA*). *RORA* encodes a protein that induces the transcriptional activation of hypoxia-inducible-factor-1alpha (*HIF-1α*) [26], thus it plays a significant role in the same pathway where adaptations were detected in Tibetans. This SNP is, therefore, an excellent candidate for variation in Hb levels. Additional SNPs of interest for follow up analyses (Table S3) include: the solute carrier family 30 member 9 (*SLC30A9*), which is regulated by hypoxia [27], the collagen type VI alpha 1 (*COL6A1*), which is associated with performance during endurance cycling [28] and is a HIF response gene [29], and the hepatocyte growth factor (*HGF)*, which is induced by hypoxia [30], activates *HIF1* DNA binding [31], plays a role in angiogenesis and protects against hypoxia induced cell injury [32,33].

**Comparing the genetic architecture of Hb levels between Ethiopians and Tibetans**

The above association analyses allow comparing the genetic architecture of Hb levels between Ethiopians and Tibetans [16,17,18].

First, we focused on the *EPAS1* and *EGLN1* SNPs that were previously associated with variation in Hb levels in Tibetans, with effect sizes of 0.8 g/dL and 1.7 g/dL, respectively [16,17]. None of these SNPs were significantly associated with Hb levels in Ethiopians (Table 1). Because we have complete or nearly complete power to detect a genotype-phenotype association in our Ethiopian samples (see Text S4 and Table S21), we infer that the SNPs associated with variation in Hb levels in the Tibetans do not make a contribution in Ethiopians.

Second, because the association signal in the Tibetans may be due to an untyped variant that is tagged by different SNPs in Tibetans and Ethiopians, we also considered all SNPs within 10kb of the *EPAS1* and the *EGLN1* genes and repeated this analysis applying a Bonferroni correction for the number of tests performed. None of the *EPAS1* or *EGLN1* SNPs was significantly associated with Hb levels in Ethiopians. Based on power analyses (Figure 3), we can exclude associated variants with the same effect size as in Tibetans if the MAF in Ethiopia is greater than 10% and 5%, respectively, for the *EPAS1* and *EGLN1* genes (see Figure S16 for Amhara and Oromo). Therefore, this more comprehensive analysis suggests that genes shown to contribute to variation in Hb levels in Tibetans either do not influence variation in the Ethiopian populations or if they do, their effect sizes are lower than those reported for the Tibetans.

Third, we considered all SNPs within 10kb of the candidate genes in the "Response to Hypoxia" Gene Ontology (GO) category (26 genes). None of these SNPs is significantly associated with Hb levels after multiple test correction ($p < 0.05/1309 = 3.81 \times 10^{-5}$). Because of the larger number of SNPs tested, this analysis has a relatively high multiple testing burden. Nonetheless, we find that we have greater than 80% power to detect a SNP significantly associated with Hb levels and effect size 0.8g/dL if its MAF is at least 20% and 100% power if its effect size is 1.7g/dL Hb (Figure 3C and Figure S16 for Oromo and Amhara). Therefore, we conclude that variation within the "Response to Hypoxia" GO category genes is unlikely to have the same effect on Hb levels in Ethiopians as that observed in Tibetans.

**Assessing genetic differences between high and low altitude populations**

A widely used family of approaches for the detection of beneficial alleles uses information about the haplotype structure around the selected site [34,35]. However, these approaches have adequate power only in the case of new advantageous alleles that were driven to high frequency by natural selection, i.e. ≥70% [34,36]. Because the largest allele frequency differences observed between HA and LA among Amhara or Oromo is less than 40%, these approaches are unlikely to be powerful in this setting. Therefore, to identify alleles that contribute to genetic adaptations to HA in Ethiopians, we used two complementary approaches that focus on the divergence of allele frequency between HA and LA populations. One of these approaches was previously used to successfully identify adaptive alleles in Tibetan highlanders [18].

The first approach is based on the population branch statistic (PBS) [18], which summarizes information about the allele frequency change ($PBS_{A\_BC}$) at a given locus in the history of a population (population A) since its divergence from two populations (population B and C) so that a high $PBS_{A\_BC}$ value represents a marked change in allele frequency on the branch leading to population A. This approach was previously used to detect advantageous alleles in Tibetans relative to Han Chinese and Europeans [18]. We tested for an excess of high allele frequency differentiation (i.e. large PBS values) on the branch leading to the Ethiopian populations in SNPs within candidate genes for response to hypoxia (i.e., genes within "Response to Hypoxia" GO category) relative to SNPs in all other genes (Table S22 lists all the population trios tested). Specifically, we calculated the ratio of the proportion of SNPs in hypoxia genes *vs*. the proportion of SNPs in all other genes in the top 0.5%, 1% and 5% of the distribution of PBS values and used bootstrap resampling to assess the significance of the excess of large PBS values. Although an excess was observed in most population trios (Table S22), this excess was rarely statistically significant; this finding suggests that levels of linkage disequilibrium in hypoxia genes tend to be higher than in other genes and that this feature may be a confounder in tests for selection [18]. A significant excess of large PBS values in hypoxia genes was observed only in the HA Amhara and the entire Amhara sample (Table S22 and Figures S17A-B), thus suggesting that HA Amhara indeed evolved genetic adaptations to hypoxic environments. When we extended this analysis of these same population trios to additional gene classifications (i.e. BioCarta, KEGG, Gene Ontology), we found significant enrichments for SNPs in gene sets related to cell cycle control, response to DNA damage and DNA repair (Table 2).

Interestingly, however, the SNPs with the highest PBS values are found in genes with a well-established role in pathogen response (Table S23-S24). More specifically, the SNPs with the highest PBS values are located within the major histocompatibility complex class II DR alpha (*HLA-DRA*). Moreover, the null allele (*FY*0*) at the Duffy blood group locus, which protects against *Plasmodium vivax* malaria [37] and predicts white blood cell and neutrophil counts [38], has the second highest value. Consistent with expectations based on the protective effects of the *FY*0* allele against malaria, its frequency is lower at HA compared to LA, where malaria is endemic (51.5% *vs.* 74.1%). Therefore, these results suggest that, in Ethiopian populations, differences in pathogen loads between LA and HA environments result in stronger selective pressures compared to differences in oxygen levels.

SNPs with large, even though not extreme PBS scores and lying within genes known to play an important role in hypoxia are of potential interest for follow up studies. These genes include: Cullin3 (*CUL3*), which potentiates *HIF-1* signaling [39], as well as adrenergic beta receptor kinase 1 (*ADRBK1*) [40], coronin actin binding protein 1B (*CORO1B*) [41], anti-silencing function 1 homolog B (*ASF1B*) [42] and MAPK-activated protein kinase MK2 (*MAPKAPK2*) [43], which are all down-regulated under hypoxia (Tables S23-24). None of those large PBS SNPs were significantly associated with Hb or $O_2$ sat (Table S25-S26), but a SNP within utrophin A *(UTRN)* - rs7753021 - reached nominal levels of significance with $O_2$ sat ($p$ = 0.005; Tables S26). *UTRN* expression correlates with oxidative capacity [44] and increases with chronic physical training [45]. Slow-twitch muscles, which are associated with endurance performance, have high levels of *UTRN* [46].

In a complementary analysis, we developed a multiple regression (MR) approach to identify SNPs that show high allele frequency differentiation in HA populations relative to predictions based on a large set of worldwide population samples. This method should also be able to predict allele frequencies appropriately in a situation where the target population is admixed. In this approach, we used allele frequency data from 61 LA populations (including the HGDP and several other populations) to predict the expected allele frequencies in the HA Amhara. We focused on the HA Amhara because they have lived at HA for a longer period of time and exhibit distinct patterns of Hb and $O_2$ sat levels compared to the Oromo (Figure 1). In addition, we omitted the LA Amhara in an attempt to reduce the effect of gene flow between altitudes, which could potentially reduce our power to detect adaptive divergence. We used all SNPs to estimate the best-fitting regression coefficients for each population: that is, these are the coefficients that generate the lowest mean square error in predicting the HA Amhara allele frequencies. The populations with the largest regression coefficients in the Amhara regression model are from geographically proximate populations in East Africa (Maasai, Luhya and LA Oromo) and from the Middle East and Southern Europe (see Figure S18). We reasoned that changes in allele frequencies due to high altitude adaptation would be detectable as departures (i.e. large residuals) from the predicted allele frequencies based on all other populations

As for the PBS analysis, we tested for an excess of SNPs with high allele frequency differentiation using the MR statistic for genes within "Response to Hypoxia" GO category relative to SNPs in all other genes and we used a bootstrap procedure to assess the significance of the observed excess. An excess was observed for all tail cut-offs, but only one reached statistical significance (Table 2). Other gene sets that showed a significant enrichment of SNPs with strong MR signals include chromosome organization and biogenesis, DNA repair, histone modification. These findings are consistent with the pattern observed in the PBS analysis, indicating that they are robust to the choice of populations used in the test (Table 2).

Among the SNPs with the largest MR scores, there are several SNPs in hypoxia genes, which may be of potential interest for follow up studies (Table S27). The SNP (rs12510722), which shows the $4^{th}$ highest MR scores, lies within the alcohol dehydrogenase 6 (*ADH6*) gene, whose expression is affected by pseudohypoxia [47]. The SNPs with the $8^{th}$ and the $15^{th}$ highest MR score (rs2660342 and rs2660343, respectively) are within 100kb from solute carrier family 30 member 9 (*SLC30A9*) and transmembrane protein 33 (*TMEM33*). *SLC30A9* is up-regulated by hypoxia [27] while *TMEM33* is down-regulated under hypoxia and up-regulated after knockdown of HIF1A [41].

**Assessing epigenetic differences between high and low altitude populations**

Methylation is an epigenetic modification that is known to play a crucial role in the cellular response to hypoxia [48]. Since HA adaptation could be in part maintained by methylation, we measured methylation levels at 27,578 CpG sites in 17 HA and 17 LA Amhara and 17 HA and 17 LA Oromo. CpG methylation levels were tested in DNA extracted from blood in the Amhara and from saliva in the Oromo. To avoid confounding due to differences in methylation across tissues, we performed the comparison across altitudes within each ethnic group.

In Oromo, four CpG sites reached significance after multiple test correction (p $<1.85 \times 10^{-6}$), but the closest genes are not known hypoxia candidate genes: apolipoprotein B mRNA editing enzyme catalytic polypeptide-like 3G (*APOBEC3G*), metallothionein 1G (*MT1G*), paired-like homeodomain 2 (*PITX2*) and olfactory receptor family 2 subfamily K member 2 (*OR2K2*) (Table S28). Interestingly, *APOBEC3G* codes for a well-established cellular antiviral protein and a specific inhibitor of human immunodeficiency virus-1 (HIV-1) infectivity [49,50]. The *MT1G* gene also has a role in HIV-1 infection because it upregulates MT1G expression in immature dendritic cells, which in turn facilitates the expansion of HIV-1 infection [51]. Although the prevalence of HIV was not surveyed in our fieldwork, HIV/AIDS is known to be a major health problem in Ethiopia [52,53]. While these functions for *APOBEC3G* and *MT1G* point to a role for methylation in defense against pathogens, *MT1G* also plays a role in the response to hypoxia as its promoter is induced by vascular endothelial growth factor (VEGF), which in turn contributes to the prosurvival and angiogenic functions of VEGF [54]. Likewise, expression of *PITX2* is required for normal hematopoiesis [55,56], raising the interesting scenario that methylation of this gene may influence beneficial phenotypes in the response to hypoxia. Some of the CpG sites with nominally significant (p$<6.7 \times 10^{-5}$) differences in methylation between HA and LA are close to genes that are differentially expressed in response to hypoxia; these genes include: toll-like receptor 6 (*TLR6*) [57], mif two 3 homolog 1 (*SUMO1*) [27]; phosphodiesterase 4A (*PDE4A*) [58] and human immunodeficiency virus type I enhancer binding protein 2 (*HIVEP2*) [42].

In Amhara, no CpG site showed a methylation difference that reached significance after multiple test correction (Table S29). However, we note that the 3rd most significant differentially methylated CpG site ($p = 8.03 \times 10^{-05}$) was closest to Glutathione-S-Transferase (*GSTP1*) whose expression is increased by prolonged hypoxia [59] and whose loss of expression correlates with methylation in prostate cancer [60]. Hypoxia also regulates the expression of genes close to other differentially methylated CpG sites in Amhara ($p<1.5 \times 10^{-3}$): protein regulator of cytokinesis 1 (*PRC1*) [42], protein tyrosine phosphatase receptor type O (*PTPRO*) [41], ring finger protein 146 (*RNF146*) [27] and Ras-related GTP binding D (*RRAGD*) [41].

Finally, no significant excess of methylation differences between LA and HA populations was observed at the genome-wide level in Oromo or Amhara (Figure S19) nor did we find a significant enrichment of methylation differences between LA and HA populations for gene sets defined by BioCarta or KEGG pathways and by Gene Ontology categories (data not shown).

### DISCUSSION

HA human populations across the world allow studying independent realizations of the adaptive process in response to the same selective pressure, i.e. hypoxia, thus providing an excellent opportunity to investigate how natural selection shapes the genetic architecture of adaptive traits. To make progress on these enduring questions, we have sampled two closely related ethnic groups in the Ethiopian highlands that include both HA and LA residents, thus allowing comparisons across altitudes within and between ethnic groups. Of these two groups, the Oromo have moved to HA only 500 years ago [61,62], thus making it unlikely that genetic adaptations evolved in this group. In contrast, the Amhara have a history of HA residence of at least 5ky and possibly as far as 70ky [24,25]. Because previously identified selection signals [63,64,65,66] occurred within a similar period of time, including HA adaptations [18], we conclude that enough time has elapsed since the Amhara moved to HA for genetic adaptations to have taken place. Consistent with this idea, we observe significant phenotypic differences between Amhara highlanders and the more recent HA residents, i.e. the Oromo. While HA Amhara are characterized by mildly elevated Hb levels (similar to Tibetans) and no or mildly reduced $O_2$ sat [13], the HA Oromo sample resembles acclimatized lowlanders with a response characterized by elevated Hb concentration and marked reduction in $O_2$ sat. Our data indicates that Amhara and Oromo are very similar at the genome-wide level, therefore, the observed phenotypic differences are likely due to the different histories of HA occupation. In addition to these phenotypic comparisons, our genomic analyses of these two ethnic groups resulted in several important observations that shed new light on the biology of HA adaptations and that are discussed in detail below.

First, in a GWAS of Hb levels in Amhara, we find a genome-wide significant signal of association as well as an excess of low p-values. In addition, the second most strongly associated SNP is found within the *RORA* gene, which belongs to HIF1 pathway and is, therefore, an excellent candidate gene for hypoxia response phenotypes. Additional strong associations were observed in other candidate hypoxia genes, such as *COL6A1, SLC30A9,* and *HGF*. We looked at the Amhara data of Scheinfeldt *et* *al* [21] to test for replication of the association signal at these genes. Two of them showed suggestive associations with Hb levels (p=0.06 and p=0.15 for *SLC30A9* and *RORA*, respectively), with β values (-0.60 and 1.31, respectively) consistent with ours (-0.67 and 0.92, respectively). It should be noted that the replication test was performed in only 21 Amhara samples in Scheinfeldt *et al* [21] for which age and BMI data were available and who had Hb levels within the normal range; thus, the lack of replication may well be due to the very low power of the replication sample. We note that, though variation in *EPAS1* and *EGLN1* has been consistently associated with Hb levels in Tibetans, no genome-wide significant association signal and no excess of low p-values were observed (see Figure S5 in Simonson et al [16,17]). While the signals we detected await replication, it is interesting to note that their effect sizes are as high as those found in Tibetans for SNPs in *EPAS1*, thus raising the interesting scenario that selection may have favored alleles with similar effect sizes on Hb levels even though the specific loci contributing to the trait are different.

Some interesting patterns are beginning to emerge with regard to the genetic contribution to variation in Hb levels and $O_2$ sat, the two phenotypes that have been most widely studied in highlander populations. No evidence of a genetic contribution to $O_2$ sat in Amhara, Oromo, and the combined Ethiopian sample could be detected (Figures S7-S15). This is true also in the Tibetans, even though segregation analysis detected a major $O_2$ sat locus, which is also associated with reproductive success [15]. Therefore, the data so far suggest that while genetic factors contribute to variation in Hb levels, their importance in $O_2$ sat is lower. This is consistent with studies in Tibetans and Andeans showing a markedly lower heritability for $O_2$ sat compared to Hb levels; indeed, the $O_2$ sat heritability in Andeans was not significantly different from zero [67,68,69]. More data, especially at the genome-wide level, are needed to elucidate the contribution of genetic factors to these two phenotypes.

Second, to compare the genetic bases of Hb variation with the Tibetans, we tested for an association between SNP genotypes and Hb levels within the Ethiopians. Although we had appropriate power, none of the SNPs within 10 kb of the *EPAS1* and *EGLN1* genes or the genes in the hypoxia pathway, including SNPs previously associated with Hb variation (and signatures of natural selection) in Tibetans, associated with Hb. Therefore, we can rule out that the SNPs and loci contributing to Hb variation and showing selection signals in Tibetans affect the same trait in Ethiopians even though Tibetans and Amhara have lower Hb levels compared to all other highlanders. Alternatively, if the same variant affects Hb levels in both populations, their effect sizes in Amhara must be markedly lower than those reported for the Tibetans.

Third, by using approaches based on allele frequency divergence, we find that outlier SNPs in the HA Amhara sample are not in hypoxia response genes, but in loci known to play a role in immune defense. These include variation in the *HLA-DRA* locus and the null allele at the Duffy blood group locus; none of these variants is associated with Hb levels or $O_2$ sat. Interestingly, malaria and schistosomiasis were prevalent in the LA, but not in the HA Amhara communities sampled in this study (Text S1), reflecting important differences in pathogens between HA *vs*. LA environments. Indeed, epidemiological studies in the areas near the LA sampling sites for the Amhara and Oromo reported malaria prevalence of 39.6% and as much as 25% of malaria morbidity is due to *P. vivax* [70,71,72]. Therefore, the immune defense variants with extreme frequency divergence represent excellent candidates as selection targets. These findings are important in several respects. First, they indicate that, because HA and LA habitats differ by multiple environmental stresses, signals of allele frequency divergence cannot be unambiguously attributed to hypoxia without additional information about the gene function or the specific phenotypic effects of the alleles. Second, they further corroborate that the Amhara populations are indeed adapted to spatially-varying selective pressures, despite likely high levels of gene flow between HA and LA communities. Third, the fact that variants in hypoxia response genes are not outliers in these analyses suggests that adaptations to pathogens and to hypoxia have a different genetic architecture or that the intensity of pathogen-related selective pressures is stronger than those due to hypoxia. Overall, these findings highlight the opportunities and challenges of ecological genomic studies and point to the power of approaches that use environmental information combined with phenotypic data collected in the field.

Although selection has not created extreme HA *vs*. LA frequency shifts in hypoxia genes, we find the SNPs within candidate genes for response to hypoxia *as a group* show an excess of allele frequency differentiation based on the PBS analysis performed in the Amhara (Table 2). This approach, however, requires specification of a set of three populations to be tested. When we used our MR approach, which uses information from all populations simultaneously, the response to hypoxia genes did not show a significant excess (Table 2). In contrast, both approaches detected a significant excess of allele frequency divergence in genes involved in cell cycle control, DNA repair, DNA damage, chromatin structure and modification, consistent with the known role of oxygen sensing in the regulation of cell proliferation [73]. Therefore, our findings raise the possibility that genetic variation in these pathways can contribute to adaptations to HA.

Fourth, though we did not observe a genome-wide excess of methylation differences between HA and LA samples, we found genome-wide significant signals in the Oromo in genes with a known function in pathogen defense or in the biology of hypoxia, i.e. VEGF signaling and hematopoiesis. Interestingly, no genome-wide significant differential methylation was observed in the Amhara; this may be due to the fact that DNA from different tissues was analyzed in the two ethnic groups. However, given the difference in the history of HA residence between Amhara and Oromo, it is also possible to speculate that epigenetic modifications play a role in the early phases of adaptations to new environments and that this role is replaced over time by adaptations at the genetic level.

It has been proposed that epigenetic modifications are important in ecological adaptations [74] and methylation is known to play a crucial role in the cellular response to hypoxia [48]. A previous study showed that gene expression differences observed between Moroccan populations were not explained by methylation differences on the tested 1,505 CpG sites [75]. Our study is more comprehensive having interrogated 27,578 CpG sites in a larger sample size. Though our results do not unambiguously point to a major role for methylation in the adaptations to HA, they suggest that further studies using DNA extracted from different tissues and including additional epigenetic modifications, in addition to methylation, are warranted.

In conclusion, studies of genetic variation in indigenous populations with long-time residence at HA are giving rise to a composite picture regarding the genes contributing to and the genetic architecture of HA adaptations. Clearly, different loci contribute to Hb levels in Ethiopians and Tibetans. Moreover, while some aspects (i.e. phenotypic effect sizes) of the genetic architecture of Hb levels may be similar, others (i.e. the allele frequency shifts due to selection) are different. Additional examples of environmental pressures that acted on different human populations include malarial endemia, low UVB radiation levels, and an adult diet rich in dairy products. Detailed genome-wide studies of parallel adaptations to these selective pressures are needed to elucidate the impact of natural selection on the genetic architecture of complex adaptive traits.

## MATERIALS AND METHODS
### Ethics statement

All participants in the study gave informed consent. The studies were approved by the Institutional Review Boards of Case Western Reserve University and of the University of Chicago and by the Ethiopian Science and Technology Council Ethics Review Board.

### Population samples

Samples were drawn from native residents of the Semien Mountains area of northern Ethiopia inhabited mainly by Christian Amhara and the Bale Mountains area of southern Ethiopia inhabited mainly by Muslim Oromo (also referred to as Galla, Boran and Gabbra). For each ethnic group, we sampled individuals at HA and LA. The LA samples were chosen to achieve the maximum altitude contrast, yet avoid confounding due to the presence of endemic malaria; all LA samples were from agropastoral people reporting the same ethnicity as the HA samples and no visits to altitudes above 2500 m in the past 6 months. For details on the sampled populations and their ecology, see Text S1.

DNA was extracted from blood samples provided by 192 Amhara individuals living at 3700 m in the Simien Mountains National Park or at 1200 m in the town of Zarima. Forty-seven of these individuals were sampled in 1995 and previously described [13,76], the remaining 145 individuals were sampled in a separate expedition in 2005. DNA was extracted from, saliva samples provided by 118 Oromo individuals and collected using Oragene DNA sample collection kits; 79 individuals lived at 4000 m in the Bale Mountains National Park while 39 individuals lived at 1560 m in the town of Melkibuta. The study participants were healthy (refer to Text S3 for details). Hb and $O_2$ sat of Hb were measured in all individuals. Hb was determined in duplicate using the cyanmethemoglobin technique (Hemocue Hemoglobinometer, Hemocue AB, Angelholm, Sweden), immediately after drawing a venous blood sample. $O_2$ sat was determined by pulse oximetry (Criticare Models 503 and SpO2) as the average of six readings taken 10 seconds apart.

### SNP Genotyping and imputation

The samples were genotyped using Hap650Yv3 (n=46), Human1M-duoV3 (n=112), and Human Omni-Quad1 (n=160) Illumina arrays at Southern California Genotyping Consortium. Nineteen samples with less than 93% genotype call rate were omitted from the analysis. We used the program RELPAIR 2.0. [77] to test for hidden relatedness in the samples by using 3 independent sets of 1800 autosomal and 200 X-linked SNPs; individuals with relationships closer than first cousins to any other individual in the sample were omitted from the analysis. After applying these filters, 260 unrelated samples remained:

102 HA Amhara, 60 LA Amhara, 63 HA Oromo and 35 LA Oromo. Principal component analysis (PCA) of the genotype data did not detect any major outlier in either population (see Text S2 and Figure S3. Due to the incomplete overlap between SNPs on the three genotyping arrays, genotypes for 1,819,369 HapMap3 SNPs were imputed by using the program IMPUTE2 [78] and the HapMap populations as a reference panel: Utah residents with Northern and Western European ancestry from the CEPH collection (CEU), Han Chinese in Beijing, China (CHB), Japanese in Tokyo, Japan (JPT), Luhya in Webuye, Kenya (LWK), Maasai in Kinyawa, Kenya (MKK), Toscani in Italy (TSI) and Yoruba in Ibadan, Nigeria (YRI) samples. A total of 1,297,134 autosomal SNPs with minor allele frequency (MAF) higher than 0 and imputation accuracy higher than 90% were used in the downstream analyses (Figure S20).

**Genotype-phenotype association**

Phenotype-genotype associations were tested at each SNP with MAF>0.1 by linear regression using the whole-genome association analysis toolset in PLINK [79]. To determine whether there was an excess of low p-values taking into account the large number of tests performed, the distribution of observed p-values from the linear regression tests were compared to a null distribution obtained by permuting 100 times the phenotype (and corresponding covariate variables – see below) value across individuals and running the same linear model. The results were visualized by means of quantile-quantile (QQ) plots and the 95% confidence interval (CI) was estimated by permutations. Gender, body mass index (BMI), altitude, ethnic group and year of sampling were used as covariates in the linear regression. As an alternative approach, we grouped the samples based on their gender, altitude, ethnic group and year of sampling and we quantile normalized the observed phenotypes within each group; these phenotypes were then pooled across groups before testing for an association with genotype by linear regression. This latter approach preserves information about the individual ranks without introducing a bias due to the effects of covariates. However, it does not preserve information about the phenotype values, thus leading to a reduction in power. The results of these two approaches correlated highly ($r^2 > 0.92$); therefore, only the results for the linear regression with the covariates are shown.

**Population genetics analyses**

The population branch statistic ($PBS_{A\_BC}$) was used to summarize the amount of allele frequency change in the history of population A since its divergence from two related populations, B and C [18]. In a complementary approach based on multiple regression (MR), we used the allele frequencies for the LA Oromo and world-wide population samples (i.e., Human Genome Diversity Project (HGDP) panel populations [80] and 4 HapMap Phase III populations -LWK, MKK, TSI and Gujarati Indians in Houston, Texas (GIH) - (www.hapmap.org)) to detect SNPs whose HA Amhara frequencies deviate most from estimated frequencies. Briefly, denoting the observed allele frequencies in the *p* populations by $X_1, X_2, \ldots, X_p$ and the expected allele frequency within the HA Amhara by *y*, we can predict the allele frequency for the $i^{th}$ SNP using the linear model $y_i = \beta_0 + \beta_1 X_{i1} + \beta_2 X_{i2} + \ldots + \beta_p X_{ip} + \varepsilon_i$. Using the data from all genotyped SNPs that overlap among the *p* populations and applying multivariate linear regression, we found $b_0, b_1, b_2, \ldots, b_p$, which are estimates of parameters $\beta_0, \beta_1, \beta_2, \ldots, \beta_p$ and estimated the magnitude of the residual for the $i^{th}$ SNP as $|\varepsilon_i| = |y_i - (b_0 + b_1X_1 + b_2X_2 + \ldots + b_pX_p)|$. We refer to this residual as the MR score.

To test for an enrichment of allele frequency differentiation in candidate hypoxia genes, we focused on the 28 genes belonging to the "Response to hypoxia" category in the Gene Ontology database (GO:0001666). We then compared the proportion of SNPs within 10 kb of each gene in this category to that of SNPs within 10 kb of all other genes in the tail of the PBS and MR score distributions. Given the arbitrary nature of choosing a single cutoff for the tail of the distribution, we set three cutoffs (5%, 1% and 0.5%). In other words, we looked at the top 5%, 1% and 0.5% of all PBS and MR values and asked whether there is an enrichment of hypoxia gene SNPs relative to all other genic SNPs for each tail cut-off. A value of 1 represents no excess and a value greater than 1 represents enrichment in the tail of the distribution. SNPs are likely to cluster along the genome due to linkage disequilibrium, thus reducing the number of independent signals contributing to an observed enrichment. To account for this possibility, we found the confidence interval for the enrichment using a bootstrap approach described in Hancock *et al* [81]. An enrichment of hypoxia SNPs was considered significant (with a one-tailed test) if at least 95% of the 1000 bootstrap replicates were enriched (*i.e.*, had a ratio above 1).

**Methylation**

Methylation levels were measured at 27,578 CpG sites in 17 HA and 17 LA Amhara and Oromo DNA samples, for a total of 68 individuals, using six Infinium HumanMethylation27 arrays at Southern California Genotyping Consortium. Two LA Amhara sample data were discarded due to low data quality. In each ethnic group, the HA methylation level of each CpG site was compared to corresponding LA levels using the following linear model correcting for age, gender, and the methylation array: lm (% methylation ~ altitude + gender + age + [array]). Two comparisons were performed: (1) HA *vs.* LA Amhara and (2) HA *vs.* LA Oromo. As for the genotype-phenotype association, excess of differential methylation between HA and LA was tested by comparing the observed p-value distribution to the null distribution obtained by permuting 100 times the altitude label across individuals and running the same linear model. Permutation based 95% CI was estimated.

ACKNOWLEDGMENTS

The authors are grateful to the study participants and communities for their hospitality and cooperation in their field work. They thank the officials of the Ethiopian Science and Technology Commission, the Amhara and the Oromia Regional Governments, and the staffs of the Semien Mountains and Bale Mountains National Parks for permissions and local arrangements. The Frankfurt Zoological Society generously allowed use of the Wolf Research Camp in the Bale Mountains. Daniel Tessema, laboratory technician, and Gezahegne Fentahun, M.D. of Addis Ababa University worked in both field sites. The authors also thank members of the Di Rienzo lab, John Marioni, Jordana T. Bell and Lucy Godley for helpful discussions during the course of this project as well as Sarah Tishkoff and Laura Scheinfeldt for sharing data on genotype-phenotype association in their samples.

Table 1. Association test of Tibetan *EGLN1* and *EPAS1* SNPs within Amhara, Oromo and combined Ethiopians.

| | | Tibetans[1] | | Amhara | | | | Oromo | | | | Ethiopians | | | |
|---|---|---|---|---|---|---|---|---|---|---|---|---|---|---|---|
| SNP | Gene | $\beta^2$ | MAF | $\beta^2$ | p | Power[3] | MAF | $\beta^2$ | p | Power[3] | MAF | $\beta^2$ | p | Power[3] |
| rs961154 | *EGLN1* | 1.70 | 0.39 | 0.08 | 0.58 | 100 | 0.41 | 0.13 | 0.57 | 100 | 0.40 | 0.12 | 0.34 | 100 |
| rs2790859 | *EGLN1* | 1.70 | 0.39 | 0.08 | 0.58 | 100 | 0.41 | 0.13 | 0.57 | 100 | 0.40 | 0.12 | 0.34 | 100 |
| rs1992846 | *EPAS1* | 0.84 | 0.36 | 0.00 | 0.99 | 100 | 0.42 | -0.39 | 0.13 | 100 | 0.38 | -0.14 | 0.29 | 100 |
| rs7594278 | *EPAS1* | 0.52 | 0.31 | 0.06 | 0.71 | 98 | 0.24 | -0.08 | 0.77 | 80 | 0.28 | -0.01 | 0.93 | 100 |
| rs6544887 | *EPAS1* | 0.79 | 0.34 | -0.05 | 0.74 | 100 | 0.34 | -0.11 | 0.63 | 100 | 0.34 | -0.07 | 0.58 | 100 |
| rs17035010 | *EPAS1* | 0.84 | 0.39 | -0.07 | 0.66 | 100 | 0.30 | -0.02 | 0.93 | 100 | 0.36 | -0.04 | 0.80 | 100 |
| rs3768729 | *EPAS1* | 0.80 | 0.44 | 0.10 | 0.52 | 100 | 0.44 | -0.08 | 0.74 | 100 | 0.44 | 0.04 | 0.77 | 100 |
| rs7583554 | *EPAS1* | 0.94 | 0.46 | 0.10 | 0.52 | 100 | 0.41 | 0.16 | 0.53 | 100 | 0.44 | 0.13 | 0.34 | 100 |
| rs7583088 | *EPAS1* | 0.92 | 0.30 | 0.10 | 0.51 | 100 | 0.22 | -0.07 | 0.79 | 100 | 0.27 | 0.02 | 0.87 | 100 |
| rs11678465 | *EPAS1* | 0.85 | 0.30 | 0.10 | 0.51 | 100 | 0.22 | -0.06 | 0.83 | 100 | 0.27 | 0.03 | 0.84 | 100 |
| rs6712143 | *EPAS1* | 0.94 | 0.40 | 0.16 | 0.28 | 100 | 0.32 | -0.09 | 0.74 | 100 | 0.37 | 0.09 | 0.52 | 100 |
| rs4953342 | *EPAS1* | 0.90 | 0.21 | -0.12 | 0.50 | 100 | 0.12 | -0.19 | 0.57 | 99 | 0.18 | -0.16 | 0.33 | 100 |
| rs2121266 | *EPAS1* | 1.02 | 0.38 | 0.14 | 0.34 | 100 | 0.30 | -0.20 | 0.43 | 100 | 0.35 | 0.01 | 0.95 | 100 |
| rs9973653 | *EPAS1* | 0.52 | 0.35 | 0.01 | 0.95 | 100 | 0.30 | -0.18 | 0.47 | 100 | 0.33 | -0.05 | 0.70 | 100 |
| rs1374749 | *EPAS1* | 0.88 | 0.47 | -0.18 | 0.18 | 100 | 0.49 | -0.29 | 0.20 | 100 | 0.48 | -0.22 | 0.07 | 100 |
| rs4953353 | *EPAS1* | 0.97 | 0.31 | -0.01 | 0.93 | 100 | 0.31 | -0.19 | 0.46 | 100 | 0.31 | -0.08 | 0.54 | 100 |
| rs6756667 | *EPAS1* | 0.93 | 0.37 | 0.02 | 0.89 | 100 | 0.32 | 0.04 | 0.86 | 100 | 0.35 | 0.03 | 0.82 | 100 |
| rs7571218 | *EPAS1* | 0.71 | 0.50 | 0.17 | 0.22 | 100 | 0.49 | -0.19 | 0.46 | 100 | 0.50 | 0.06 | 0.64 | 100 |

[1] Genotype-phenotype association β coefficients for *EGLN1* were obtained from Simonson *et al* [16] while those for *EPAS1* were obtained from Beall *et al* [17].
[2] β indicates the observed linear coefficient for the relationship between SNP genotype and Hb levels.
[3] Power refers to the probability of detecting a significant association (p<0.05) between SNP genotype and Hb level given the MAF and the sample size in the Ethiopian populations assuming that the β coefficient is as high or higher as that observed in Tibetans.

Table 2. Biological pathways for which a significant excess of genic relative to all other genic SNPs are observed for all three tail cut-offs of the Amhara PBS, Amhara High PBS, or Amhara High MR distributions, respectively. * and ** denote support from ≥ 95% and 99% of bootstrap replicates, respectively.

| Classification | Gene Set | Amhara PBS tail cut-off | | | Amhara High PBS tail cut-off | | | Amhara High MR tail cut-off | | |
|---|---|---|---|---|---|---|---|---|---|---|
| | | 0.01 | 0.01 | 0.05 | 0.01 | 0.01 | 0.05 | 0.01 | 0.01 | 0.05 |
| GO BP | RESPONSE TO HYPOXIA | 3.40* | 3.41** | 1.84* | 3.31 | 2.9 | 1.92* | 2.58 | 2.2 | 1.75** |
| GO BP | CELL CYCLE CHECKPOINT | 5.72** | 3.40** | 1.83** | 0.48 | 0.96 | 0.96 | 0.68 | 0.36 | 0.9 |
| GO BP | DNA INTEGRITY CHECKPOINT | 6.30** | 3.86** | 2.00* | 0 | 0.82 | 0.94 | 0 | 0.07 | 0.85 |
| GO BP | DNA DAMAGE RESPONSE AND SIGNAL TRANSDUCTION | 4.83** | 2.91** | 1.59* | 2.02 | 1.79 | 1.07 | 1.12 | 1.26 | 1.18 |
| GO BP | DNA DAMAGE CHECKPOINT | 6.47** | 4.00** | 1.97** | 0 | 0.9 | 0.94 | 0 | 0.09 | 0.81 |
| GO BP | SENSORY ORGAN DEVELOPMENT | 10.29* | 6.08* | 2.31* | 2.72 | 1.36 | 0.87 | 1.9 | 0.95 | 1.04 |
| BioCarta | SALMONELLA PATHWAY | 15.61* | 9.63** | 2.89* | 0 | 0 | 0.91 | 0 | 0 | 0.31 |
| KEGG | CELL CYCLE | 1.1 | 1.1 | 1.39* | 2.93** | 2.70** | 1.60** | 4.39* | 3.35* | 1.34 |
| KEGG | CHRONIC MYELOID LEUKEMIA | 1.23 | 1.24 | 1.34 | 2.72* | 2.02** | 1.44* | 2.67* | 2.09* | 1.1 |
| GO BP | CELLULAR RESPIRATION | 0 | 1.01 | 0.69 | 4.55* | 3.07* | 2.20** | 1.11 | 2.25 | 2.42* |
| GO BP | AXON GUIDANCE | 0.35 | 0.62 | 0.8 | 1.97* | 1.72** | 1.28* | 0.47 | 0.95 | 0.93 |
| GO BP | CHROMOSOME ORGANIZATION AND BIOGENESIS | 0.69 | 0.77 | 0.87 | 1.46 | 2.22* | 1.46* | 3.93** | 3.17** | 1.99** |
| GO BP | DNA REPAIR | 1.76 | 1.2 | 1.22 | 1.83 | 1.6 | 1.36* | 2.23* | 2.41** | 1.64** |
| GO BP | ESTABLISHMENT AND OR MAINTENANCE OF CHROMATIN ARCHITECTURE | 0.73 | 0.91 | 0.83 | 1.78 | 2.09 | 1.39* | 3.31* | 2.57* | 1.74** |
| GO BP | HISTONE MODIFICATION | 1.76 | 1.76 | 0.95 | 1.58 | 2.28 | 1.26 | 5.58* | 4.57* | 2.34** |
| GO BP | CHROMATIN MODIFICATION | 0.96 | 0.97 | 0.84 | 2.4 | 2.4 | 1.37 | 4.29* | 3.34* | 1.75* |
| GO BP | ALCOHOL METABOLIC PROCESS | 1.44 | 1.18 | 0.93 | 2.1 | 1.84 | 1.15 | 3.48** | 2.51* | 1.53* |

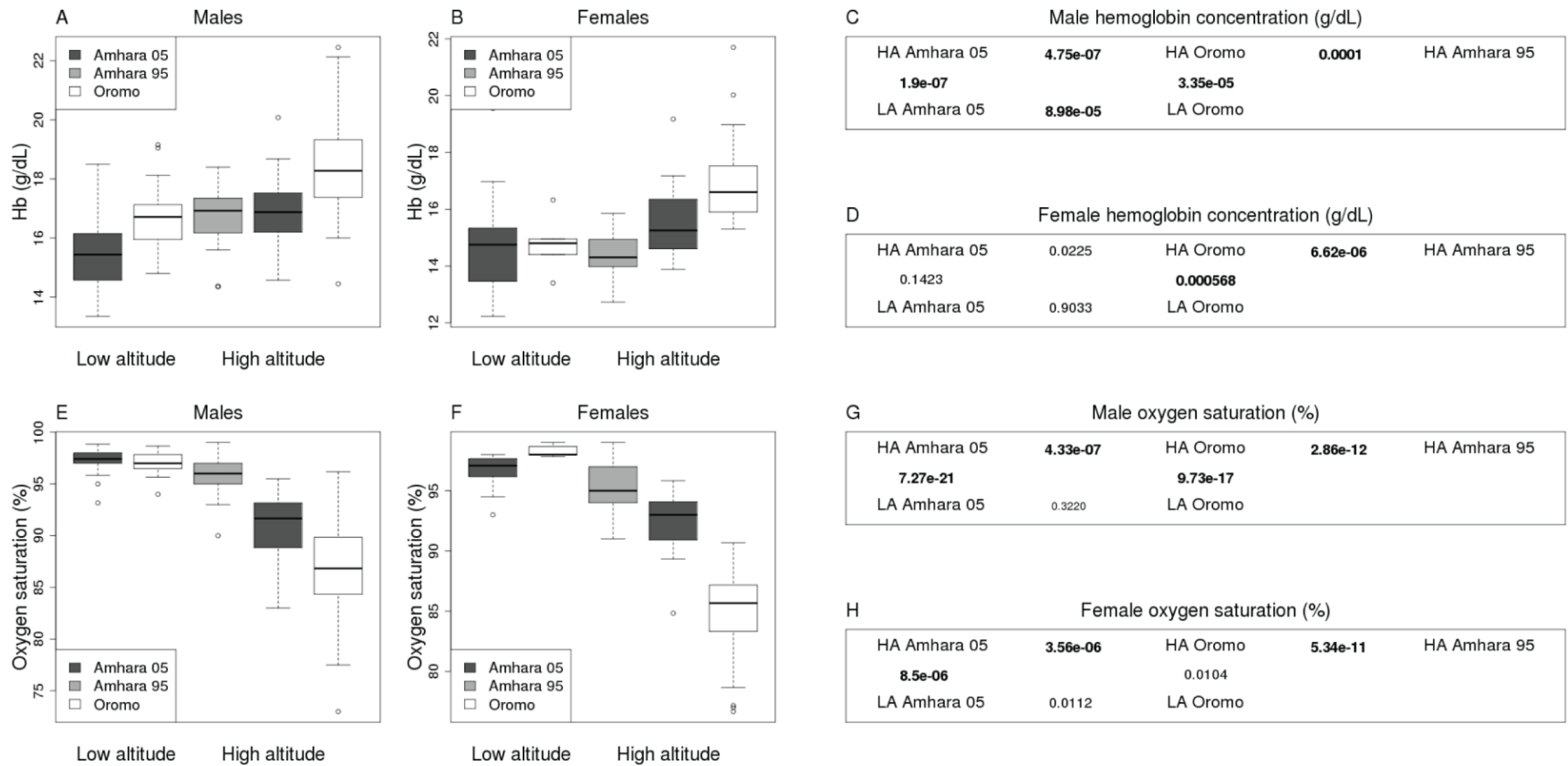

Figure 1. **Hemoglobin and oxygen saturation measurements**. Box plots describe variation in the Amhara 05 (dark grey boxes), Amhara 95 (grey boxes) and Oromo (white boxes) for Hb concentration (g/dL) among males (A) and females (B) and for $O_2$ sat also among males (E) and females (F). Box plots show the median (horizontal line), interquartile range (box), and range (whiskers), except the extreme values represented by circles. Statistically significant differences after multiple test correction between groups (unpaired two-sided two-sample t-test) are bolded in C, D, G, and H.

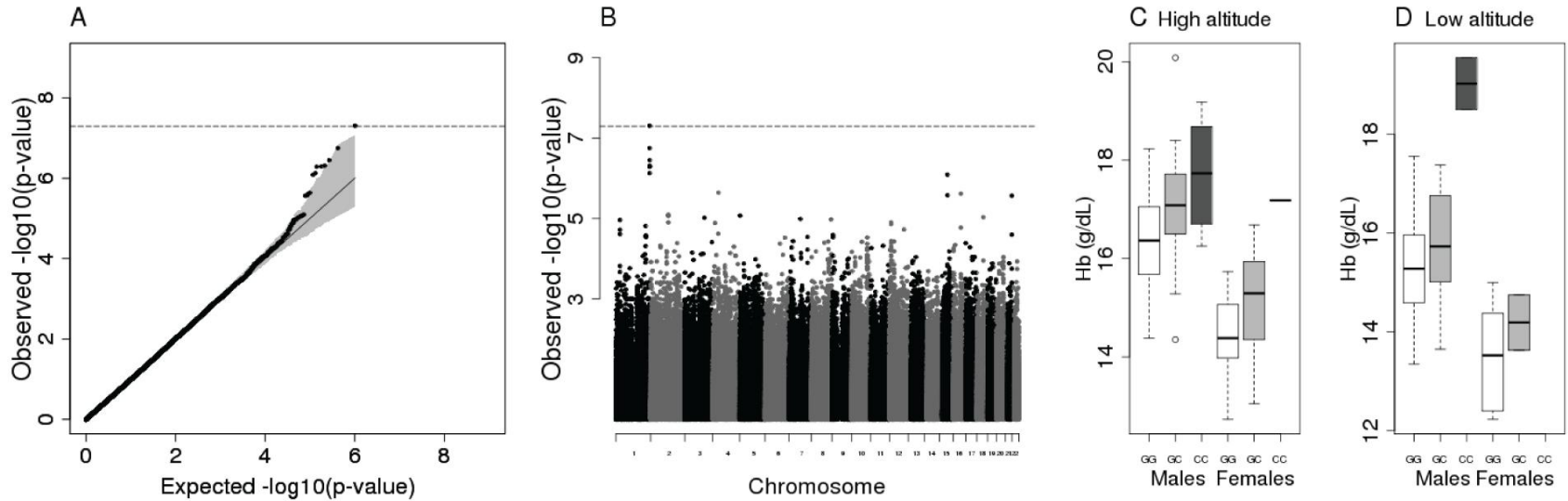

Figure 2. **Hemoglobin association test within Amhara**. The QQplot represents the excess of strong association with Hb among Amhara individuals (A). The observed –log10 p-value distribution is ranked from smallest to largest and plotted (y-axis) against the expected –log10 p-value (y-axis) in black. The grey area indicates the 95% confidence interval (see methods). Genome-wide (GW) significance level (after multiple test correction) is indicated by the dashed line. The Manhattan plot (B) shows the GW significance achieved by a set of high-LD SNPs in chromosome 1. The box plots describe the correlation between hemoglobin levels and the 3 genotypes of the top and GW significant SNP (rs10803083) among high (C) and low altitude (D) Amhara.

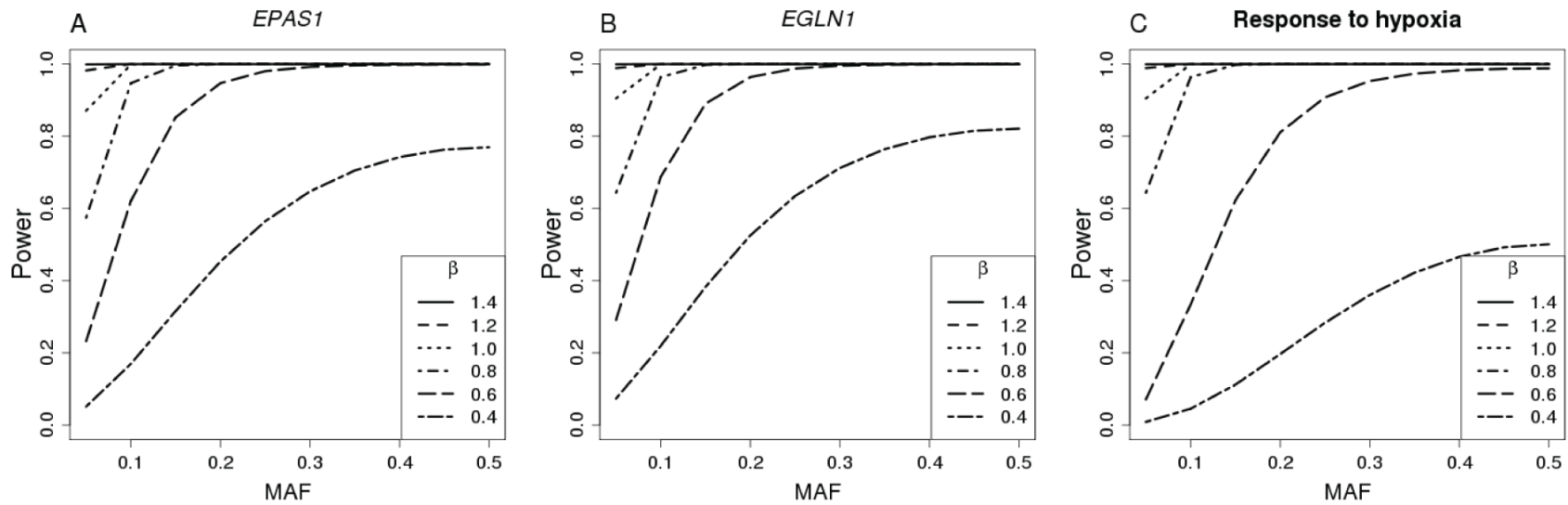

Figure 3. **Power plots**. The effect of *β* and MAF on the power of association tests based on the Ethiopian sample size (corrected for the number of SNPs tested within 10kb from gene) is illustrated for *EPAS1* (A, 72 SNPs)), *EGLN1* (B, 38 SNPs) and any gene within the Response to Hypoxia gene ontology category (C, 1309 SNPs).

# SUPPORTING TEXT

Text S1. **Sampled populations and their ecology**

    The Simien Plateau in Northwest Ethiopia is cut by streams and rivers and has cliffs and land areas sloping down to canyons. The high altitude (HA) Amhara are agropastoralists living in a temperate Afro-alpine ecosystem in the Simien Mountains National Park at altitudes ranging from 3500-4100 meters (m). Altitudes above 2500m on the East African Plateau have been inhabited for at least 5 thousand years (ky) and altitudes around 2300-2400m for more than 70ky [24,25]. Linking historical and modern ethnic groups in Ethiopia with prehistoric sites is not feasible with current knowledge. This paper considers that 5ky is a reasonable lower estimate for human habitation of this area. The Bale Plateau in Southeast Ethiopia is a flat mesa with sharp relief down to the lowlands. The HA Oromo are pastoralists herding cattle, sheep and goats and living in a temperate Afro-alpine ecosystem in the Bale Mountains National park and reside on the Sanetti Plateau at 4000-4100m. The HA areas of the Bale Plateau have been inhabited by Oromo since the early 1500s according to historical records [22,23]. The two plateaus are about 500 miles apart and the intervening terrain is not a continuous plateau as the Tibetan or Andean, instead it is a mosaic of terrain generally above 1500m and punctuated by lowlands including the Great Rift Valley. The Amhara language is sub-classified as Ethio-Semitic and Oromo as Eastern Cushitic in the Afroasiatic language group [82] (http://www.ethnologue.com/ **- accessed February 27, 2012).**

    Data were collected in field laboratories in the four communities. The Amhara samples provided data during December 1995, April 2005, December 2005 and January 2006. In 2005-6, they were studied at a field laboratory established in a park guard camp at 3700m. In 1995, they were studied at an altitude of 3530m [13,83]. Ambient conditions at the time of morning calibrations in 2005 and 2006 averaged 498 mmHg, 6 °C and 36% relative humidity at the HA site, 659 mmHg, 22 °C, and 36% relative humidity at the LA site. The Oromo samples provided data during December 2005 and January 2006 when ambient conditions averaged 469mm Hg and 3°C at the HA site and 635 mmHg, 15°C at the LA site. They were studied at a field laboratory established at the Wolf Research Camp at 4000m and in Melkibuta town at 1500m.

    The Amhara reported occupations as farmers and, at LA, farmers and traders. The HA sample reported eating a diet based on barley injera (a flat crepe-like bread made of fermented dough) with sauce (wot) of potato or beans. The LA Amhara community was exposed to malaria and schistosomiasis and had a high prevalence of visible goiter and iron deficiency. Such individuals were excluded from the sample. The HA Oromo sample reported eating a diet of teff injera with potato or bean wot and some meat and dairy products. The LA Oromo were farmers, traders, and government officials such as teachers; they were also exposed to malaria.

Text S2. **The genetic structure of the Amhara and Oromo populations**

In order to investigate the genetic relationship between Amhara and Oromo and other populations, we used genotype data in the two Ethiopian populations for SNPs that had previously been typed also in a set of 60 worldwide populations, namely 52 Human Genome Diversity Project (HGDP) populations, 4 HapMap populations (Gujarati Indians in Houston, Texas (GIH); Maasai in Kinyawa, Kenya (MKK); Luhya in Webuye, Kenya (LWK); Toscani in Italia (TSI)), and 4 populations from Hancock et al [84]. A neighbor-joining tree shows the relationships of the Ethiopian populations to the worldwide populations (Figure S1). As expected, Oromo and Amhara cluster closely together in the tree and occupy an intermediate position between African (with the exception of the Mozabites) and non-African populations. This position is further supported by the Bayesian clustering analysis performed using the program STRUCTURE [85]. In this analysis, 3 different sets of 57652 SNPs were used to infer the ancestral composition of each population assuming 7 ancestral groups. The STRUCTURE plots clearly show that Ethiopian populations share ancestral components with sub-Saharan African and Middle Eastern populations falling in the middle of the ancestry gradient between these two groups of populations (Figure S2. We also calculated the haplotype diversity and compared it to that observed in the worldwide populations. Interestingly, the Oromo (0.822) and Amhara (0.810) haplotype diversity values are as high as or higher than the highest values [80] observed in the HGDP, i.e. Bantu (0.818), Biaka Pygmies (0.815), Yoruba (0.815) and Mandenka (0.807); this is true regardless of altitude (0.798 for HA Amhara; 0.803 for LA Amhara, 0.813 for HA Oromo, and 0.813 for LA Oromo).

To look more closely at the relationship between the Amhara and the Oromo, we performed principal component analysis (PCA) using the genotype data from 13,000 random autosomal SNPs for all Ethiopian individuals. In this analysis, PC1, which explains 0.78% of the variance, separates the majority of the Amhara from the majority of the Oromo individuals; in addition, within each ethnic group, the HA and LA individuals tend to cluster separately (Figure S3. Accordingly, the mean $F_{ST}$ between LA Amhara and Oromo is very low (0.0098), though it is higher than the mean $F_{ST}$ between LA and HA populations within the same ethnic group (HA *vs.* LA Amhara: $F_{ST}$=0.0047; HA *vs.* LA Oromo high $F_{ST}$=0.0074). The STRUCTURE analysis did not reveal detectable differences between the HA and LA populations ($10^6$ iterations; following a burn-in period of 30,000 iterations; k=3) (Figure S4. However, when we compared the $F_{ST}$ distribution between HA and LA populations (Amhara or Oromo) to the distribution obtained by permuting the altitude labels (*i.e.* under random mating for HA and LA populations within each ethnic group) we detected a significant excess of high $F_{ST}$ values, indicating that some population structure between altitudes does exist (Figure S5). In order to test for an excess of allele frequency divergence between populations, the observed distribution of $F_{ST}$ was compared to a null distribution obtained by permuting the population labels 100 times, and for each comparison, a different permuted $F_{ST}$ distribution was created. Permutations were used to estimate the 95% confidence interval.

Text S3. **Phenotypic variation in Amhara and Oromo**

260 unrelated people provided the phenotype and genotype data presented here. HA Amhara were sampled in 1995 (n=47) and 2005 (n=65) and the LA Amhara in 2005-6 (n=59). All Oromo were sampled in 2007 (n=88). These unrelated people were identified from a larger sample of partially related individuals.

In the Semien Mountain area, two models of Criticare oximeter were used, the Criticare Model 503 was used for 152 people and the SpO2 was used for 150 people. Twenty-five people were measured using both instruments and the average difference (SpO2 minus 503) was =0.031 $\pm$ 2 (SD) %. One observation was just outside the limits of agreement of the two models ($\pm$ 2 SDs). The Model 503 values were used as the $O_2$ sat measured by Criticare oximeter for the 127 who did not have SpO2 measurements.

Table S1 summarizes height, weight, body mass index (BMI), and pulse. Considering altitude differences, HA Amhara native men and women were shorter, lighter and had lower BMI than their lowland Amhara counterparts. HA Oromo native men had higher BMI than their lowland Oromo counterparts although there were no significant differences in height or weight. HA Oromo native women were shorter but did not differ in weight or BMI from their lowland counterparts. Considering ethnic differences within altitude zones, HA Amhara men and women were shorter, lighter and had lower BMI than Oromo. LA Amhara men were heavier and had higher BMIs. Amhara lowland women did not differ from Oromo lowland women in these characteristics although there was a trend toward shorter height.

Table S2 summarizes the three HA adaptation phenotypes, hemoglobin (Hb) concentration, percent of oxygen saturation ($O_2$ sat) of Hb and calculated arterial oxygen content (AOC). Arterial oxygen content in ml $O_2$/dL was calculated as 1.39*(Hb*$O_2$ sat)/100 [86]. Phenotypes were assessed using the same equipment and protocols in all areas [13]. To remove many possible confounding factors that could have added spurious variation to the phenotypes, the analyses were confined to healthy people (based on self-report and a review of systems by an Ethiopian physician), who were free of respiratory symptoms, not hypertensive, not pregnant, had not delivered an infant in the past year and had not visited at another altitude (below 2500m for highlanders or above 2500m for lowlanders) in the past six months. Samples collected in 2005, 2006 and 2007 were also screened for normal lung function and for infection with all four species of human-associated malaria [87]. Furthermore, because poor iron status could limit the Hb response to HA hypoxia, the analyses were limited to those with normal iron status. Iron deficiency in the 1995 sample was identified on the bases of zinc erythrocyte protoporphyrin $\geq$ 70 gm/dL, plasma ferritin concentration <12 ng/ml or transferrin receptor concentration <8.3 g/L [13]. Iron deficiency in the samples collected later was identified on the bases of body iron stores calculated using the log of the ratio of transferrin receptor to ferritin concentration [88]. If an individual were missing data for any of these variables, then his or her Hb concentration data were not analyzed for association with genetic variants. Stringent criteria for phenotypes improve the likelihood of finding phenotypic variation associated with health, normal genetic variation and add confidence to findings of a lack of association.

With respect to altitude differences, Hb was significantly higher among all four HA samples compared with the age-sex groups at LA but size of the effect was smaller among the Amhara (Table S2 and Figure S6. Similarly, the percent of $O_2$ sat was lower among all four HA samples compared with the age-sex groups at LA, but the size of the effect was smaller among the Amhara (Table S2 and Figure S6. The altitude sub-samples did not differ from one another in calculated AOC. With respect to ethnic group differences within altitude zones, Amhara had lower Hb concentrations at both HA and LA (Table S2 and Figure S6 show that the differences were more than one gm/dL of Hb and more than one standard deviation). In contrast, Amhara had higher percent of $O_2$ sat at HA than their Oromo counterparts. The result was similar for AOC among highlanders but a trend toward lower AOC among LA Amhara males that was statistically significant among Amhara females was observed.

In summary, there were altitude and ethnic group differences in phenotypes with the Oromo showing a larger altitude response than the Amhara.

Text S4. **Comparing the genetic architecture of Hb levels between Ethiopian and Tibetans**

To compare the genetic architecture of Hb levels between Ethiopians and Tibetans [16,17,18], we performed association tests and power analyses at three levels: 1) the SNPs in *EPAS1* and *EGLN1* associated with the trait in Tibetans, 2) all the SNPs within the same genes, and 3) all the SNPs within all the genes in the same pathway, i.e. response to hypoxia.

The *EPAS1* and *EGLN1* SNPs that were previously associated with variation in Hb levels in Tibetans have effect sizes of 0.8 g/dL and 1.7 g/dL, respectively [16,17]. Of these SNPs, we considered those that were genotyped or imputed with greater than 90% accuracy in our Ethiopian data. Because the lack of association in the Ethiopians could be due to incomplete power, we calculated the probability of observing a significant association between SNP genotype and Hb levels in our data, assuming that the β coefficient for each SNP is as high as or higher than that observed in Tibetans and using the corresponding MAF in the Ethiopians. As shown in Table 1, we have complete or nearly complete power to detect a genotype-phenotype association in our Ethiopian samples. This suggests that either the SNPs associated with variation in Hb levels in the Tibetans do not make a contribution in Ethiopians or if they do their phenotypic effect is smaller in Ethiopia. However, given their corresponding MAFs in Amhara, Oromo and Ethiopian, we have nearly complete power to detect a significant association even if the phenotypic effect was half as large as that observed in the Tibetan studies (Table S21).

Next, we considered all SNPs within 10kb of the *EPAS1* gene and repeated this analysis applying a Bonferroni correction for the number of tests performed (*i.e*., 72 SNPs). We find that we have greater than 95% power to detect a SNP with an effect size of at least 0.8 g/dL Hb concentration if the MAF in Ethiopia (Figure 3A) is greater than 10% (see Figure S16 for Amhara and Oromo). When we performed the same analysis for SNPs within 10kb of the *EGLN1* gene (38 SNPs), we find that we had 100% power to detect SNPs with effect size of at least 1.7 g/dL Hb concentration if the SNP MAF in Ethiopia is greater than 5% (Figure 3B and Figure S16 for Oromo and Amhara). Because none of the *EPAS1* or *EGLN1* SNPs was significantly associated with Hb levels in Ethiopians after multiple test correction, this analysis suggests that genes shown to contribute to variation in Hb levels in Tibetans either do not influence variation in the Ethiopian populations or if they do, their effect sizes are lower than those reported for the Tibetans.

Finally, we considered the SNPs within 10kb of the candidate genes in the "Response to Hypoxia" GO category (26 genes). Because of the larger number of SNPs tested, this analysis has a relatively high multiple testing burden. Nonetheless, we find that we have greater than 80% power to detect a SNP significantly associated with Hb levels and effect size 0.8g/dL if its MAF is at least 20% and 100% power if its effect size is 1.7g/dL Hb (Figure 3C and Figure S16 for Oromo and Amhara).

**SUPPORTING FIGURES**

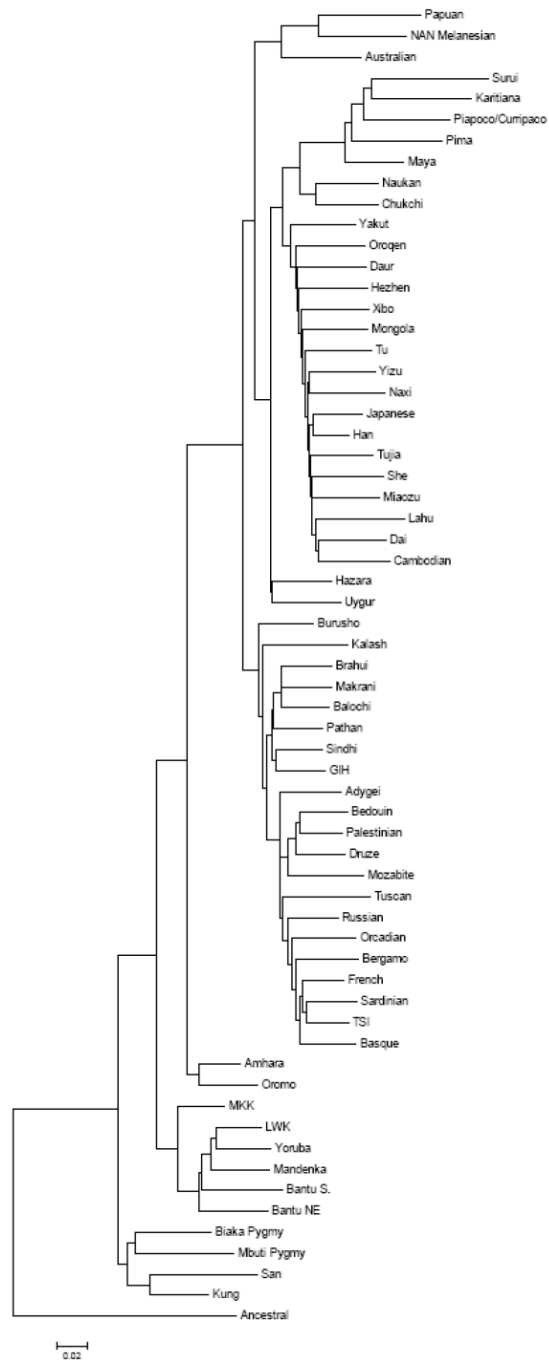

Figure S1. **F$_{ST}$-based neighbor-joining tree** showing the relationships of the Ethiopians to the worldwide populations. Oromo and Amhara cluster closely together in the tree and occupy an intermediate position between African (with the exception of the Mozabites) and non-African populations. An ancestral population fixed for ancestral alleles at all SNPs was used.

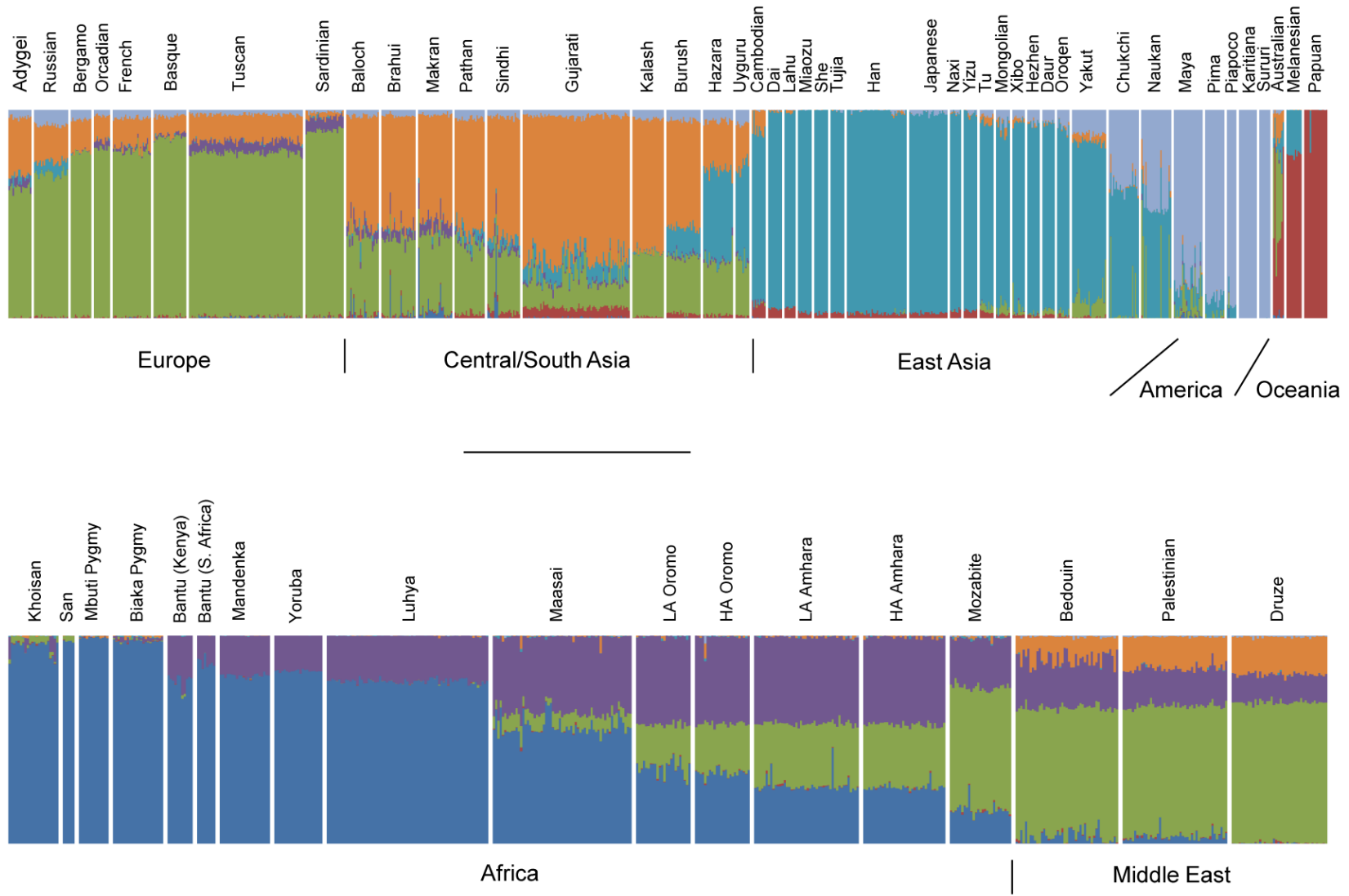

Figure S2. **Worldwide STRUCTURE plot**. Each vertical line represents an individual, and the colors comprising each line correspond to the inferred proportion of ancestry from seven ancestral populations using 57652 random autosomal SNPs.

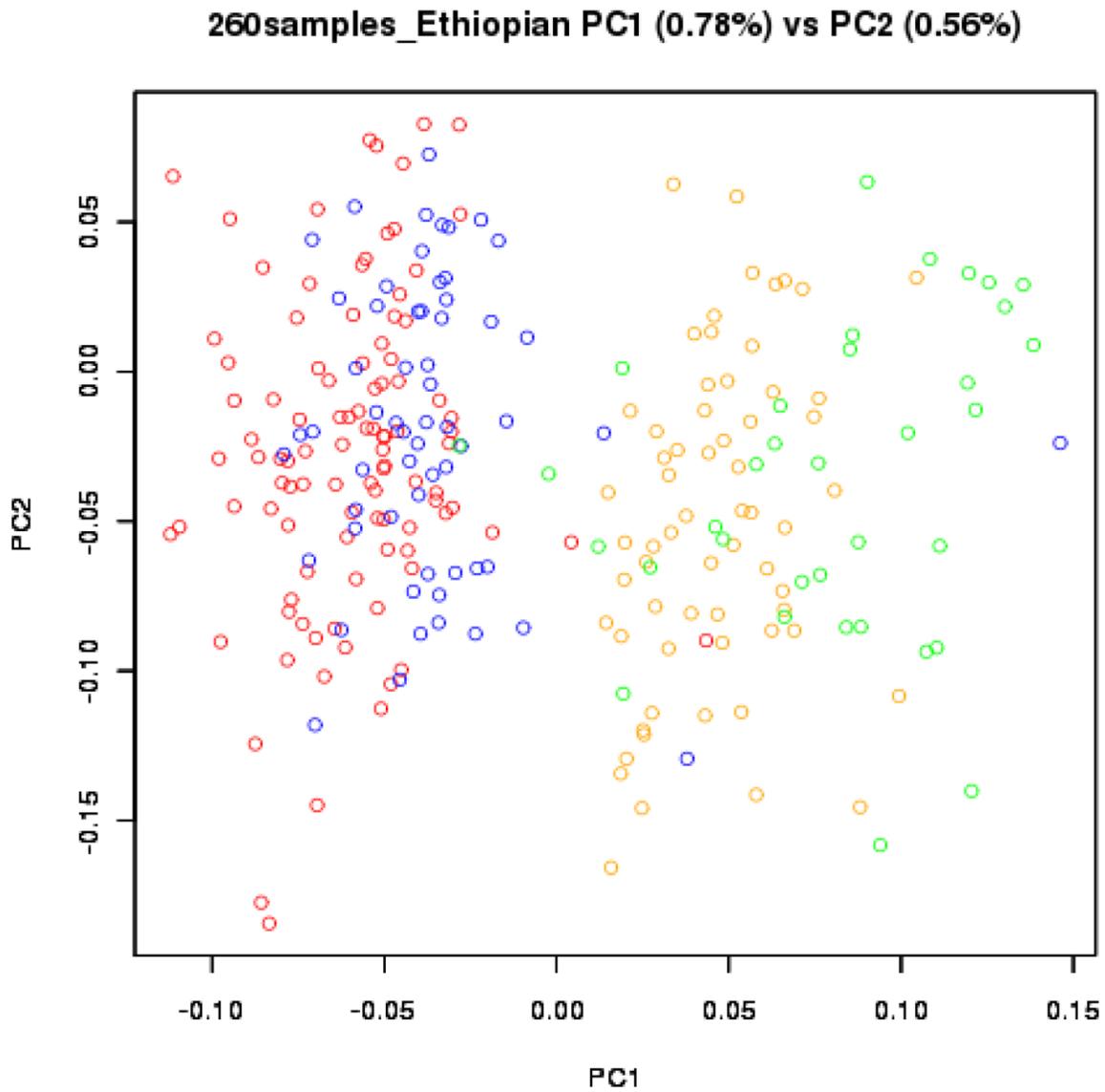

Figure S3. **Scatter plot** of the first two coordinates obtained by principal component analysis. A set of 13,000 random autosomal SNPs were used. HA Amhara individuals are represented in red, LA Amhara in blue, HA Oromo in orange and LA Oromo in green.

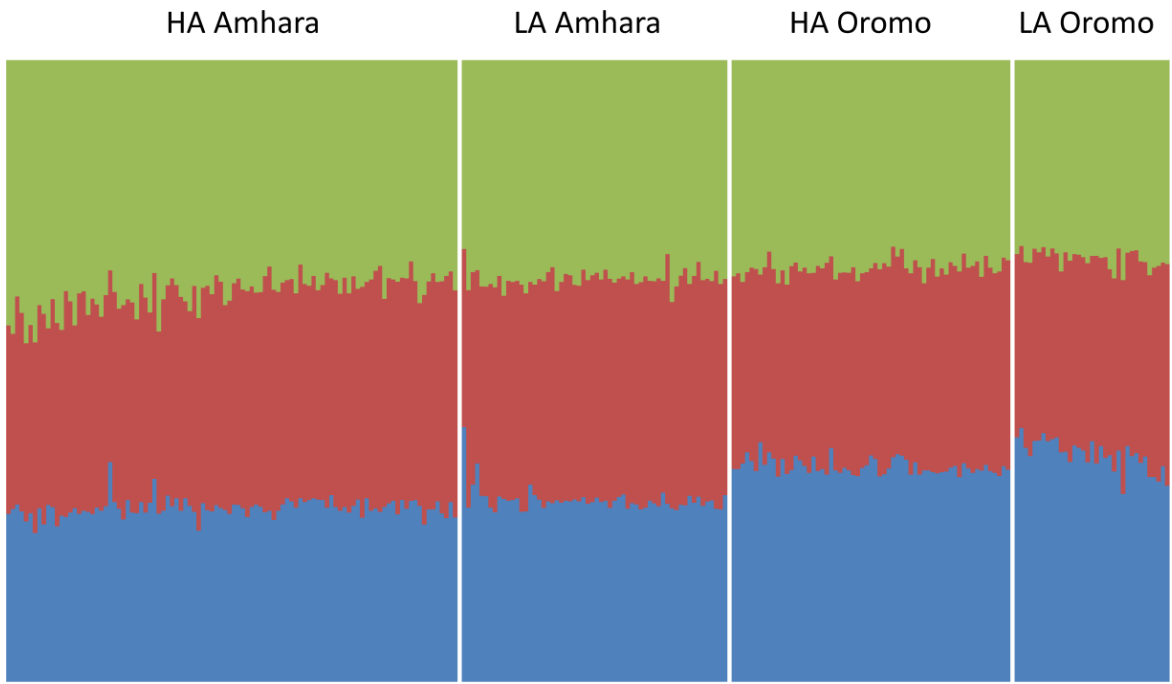

Figure S4. **Ethiopian STRUCTURE plot**. Each vertical line represents an individual, and the colors comprising each line correspond to the inferred proportion of ancestry from three ancestral populations using 57652 random autosomal SNPs. Samples were ordered from left to right as follows: HA Amhara, LA Amhara, HA Oromo and LA Oromo.

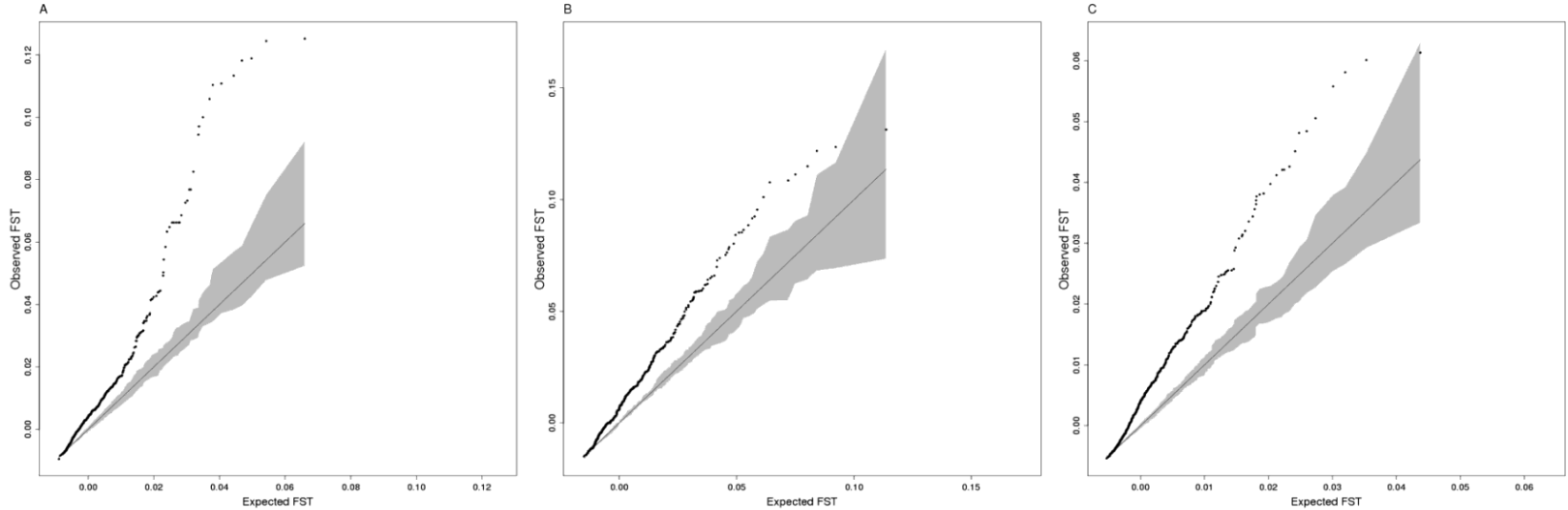

Figure S5. **Population differentiation among Ethiopian subgroups**. The QQplots represent difference in allele frequency as summarized by $F_{ST}$ between HA and LA Amhara (A), Oromo (B) and Ethiopia (C). The observed $F_{ST}$ distribution is ranked from smallest to largest and plotted against the expected $F_{ST}$ in black. The expected $F_{ST}$ distribution was obtained by permuting the subgroup labels mimicking random mating. The grey area indicates the 95% confidence interval of the expected distribution (see Methods).

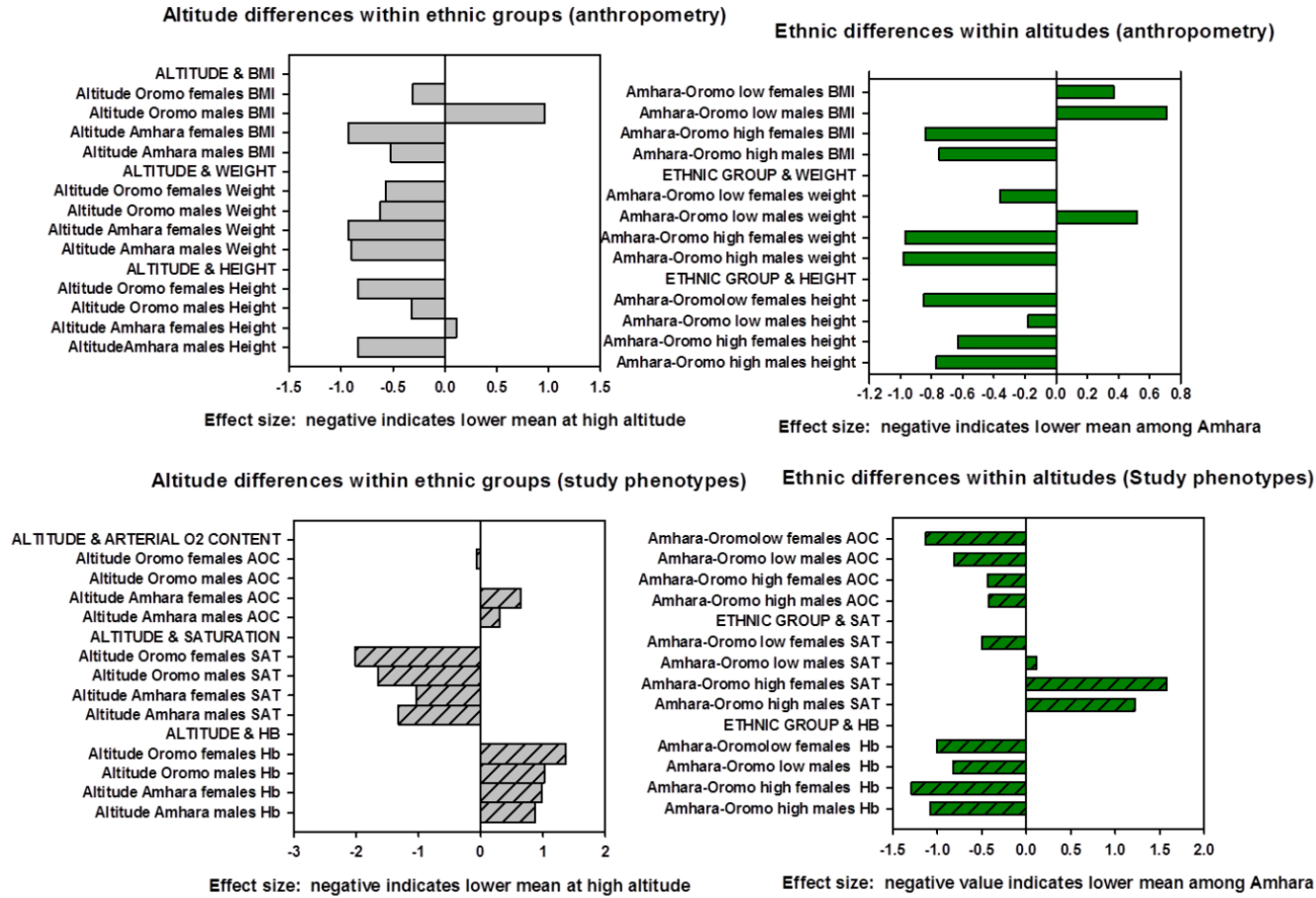

Figure S6. **Altitude differences in anthropometric and phenotypic characteristics are summarized in the left two (gray) panels on terms of effect size d**. Ethnic differences are summarized in the right two (green) panels the same way. D is dimensionless and is calculated as the difference between two sample means divided by their pooled standard deviation. Comparison based on d values allows contrasting the altitude and ethnic-group differences in phenotypes independent of the units of measurement. By convention effect sizes of 0.8 or more are considered to be 'large' [89].

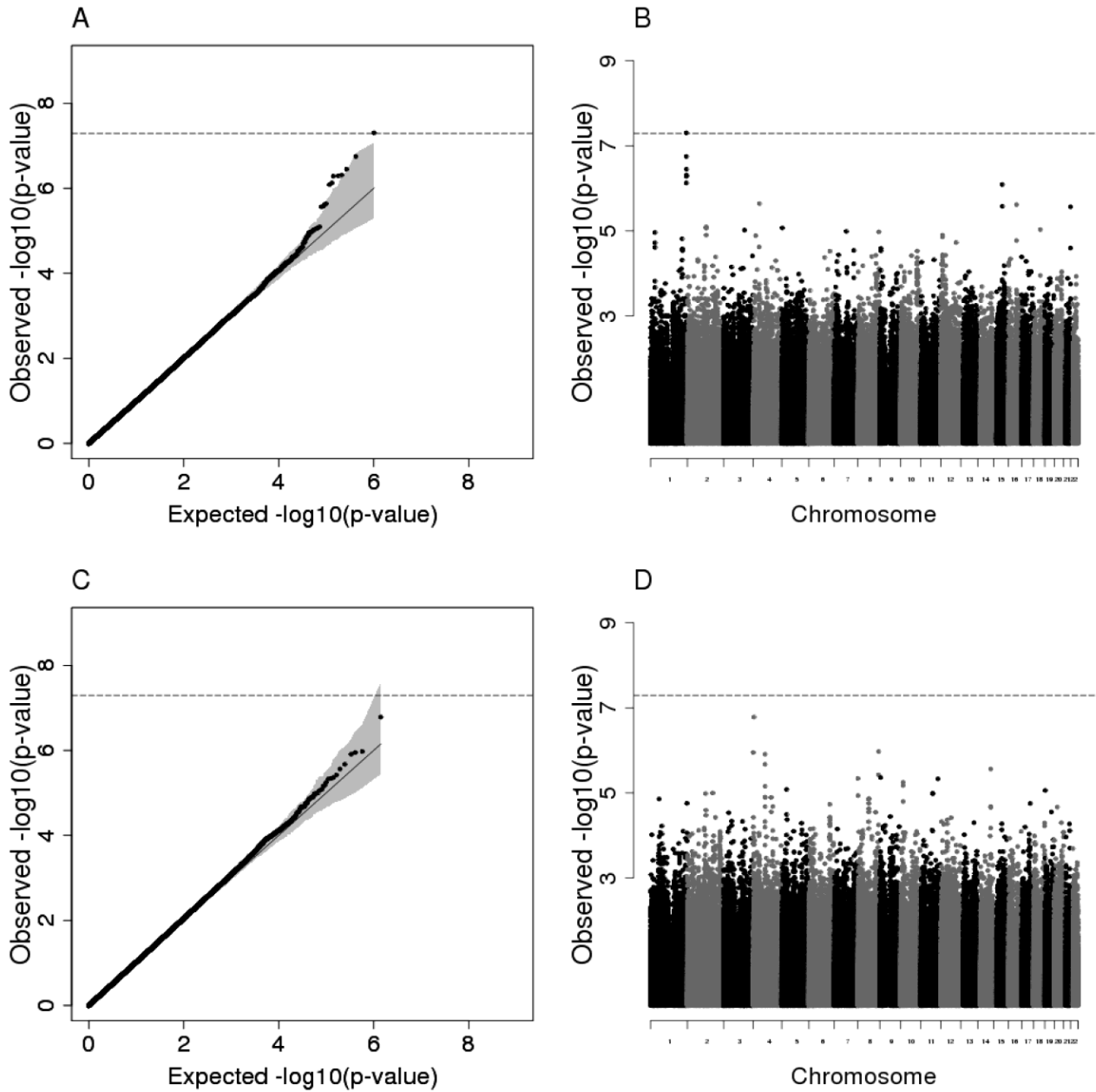

Figure S7. **Amhara Hb level and $O_2$ sat GWAS results**. The QQplot compares the observed –log10 association p-value distribution (y-axis) with an expected distribution (x-axis) in black (see Methods) for Hb (A) and $O_2$ sat (C). The grey area represents the 95% confidence interval. The Manhattan plot shows the observed –log10 association p-value of SNPs for Hb (B) and $O_2$ sat (D).

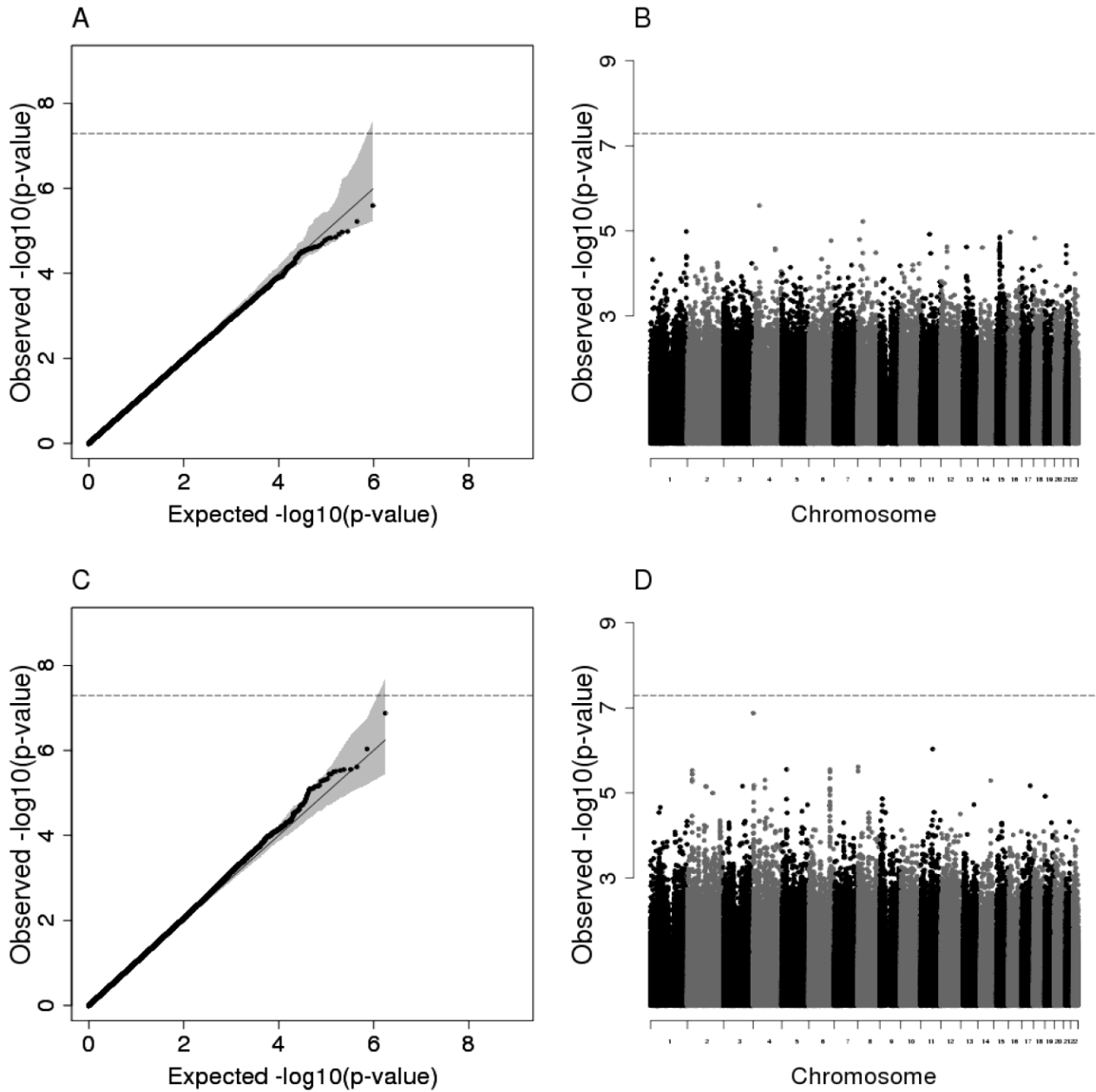

Figure S8. **HA Amhara Hb level and $O_2$ sat GWAS results.** The QQplot compares the observed –log10 association p-value distribution (y-axis) with an expected distribution (x-axis) in black (see Methods) for Hb (A) and $O_2$ sat (C). The grey area represents the 95% confidence interval. The Manhattan plot shows the observed –log10 association p-value of SNPs for Hb (B) and $O_2$ sat (D).

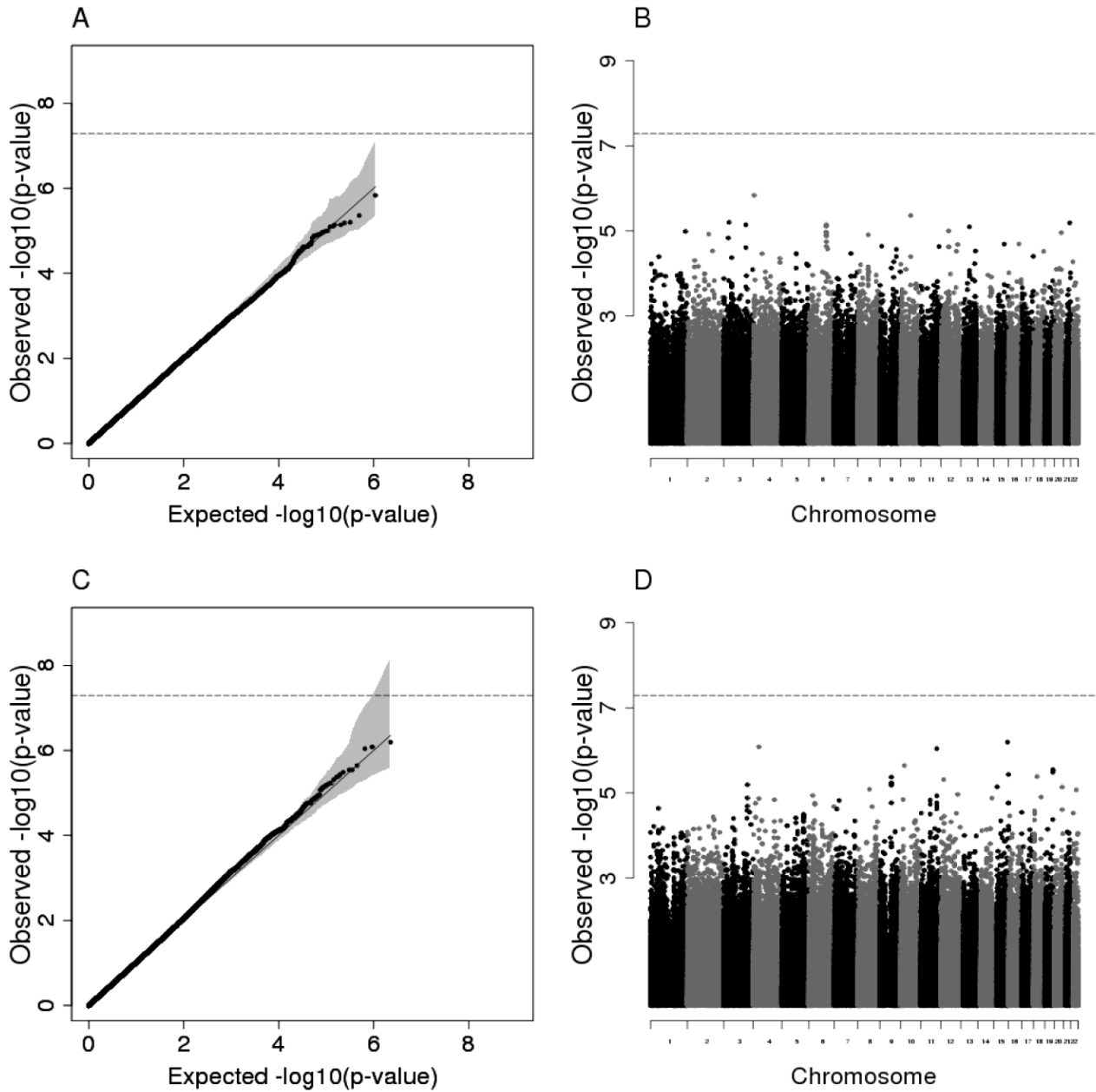

Figure S9. **LA Amhara Hb level and O$_2$ sat GWAS results.** The QQplot compares the observed –log10 association p-value distribution (y-axis) with an expected distribution (x-axis) in black (see Methods) for Hb (A) and O$_2$ sat (C). The grey area represents the 95% confidence interval. The Manhattan plot shows the observed –log10 association p-value of SNPs for Hb (B) and O$_2$ sat (D).

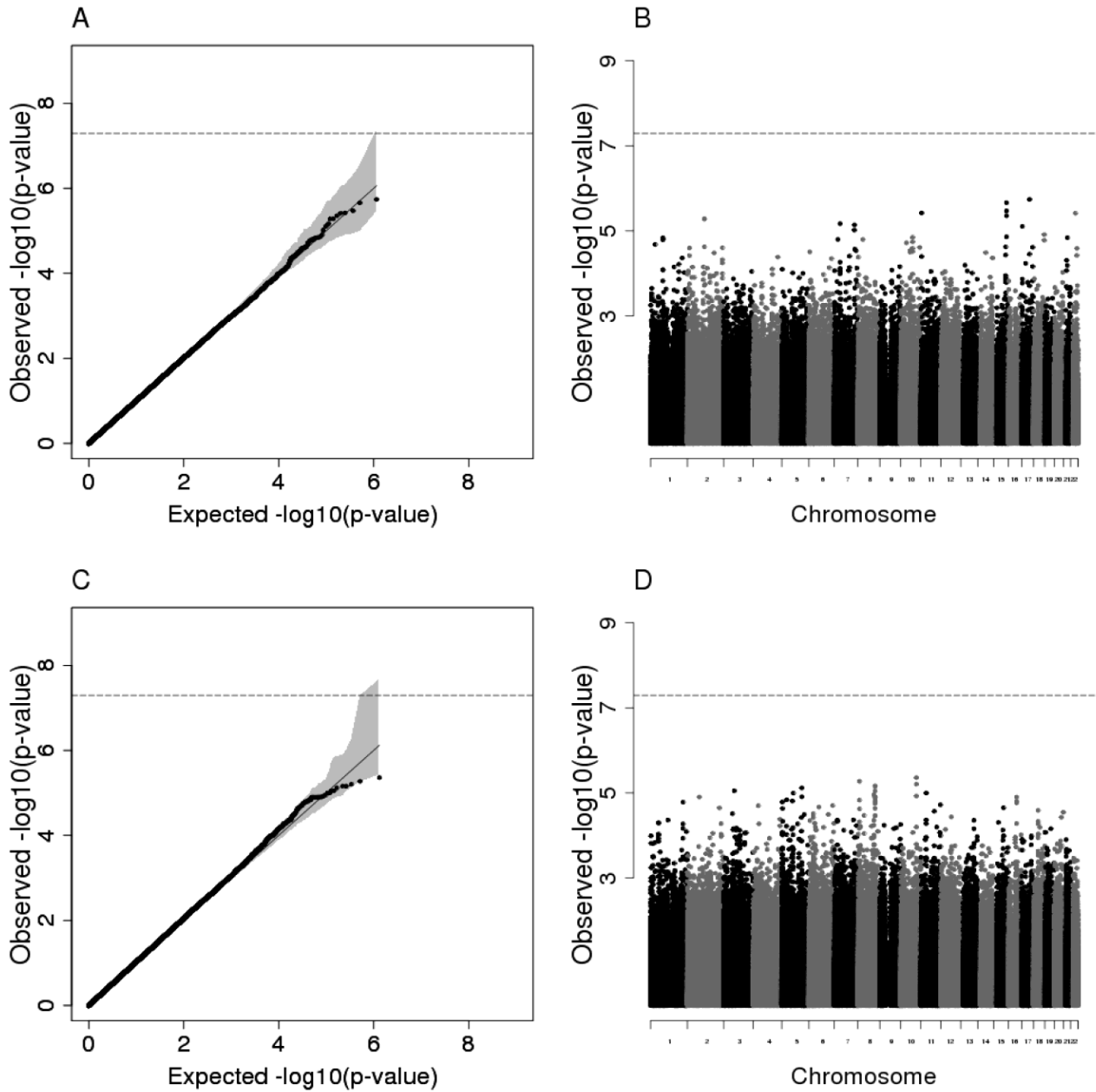

Figure S10. **Oromo Hb level and $O_2$ sat GWAS results.** The QQplot compares the observed –log10 association p-value distribution (y-axis) with an expected distribution (x-axis) in black (see Methods) for Hb (A) and $O_2$ sat (C). The grey area represents the 95% confidence interval. The Manhattan plot shows the observed –log10 association p-value of SNPs for Hb (B) and $O_2$ sat (D).

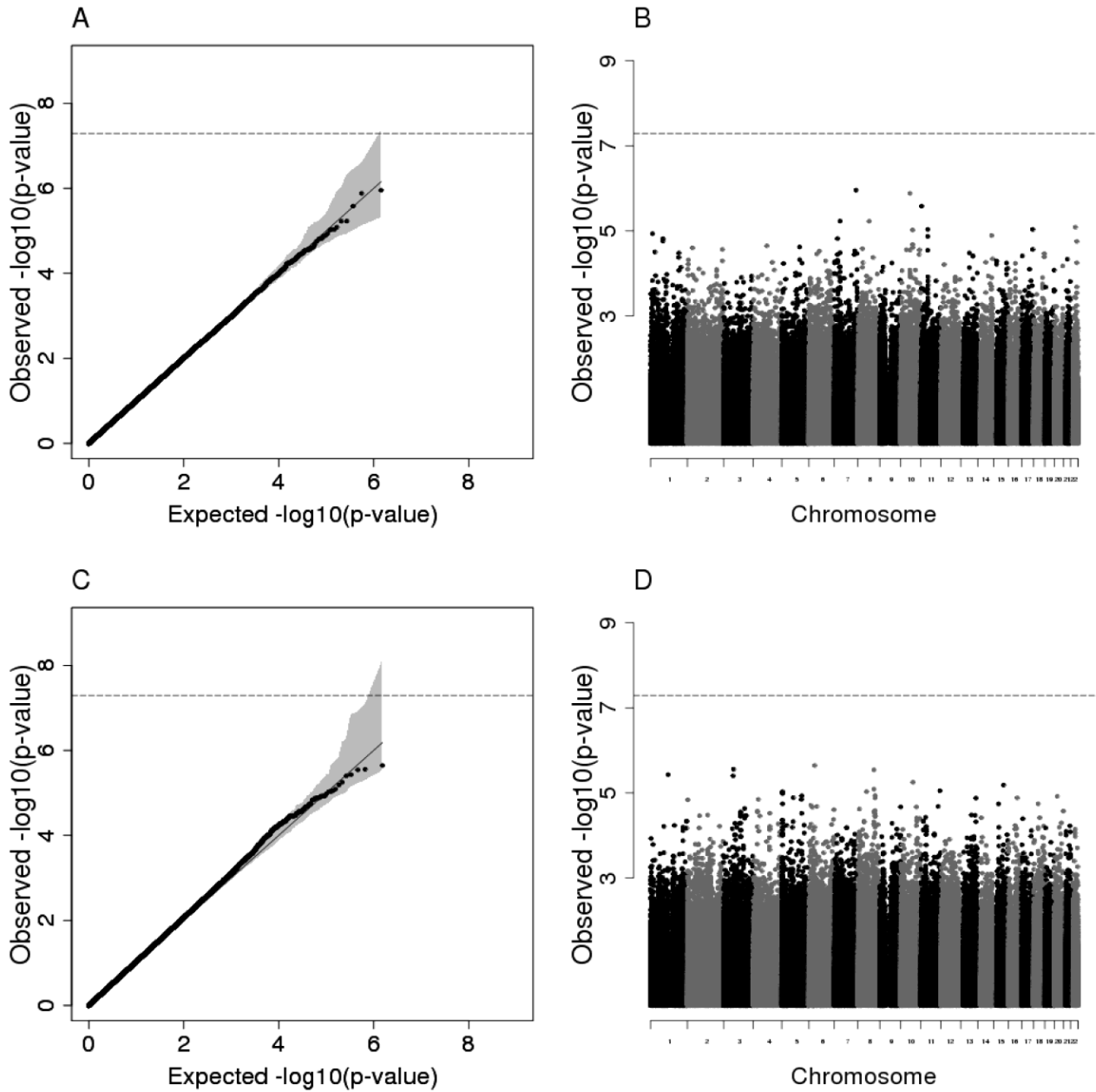

Figure S11. **HA Oromo Hb level and $O_2$ sat GWAS results.** The QQplot compares the observed −log10 association p-value distribution (y-axis) with an expected distribution (x-axis) in black (see Methods) for Hb (A) and $O_2$ sat (C). The grey area represents the 95% confidence interval. The Manhattan plot shows the observed −log10 association p-value of SNPs for Hb (B) and $O_2$ sat (D).

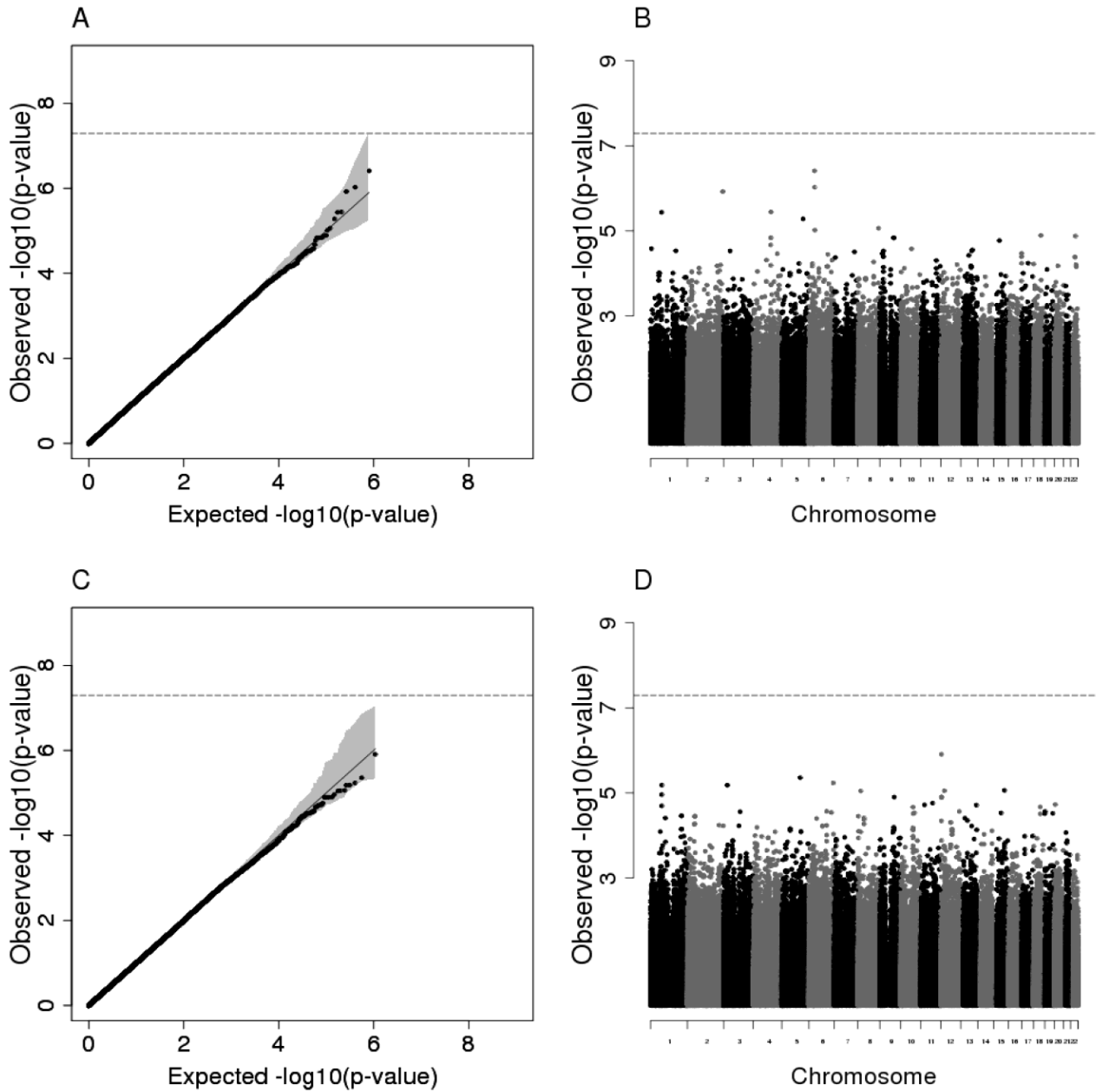

Figure S12. **LA Oromo Hb level and $O_2$ sat GWAS results.** The QQplot compares the observed –log10 association p-value distribution (y-axis) with an expected distribution (x-axis) in black (see Methods) for Hb (A) and $O_2$ sat (C). The grey area represents the 95% confidence interval. The Manhattan plot shows the observed –log10 association p-value of SNPs for Hb (B) and $O_2$ sat (D).

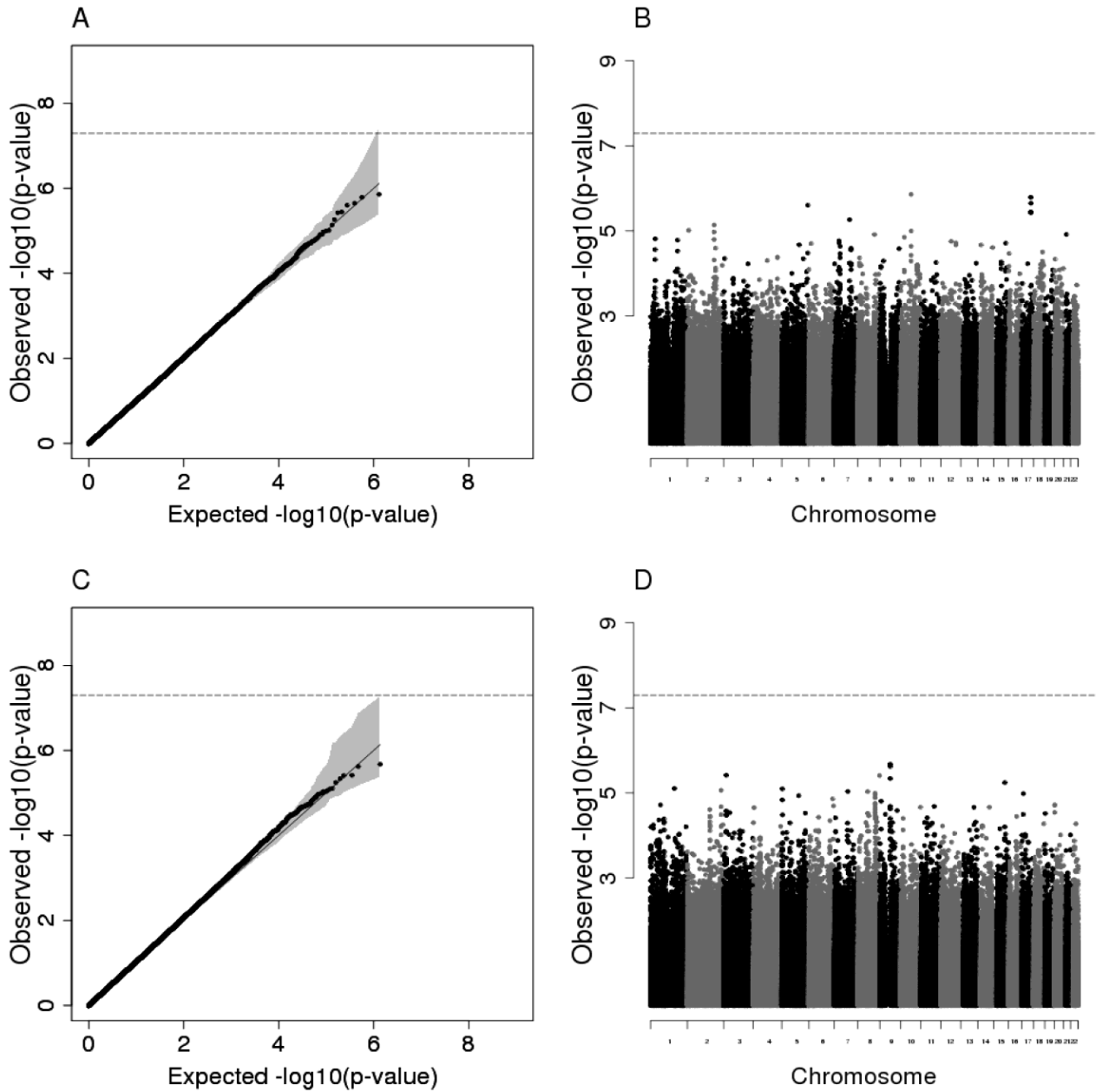

Figure S13. **Ethiopia Hb level and $O_2$ sat GWAS results.** The QQplot compares the observed –log10 association p-value distribution (y-axis) with an expected distribution (x-axis) in black (see Methods) for Hb (A) and $O_2$ sat (C). The grey area represents the 95% confidence interval. The Manhattan plot shows the observed –log10 association p-value of SNPs for Hb (B) and $O_2$ sat (D).

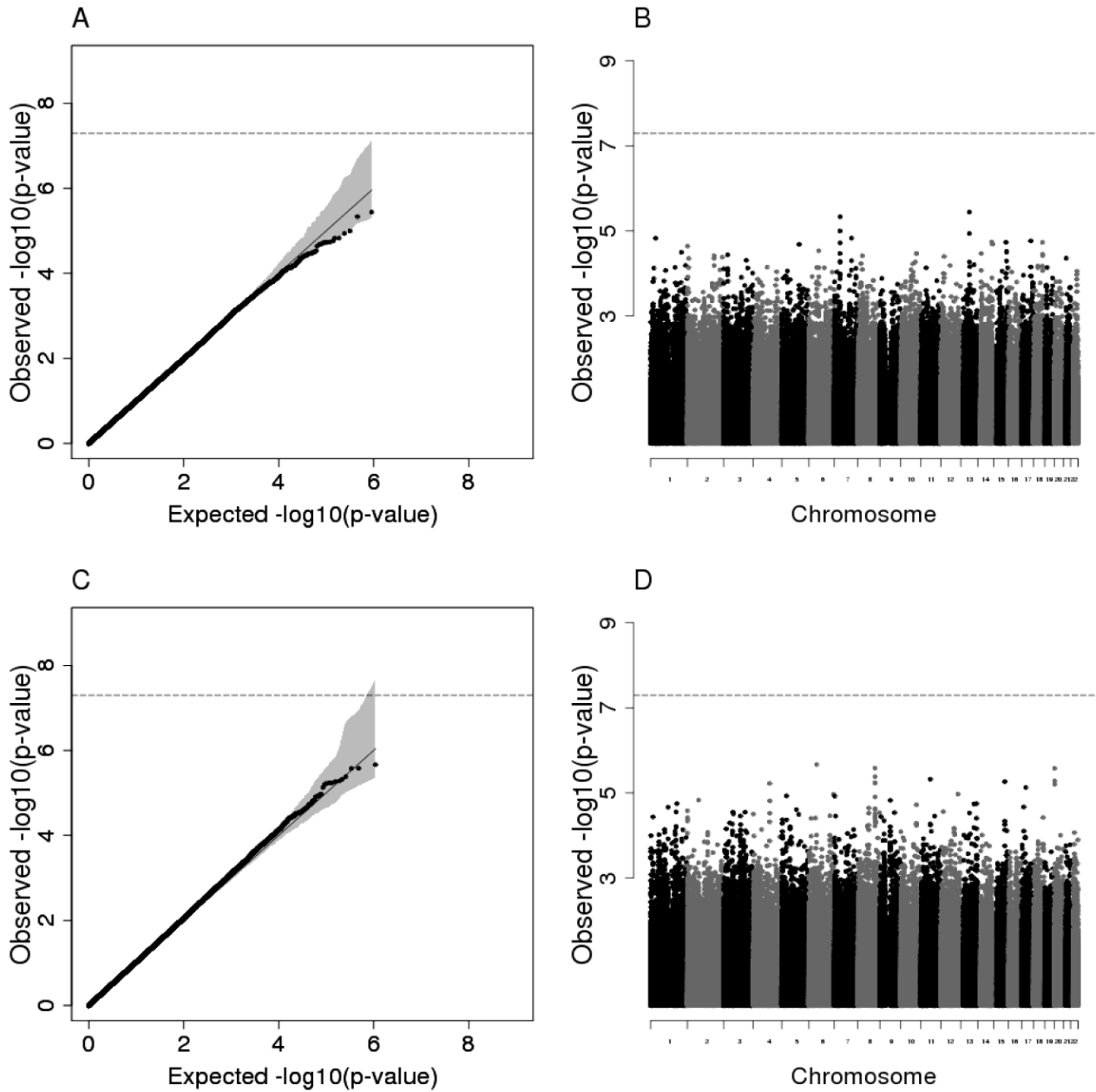

Figure S14. **HA Ethiopia Hb level and $O_2$ sat GWAS results.** The QQplot compares the observed –log10 association p-value distribution (y-axis) with an expected distribution (x-axis) in black (see Methods) for Hb (A) and $O_2$ sat (C). The grey area represents the 95% confidence interval. The Manhattan plot shows the observed –log10 association p-value of SNPs for Hb (B) and $O_2$ sat (D).

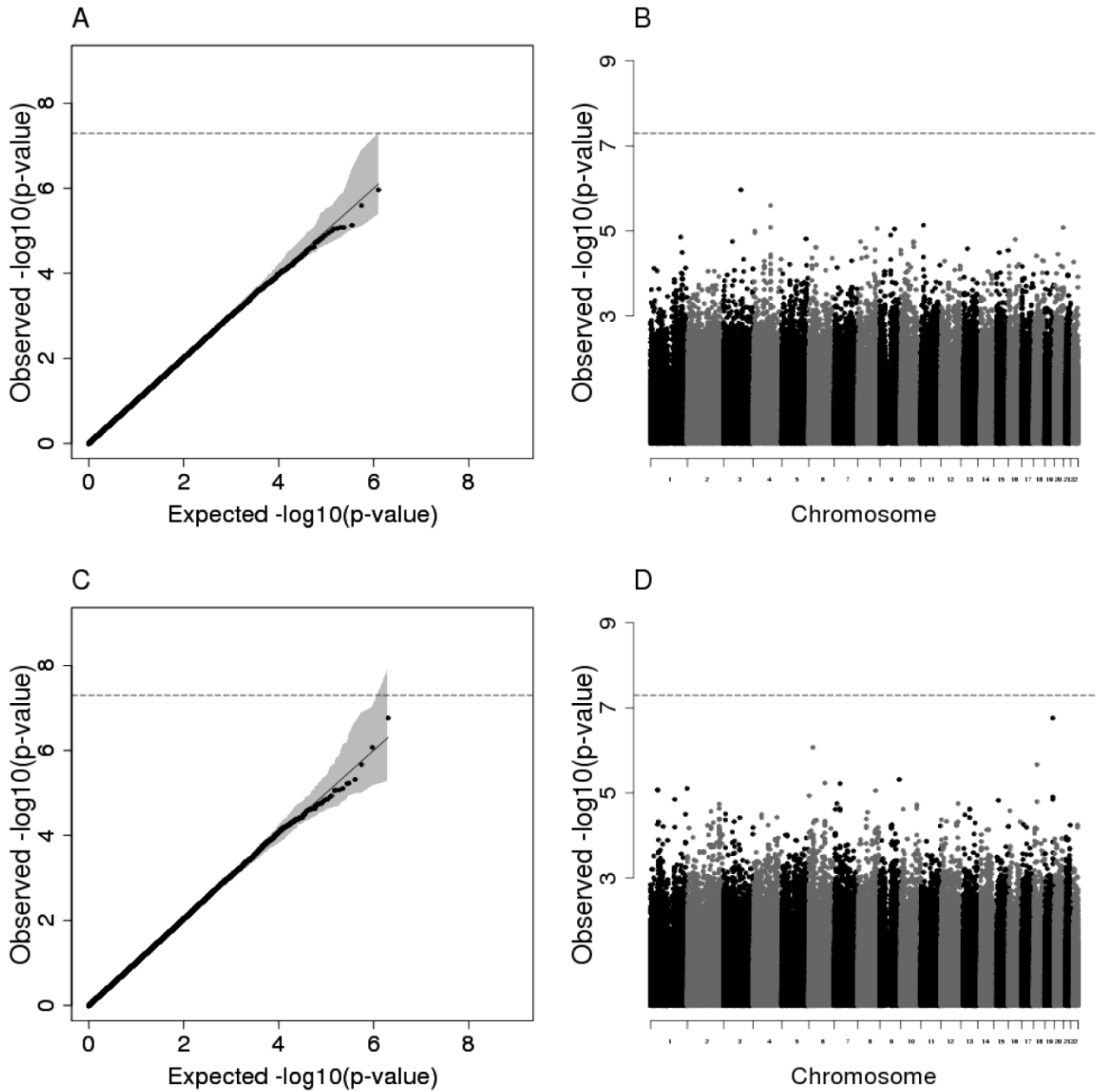

Figure S15. **LA Ethiopia Hb level and $O_2$ sat GWAS results.** The QQplot compares the observed – log10 association p-value distribution (y-axis) with an expected distribution (x-axis) in black (see Methods) for Hb (A) and $O_2$ sat (C). The grey area represents the 95% confidence interval. The Manhattan plot shows the observed –log10 association p-value of SNPs for Hb (B) and $O_2$ sat (D).

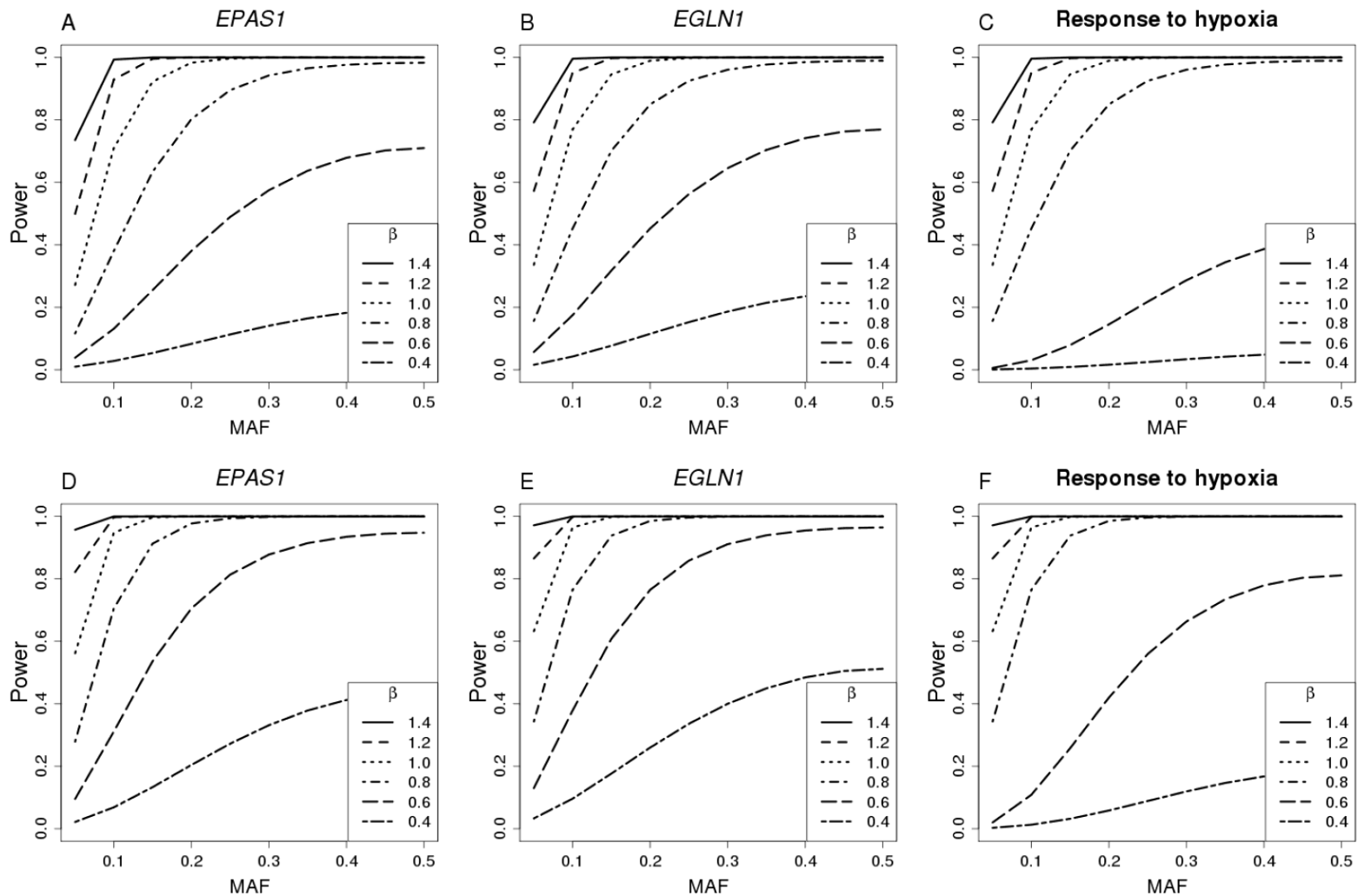

Figure S16. **Power plots**. The effect of *β* and MAF on the power of association tests based on the Oromo (A-C) and Amhara (D-F) sample sizes (corrected for the number of SNPs tested within 10kb from gene) is illustrated for *EPAS1* (A and D, 72 SNPs), *EGLN1* (B and E, 38 SNPs) and any gene within the Response to Hypoxia gene ontology category (C and F, 1309 SNPs).

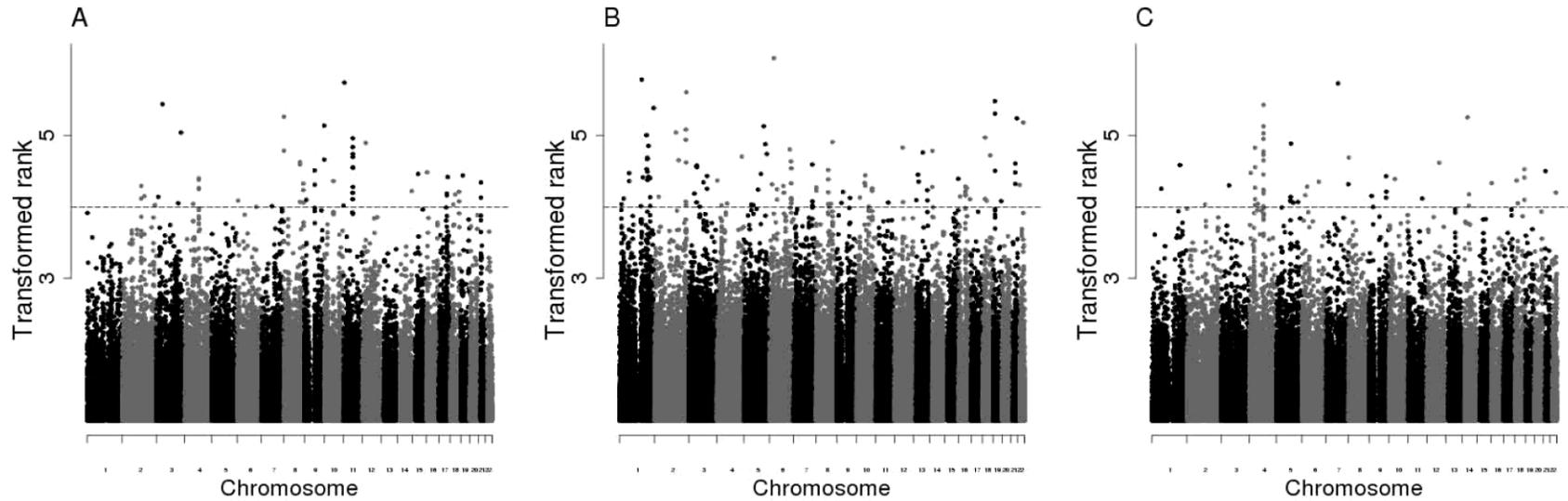

Figure S17. **Manhattan plots of PBS and MR transformed rank.** Manhattan plots with SNPs with transformed rank less than 0.1 are shown for Amhara PBS (*vs*. Maasai and Luyha) (A), for high altitude Amhara PBS (*vs*. low altitude Amhara and low altitude Oromo) (B), and high altitude Amhara MR (C). The transformed rank is the rank of the SNP in the corresponding distribution divided by the total number of SNPs. The horizontal line represents 0.001 transformed rank.

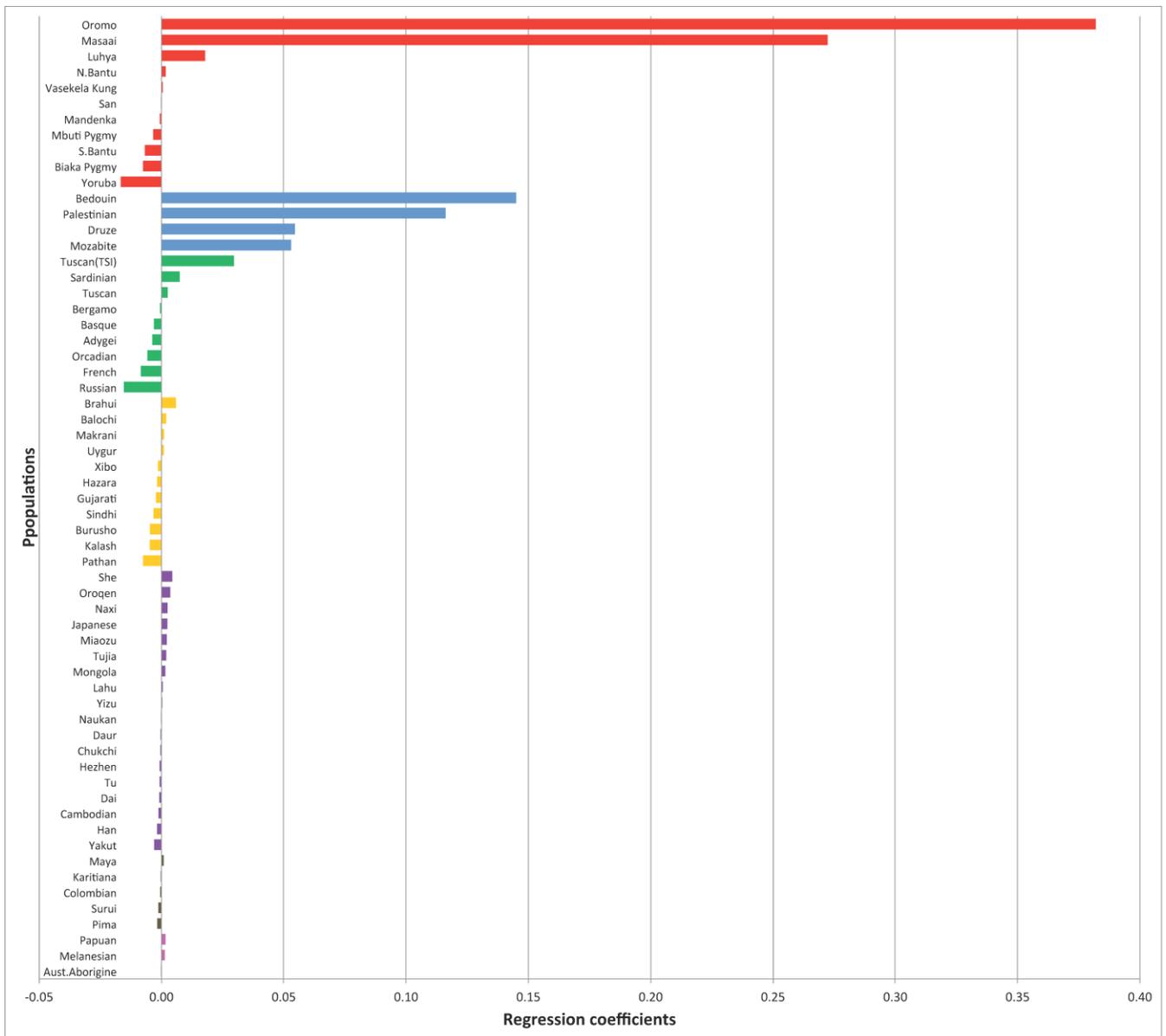

Figure S18. **Multiple linear regression coefficients**. This plot shows the regression coefficients of each of the 61 populations used to predict the expected allele frequencies in the HA Amhara in the multiple linear regression analysis. Populations have been grouped in Africa, Middle East, Europe, Southwest Asia, East Asia, America and Oceania. In each group, populations have been ordered from larger to smaller coefficients.

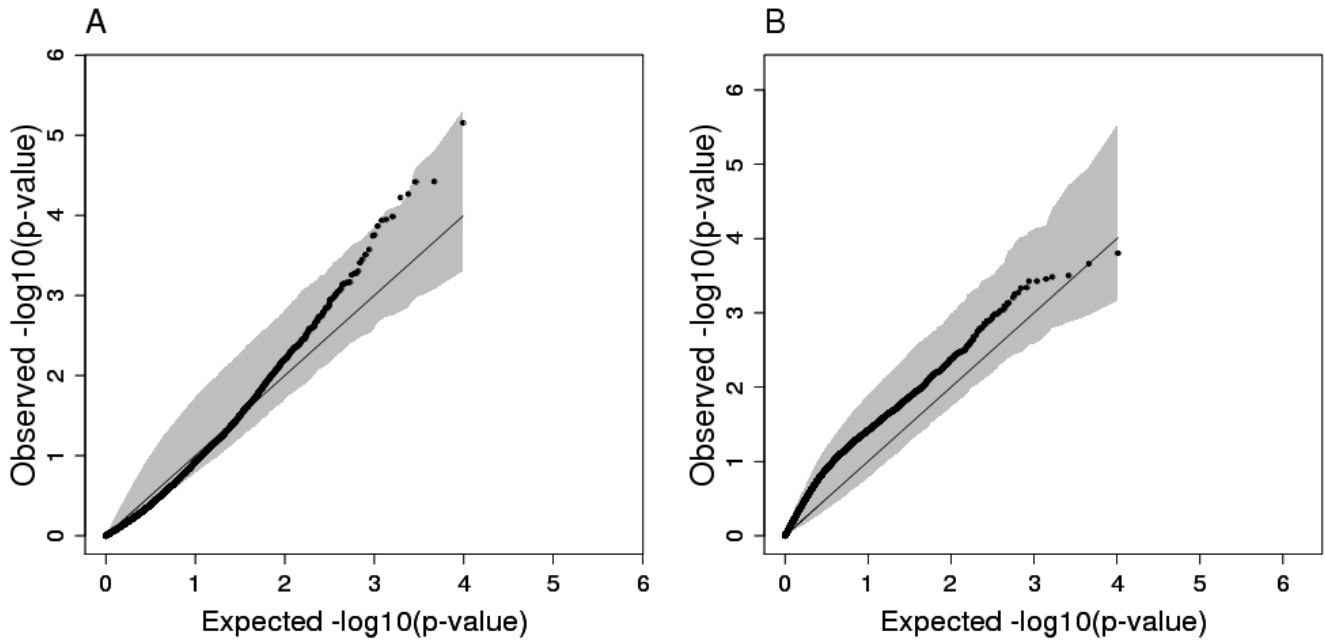

Figure S19. **Quantile-quantile (QQ) plots of methylation differentiation p-value**. Each QQplot compares the observed methylation differentiation p-value distribution (y-axis) with the corresponding null distribution (x-axis) – which was created by reshuffling 100 times the altitude status among the compared samples and running the same lineal model. The grey area represents the 95 % confidence interval. No evidence of genome-wide HA *vs*. LA epigenetic adaptation was observed within Oromo (A) and Amhara (B).

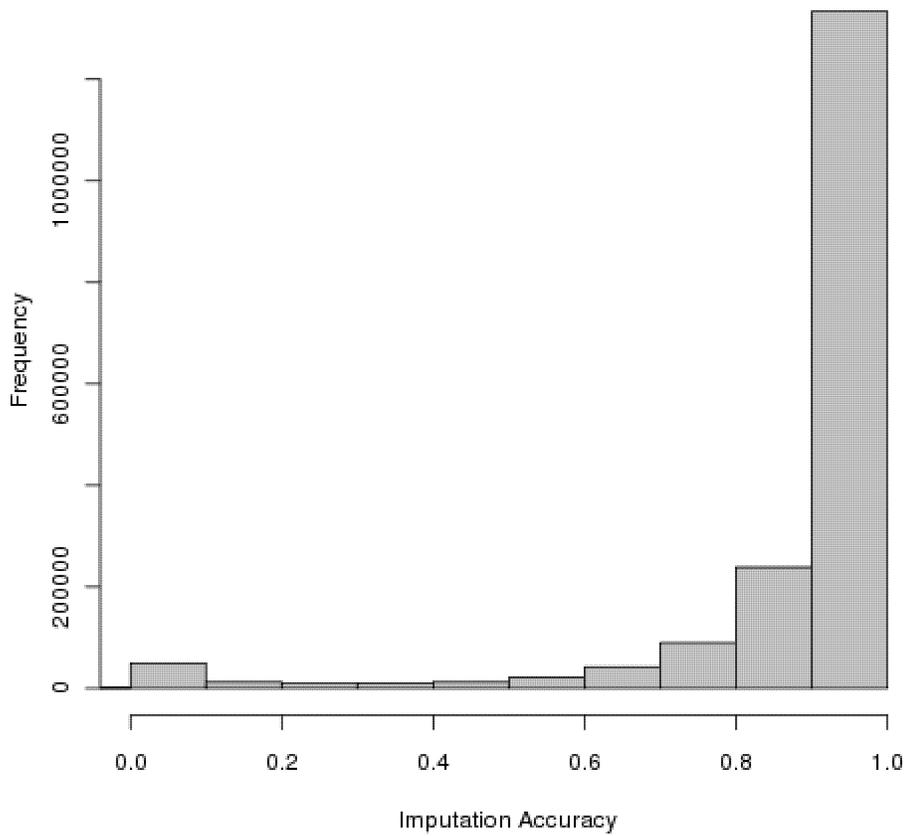

Figure S20. **Frequency distribution of imputation accuracy**. SNPs with accuracy values less than 0.9 were removed from further analyses.

## SUPPORTING TABLES

## SUPPLEMENTARY TABLES

Table S1. Sample description for Amhara and Oromo high and low altitude (HA and LA) males and females (mean ± SEM).

| Sample Subset | N | AGE (years) | Height (cm) | Weight (kg) | BMI (kg/m$^2$) | Pulse (f/min) |
|---|---|---|---|---|---|---|
| HA Amhara males | 78 | 31 + 1 | 164.8 + 0.8 **,b | 52.0 + 0.8 **,b | 19.1 + 0.17 **,b | 72.7 + 1.5 |
| LA Amhara males | 48 | 34 + 1 b | 171.1 + 0.9 | 58.3 + 0.8 a | 19.9 + 0.24 b | 75.7 + 1.5 |
| HA Amhara females | 24 | 30 + 2 | 153.4 + 1.5 a | 45.2 + 1.5 **,b | 19.1 + 0.4 +,c | 84.7 + 3.7 |
| LA Amhara females | 12 | 29 + 3 | 152.3 + 3.7 c | 52.8 + 2.3 | 23.4 + 1.9 | 84.1 + 3.5 |
| HA Oromo males | 35 | 30 + 1 ** | 170.4 + 1.0 | 58.9 + 1.0 * | 20.3 + 0.25 ** | 74.5 + 2.5 |
| LA Oromo males | 27 | 25 + 1 | 172.2 + 1.1 | 55.4 + 1.0 | 18.7 + 0.28 | 73.6 + 2.2 |
| HA Oromo females | 28 | 27 + 1 | 157.5 + 1.0 * | 52.1 + 1.0 | 21.0 + 0.39 | 86.8 + 2.9 |
| LA Oromo females | 8 | 26 + 2 | 161.9 + 1.6 | 55.7 + 2.9 | 21.3 + 1.33 | 86.5 + 2.6 |

[*] $p < 0.05$ t-test comparing same sex and ethnic group at high and low altitude
[**] $P < 0.01$ t-test comparing same sex and ethnic group at high and low altitude
[+] $0.05 < p < 0.01$ t-test comparing same sex and ethnic group at high and low altitude
[a] $p < 0.05$ t-test comparing Amhara and Oromo of the same sex at one altitude
[b] $P < 0.01$ t-test comparing Amhara and Oromo of the same sex and ethnic group at one altitude
[c] $0.05 < p < 0.01$ t-test comparing Amhara and Oromo of the same sex and ethnic group at one altitude

Table S2. Study phenotypes for Amhara and Oromo high and low altitude (HA and LA) males and females (mean $\pm$ SEM).

| Sample Subset | N | Hb (gm/dL) | O$_2$ Sat (%) | Arterial O$_2$ content (mlO$_2$/dL) |
|---|---|---|---|---|
| HA Amhara males | 72 | 16.8 + 0.13 *,b | 92.4 + 0.4 **,b | 21.5 + 0.19 |
| LA Amhara males | 41 | 15.6 + 0.18 b | 97.2 + 0.2 | 21.1 + 0.27 c |
| HA Amhara females | 21 | 14.8 + 0.24 **,b | 93.8 + 0.6 **,b | 19.4 + 0.32 |
| LA Amhara females | 12 | 13.7 + 0.36 a,+ | 96.9 + 0.4 | 18.4 + 0.41 a |
| HA Oromo males | 35 | 18.5 + 0.31 **,+ | 86.7 + 0.8 ** | 21.7 + 0.39 |
| LA Oromo males | 27 | 16.7 + 0.22 | 97.1 + 0.2 | 21.7 + 0.33 |
| HA Oromo females | 27 | 17.1 + 0.38 ** | 84.9 + 0.7 ** | 20.1 + 0.34 |
| LA Oromo females | 6 | 14.8 + 0.34 | 97.6 + 0.5 | 20.2 + 0.50 |

*$p < 0.05$ t-test comparing same sex and ethnic group at high and low altitude
**$P < 0.01$ t-test comparing same sex and ethnic group at high and low altitude
+ $0.05 < p < 0.01$ t-test comparing same sex and ethnic group at high and low altitude
a $p < 0.05$ t-test comparing Amhara and Oromo of the same sex at one altitude
b $P < 0.01$ t-test comparing Amhara and Oromo of the same sex and ethnic group at one altitude
c $0.05 < p < 0.01$ t-test comparing Amhara and Oromo of the same sex and ethnic group at one altitude

Table S3. 20 SNPs with lowest Hb association p-values within Amhara.

| SNP | Chr | N | A1 | β | P | Rank | Genes (within 10kb) | Genes (within 100kb) |
|---|---|---|---|---|---|---|---|---|
| rs4949254 | 1 | 137 | G | -0.62 | 1.09E-05 | 20 | | |
| rs2019842 | 1 | 145 | G | 0.80 | 3.55E-07 | 3 | | |
| rs6700249 | 1 | 145 | A | 0.74 | 4.88E-07 | 4 | | |
| rs2184186 | 1 | 144 | A | 0.80 | 5.16E-07 | 6 | | |
| rs10737884 | 1 | 144 | G | 0.74 | 7.45E-07 | 7 | | |
| rs1855368 | 1 | 146 | A | 0.74 | 5.10E-07 | 5 | | |
| rs10803083 | 1 | 147 | A | 0.84 | 4.96E-08 | 1 | | |
| rs876912 | 1 | 143 | G | 0.78 | 1.79E-07 | 2 | | |
| rs1378539 | 2 | 131 | A | 0.65 | 8.02E-06 | 13 | | |
| rs12711718 | 2 | 147 | A | 0.62 | 8.69E-06 | 15 | | |
| rs4894345 | 3 | 146 | G | -0.58 | 9.60E-06 | 17 | | *MRPS22,COPB2* |
| rs16854020 | 4 | 140 | A | -0.66 | 2.29E-06 | 9 | *CCDC4* | *SLC30A9* |
| rs16871658 | 5 | 132 | A | -0.85 | 8.56E-06 | 14 | | |
| rs5745616 | 7 | 147 | A | 0.64 | 1.03E-05 | 18 | *HGF* | |
| rs2873896 | 8 | 133 | A | -0.74 | 1.06E-05 | 19 | | |
| rs2899662 | 15 | 147 | A | 0.91 | 8.16E-07 | 8 | *RORA* | |
| rs17204475 | 15 | 139 | A | 0.89 | 2.65E-06 | 11 | *RORA* | |
| rs13331158 | 16 | 136 | A | -0.66 | 2.42E-06 | 10 | | *IRX5* |
| rs17805042 | 18 | 144 | A | 0.90 | 9.36E-06 | 16 | *MRO* | *ME2,MAPK4* |
| rs13051923 | 21 | 147 | G | 0.68 | 2.72E-06 | 12 | | *COL6A2,FTCD,C21orf56,COL6A1* |

Only SNPs with MAF <10% and imputation accuracy > 0.9 were tested. In addition to age, sex, BMI (body mass index) and altitude, collection year was also used as covariate since samples were collected ten years apart.

Table S4. 20 SNPs with lowest oxygen saturation association p-values within Amhara.

| SNP | Chr | N | A1 | β | P | Rank | Genes (within 10kb) | Genes (within 100kb) |
|---|---|---|---|---|---|---|---|---|
| rs10496577 | 2 | 149 | G | -1.71 | 1.04E-05 | 18 | | |
| rs17279667 | 2 | 161 | A | -1.40 | 1.00E-05 | 16 | MYO3B | |
| rs7659929 | 4 | 143 | A | -2.13 | 1.12E-06 | 3 | ZNF721, PIGG | ABCA11 |
| rs6446322 | 4 | 146 | G | -1.92 | 1.64E-07 | 1 | STK32B | CYTL1 |
| rs10518184 | 4 | 161 | A | 1.32 | 1.23E-06 | 4 | FRAS1 | |
| rs4859905 | 4 | 161 | A | 1.31 | 2.12E-06 | 5 | FRAS1 | |
| rs873455 | 4 | 160 | A | -1.08 | 6.97E-06 | 13 | FRAS1 | |
| rs157492 | 5 | 153 | G | -1.43 | 8.28E-06 | 14 | GOLPH3 | MTMR12,PDZD2 |
| rs11774254 | 8 | 154 | G | -1.86 | 4.57E-06 | 9 | | |
| rs17065459 | 8 | 161 | G | -1.70 | 1.15E-05 | 20 | | |
| rs10106108 | 8 | 161 | A | -1.26 | 1.06E-06 | 2 | | |
| rs11784218 | 8 | 159 | A | -1.60 | 3.77E-06 | 7 | | |
| rs4740840 | 9 | 151 | G | -1.13 | 4.34E-06 | 8 | IL33 | TPD52L3 |
| rs10904733 | 10 | 154 | G | -1.26 | 6.66E-06 | 12 | | PTER,C1QL3 |
| rs11253992 | 10 | 158 | A | 1.20 | 5.64E-06 | 11 | PTER | C1QL3 |
| rs12293066 | 11 | 161 | A | -1.68 | 1.02E-05 | 17 | | |
| rs1528635 | 11 | 161 | G | -1.32 | 1.05E-05 | 19 | | |
| rs2462165 | 11 | 146 | A | 1.20 | 4.66E-06 | 10 | | |
| rs416542 | 14 | 159 | A | 1.24 | 2.74E-06 | 6 | GALC | GPR65 |
| rs426633 | 19 | 161 | A | -1.09 | 8.71E-06 | 15 | BRUNOL5 | EDG6,NCLN,GNA15 |

Only SNPs with MAF <10% and imputation accuracy > 0.9 were tested. In addition to age, sex, BMI (body mass index) and altitude, collection year was also used as covariate.

Table S5. 20 SNPs with lowest hemoglobin association p-values within high altitude Amhara.

| SNP | Chr | N | A1 | β | P | Rank | Genes (within 10kb) | Genes (within 100kb) |
|---|---|---|---|---|---|---|---|---|
| rs16854020 | 4 | 92 | A | -0.76 | 2.54E-06 | 1 | CCDC4 | SLC30A9 |
| rs13125215 | 4 | 94 | A | -0.66 | 2.63E-05 | 18 | | |
| rs4897037 | 6 | 86 | G | -0.82 | 1.71E-05 | 10 | | UST |
| rs9329248 | 8 | 92 | A | 0.86 | 1.61E-05 | 9 | GATA4,NEIL2 | FDFT1,CTSB |
| rs7829680 | 8 | 79 | G | -0.83 | 6.02E-06 | 2 | | |
| rs643273 | 11 | 86 | A | 1.12 | 1.20E-05 | 5 | | PLAC1L,MS4A2,MS4A3 |
| rs2730985 | 12 | 92 | A | -0.73 | 2.38E-05 | 14 | | |
| rs9315776 | 13 | 94 | A | -0.73 | 2.40E-05 | 16 | LOC646982 | LOC646982,FOXO1 |
| rs7494316 | 14 | 94 | A | -0.68 | 2.46E-05 | 17 | AKAP6 | |
| rs514438 | 15 | 92 | G | -0.85 | 1.47E-05 | 7 | LCMT2, ADAL | TUBGCP4,TP53BP1,ZSCAN29,TGM7,TGM5 |
| rs12911740 | 15 | 89 | A | -0.84 | 2.65E-05 | 19.5 | TP53BP1 | HISPPD2A,MAP1A,CKMT1B,TUBGCP4 |
| rs7177146 | 15 | 89 | G | -0.84 | 2.65E-05 | 19.5 | TP53BP1 | HISPPD2A,CKMT1B,STRC,MAP1A |
| rs6493090 | 15 | 92 | A | -0.89 | 1.40E-05 | 6 | HYPK,SERINC4, SERF2, ELL3 | hCG_1789710,PDIA3,WDR76,FRMD5, CATSPER2P1,MFAP1,CKMT1A |
| rs678084 | 15 | 94 | A | -0.86 | 1.99E-05 | 11 | MFAP1, WDR76 | FRMD5,CATSPER2P1,SERF2,PDIA3, hCG_1789710,SERINC4,HYPK,ELL3 |
| rs7174241 | 15 | 86 | A | -0.85 | 2.30E-05 | 13 | WDR76 | HYPK,CATSPER2P1,SERF2,ELL3,PDIA3, SERINC4,FRMD5,MFAP1,hCG_1789710 |
| rs495880 | 15 | 82 | G | -0.89 | 2.39E-05 | 15 | WDR76 | SERF2,ELL3,FRMD5,hCG_1789710, SERINC4,HYPK,MFAP1,PDIA3 |
| rs16962916 | 16 | 80 | A | 0.78 | 1.06E-05 | 4 | | |
| rs2298503 | 18 | 94 | A | 0.82 | 1.48E-05 | 8 | RAB31 | |
| rs1505267 | 21 | 83 | G | 0.70 | 2.23E-05 | 12 | CHODL | |

Only SNPs with MAF <10% and imputation accuracy > 0.9 were tested. In addition to age, sex and BMI (body mass index), collection year was also used as covariate.

Table S6. 20 SNPs with lowest oxygen saturation association p-values within high altitude Amhara.

| SNP | Chr | N | A1 | β | P | Rank | Genes (within 10kb) | Genes (within 100kb) |
|---|---|---|---|---|---|---|---|---|
| rs17011982 | 2 | 98 | G | -2.11 | 5.38E-06 | 15 | *BIRC6* | |
| rs3731579 | 2 | 100 | G | -2.15 | 2.97E-06 | 6 | *BIRC6* | |
| rs17429032 | 2 | 100 | G | -2.26 | 3.62E-06 | 9.5 | *BIRC6* | |
| rs11683905 | 2 | 100 | G | -2.26 | 3.62E-06 | 9.5 | *BIRC6* | |
| rs12465072 | 2 | 98 | G | -2.36 | 4.86E-06 | 12 | *BIRC6* | *TTC27* |
| rs17321135 | 2 | 98 | G | -1.89 | 7.15E-06 | 20 | *CNTNAP5* | |
| rs7559515 | 2 | 99 | A | -1.88 | 7.02E-06 | 19 | *CNTNAP5* | |
| rs1873393 | 3 | 92 | G | -3.05 | 6.95E-06 | 18 | *CHST13, C3orf22* | *UROC1,TR2IT1,ZXDC* |
| rs7659929 | 4 | 92 | A | -3.45 | 1.33E-07 | 1 | *PIGG, ZNF721* | *ABCA11* |
| rs6446322 | 4 | 94 | G | -2.38 | 6.68E-06 | 16 | *STK32B* | *CYTL1* |
| rs10518184 | 4 | 101 | A | 1.80 | 4.96E-06 | 13 | *FRAS1* | |
| rs157492 | 5 | 94 | G | -2.29 | 2.81E-06 | 4 | *GOLPH3* | *MTMR12,PDZD2* |
| rs9403279 | 6 | 98 | G | 1.79 | 3.21E-06 | 8 | | |
| rs719196 | 6 | 88 | A | 1.92 | 2.84E-06 | 5 | | |
| rs9376611 | 6 | 93 | A | 1.94 | 4.56E-06 | 11 | | |
| rs11774254 | 8 | 96 | G | -2.93 | 2.46E-06 | 3 | | |
| rs17065459 | 8 | 101 | G | -2.74 | 3.10E-06 | 7 | | |
| rs1528635 | 11 | 101 | G | -2.29 | 9.28E-07 | 2 | | |
| rs416542 | 14 | 99 | A | 1.80 | 5.13E-06 | 14 | *GALC* | *GPR65* |
| rs8081452 | 17 | 101 | A | 1.85 | 6.78E-06 | 17 | | *EFCAB3* |
| rs876912 | 1 | 91 | G | 0.77 | 1.04E-05 | 3 | | |

Only SNPs with MAF <10% and imputation accuracy > 0.9 were tested. In addition to age, sex and BMI (body mass index), collection year was also used as covariate.

Table S7. 20 SNPs with lowest hemoglobin association p-values within low altitude Amhara.

| SNP | Chr | N | A1 | β | P | Rank | Genes (within 10kb) | Genes (within 100kb) |
|---|---|---|---|---|---|---|---|---|
| rs6657602 | 1 | 49 | G | 2.05 | 1.03E-05 | 10 | RYR2 | |
| rs12105739 | 2 | 53 | A | -1.73 | 1.20E-05 | 13 | ARHGAP15 | |
| rs7612780 | 3 | 52 | A | 1.61 | 1.47E-05 | 18 | OSBPL10 | |
| rs6550509 | 3 | 53 | G | -1.12 | 6.29E-06 | 3 | ITGA9 | CTDSPL |
| rs12152212 | 3 | 53 | A | 1.46 | 7.21E-06 | 6 | | |
| rs6813176 | 4 | 53 | G | 1.55 | 1.47E-06 | 1 | FLJ46481 | CRMP1,JAKMIP1 |
| rs210617 | 6 | 52 | G | 0.99 | 1.30E-05 | 16 | DCBLD1, GOPC | |
| rs2071825 | 6 | 53 | A | 1.01 | 7.20E-06 | 5 | DCBLD1,GOPC | |
| rs10457315 | 6 | 53 | G | 0.98 | 1.32E-05 | 17 | DCBLD1,GOPC | |
| rs10872153 | 6 | 52 | G | 0.93 | 1.23E-05 | 14 | | DCBLD1, GOPC,NUS1 |
| rs4946273 | 6 | 53 | G | 0.92 | 1.80E-05 | 19 | | DCBLD1, GOPC,NUS1 |
| rs9489238 | 6 | 53 | A | 0.99 | 7.79E-06 | 7 | | DCBLD1, GOPC,NUS1 |
| rs1967194 | 6 | 53 | G | 0.92 | 1.08E-05 | 11 | | DCBLD1, GOPC,NUS1 |
| rs10087150 | 8 | 49 | G | 1.22 | 1.24E-05 | 15 | | CPA6 |
| rs7076094 | 10 | 53 | G | 1.18 | 4.35E-06 | 2 | CTNNA3 | |
| rs1971762 | 12 | 52 | A | 1.08 | 9.97E-06 | 9 | ATP5G2 | ATF7,CALCOCO1 |
| rs17056224 | 13 | 53 | A | 1.79 | 8.04E-06 | 8 | | |
| rs10852511 | 16 | 42 | G | 1.05 | 2.04E-05 | 20 | PMFBP1 | HP,HPR,TXNL4B,DHX38 |
| rs6063285 | 20 | 50 | G | -1.74 | 1.10E-05 | 12 | | PREX1 |
| rs17000823 | 21 | 42 | G | 1.04 | 6.51E-06 | 4 | FAM3B | MX2,BACE2,MX1 |

Only SNPs with MAF <10% and imputation accuracy > 0.9 were tested. Age, sex and BMI (body mass index) were used as covariates.

Table S8. 20 SNPs with lowest oxygen saturation association p-values within low altitude Amhara.

| SNP | Chr | N | A1 | β | P | Rank | Genes (within 10kb) | Genes (within 100kb) |
|---|---|---|---|---|---|---|---|---|
| rs1020470 | 3 | 55 | A | -1.04 | 6.40E-06 | 14 | SCHIP1,IQCJ | |
| rs1873515 | 4 | 59 | G | -1.48 | 8.25E-07 | 2 | | |
| rs4735729 | 8 | 60 | A | -1.35 | 8.21E-06 | 18 | | |
| rs2490583 | 9 | 58 | A | -1.52 | 5.85E-06 | 12 | RFK | GCNT1 |
| rs2501920 | 9 | 60 | A | -1.45 | 5.95E-06 | 13 | | GCNT1,RFK |
| rs7847608 | 9 | 58 | G | -1.50 | 6.60E-06 | 15 | GCNT1 | RFK |
| rs6560519 | 9 | 60 | G | -1.51 | 4.23E-06 | 10 | GCNT1 | RFK |
| rs12221280 | 10 | 47 | G | -1.20 | 2.28E-06 | 4 | GPR158 | |
| rs1352735 | 11 | 58 | A | -1.44 | 9.09E-07 | 3 | | |
| rs1982849 | 12 | 58 | G | -1.50 | 4.85E-06 | 11 | | PLCZ1,CAPZA3 |
| rs7980422 | 12 | 60 | A | -1.18 | 1.09E-05 | 20 | | |
| rs17116497 | 15 | 58 | G | -1.64 | 7.19E-06 | 16 | | |
| rs12323993 | 15 | 50 | A | -1.56 | 6.39E-07 | 1 | | |
| rs7172235 | 15 | 53 | G | -1.12 | 3.69E-06 | 8 | PCSK6 | |
| rs4502305 | 18 | 60 | A | -1.04 | 4.11E-06 | 9 | | |
| rs183049 | 19 | 59 | A | -1.16 | 2.88E-06 | 5.5 | SLC6A16 | FLJ32658,DKKL1,CD37,TEAD2 |
| rs659555 | 19 | 59 | A | -1.16 | 2.88E-06 | 5.5 | CD37,SLC6A16 | FLJ32658,TEAD2,DKKL1,PTH2 |
| rs35270176 | 19 | 54 | A | -1.36 | 3.27E-06 | 7 | LOC199800,CPT1C,PRMT1 | AP2A1,SCAF1,IRF3,PRR12,RRAS,BCL2L12,TSKS,PRRG2 |
| rs159095 | 20 | 60 | A | -1.48 | 7.29E-06 | 17 | | |
| rs137994 | 22 | 59 | G | -1.21 | 8.50E-06 | 19 | GRAP2 | TNRC6B,ENTHD1,FAM83F |

Only SNPs with MAF <10% and imputation accuracy > 0.9 were tested. Age, sex and BMI (body mass index) were used as covariates.

Table S9. 20 SNPs with lowest hemoglobin association p-values within Oromo.

| SNP | Chr | N | A1 | β | P | Rank | Genes (within 10kb) | Genes (within 100kb) |
|---|---|---|---|---|---|---|---|---|
| rs323937 | 1 | 94 | A | -0.98 | 1.46E-05 | 17 | | |
| rs6682943 | 1 | 94 | A | -0.98 | 1.56E-05 | 18 | | |
| rs1053413 | 2 | 94 | G | 1.31 | 5.22E-06 | 7.5 | *SLC35F5* | *RABL2A,RPL23AP7* |
| rs12624301 | 2 | 94 | G | 1.31 | 5.22E-06 | 7.5 | *SLC35F5* | *RABL2A* |
| rs4722321 | 7 | 91 | A | 1.08 | 1.59E-05 | 19 | | |
| rs11765705 | 7 | 94 | A | 1.42 | 6.77E-06 | 9 | *POU6F2* | |
| rs834781 | 7 | 91 | G | -0.99 | 7.26E-06 | 10 | | |
| rs834787 | 7 | 93 | G | -1.19 | 9.61E-06 | 12 | | |
| rs2643268 | 8 | 81 | G | -1.30 | 1.60E-05 | 20 | | |
| rs780159 | 10 | 94 | A | 1.30 | 1.41E-05 | 15 | *ZMIZ1* | *LOC283050* |
| rs7120319 | 11 | 94 | A | 1.55 | 3.80E-06 | 4 | *OR51B6,OR51B5* | *HBG1,HBBP1,OR51Q1,OR51I1, HBG2,OR51M1,OR51B2,HBE1,OR51B4* |
| rs12904003 | 15 | 94 | G | 0.96 | 2.19E-06 | 2 | | *MFGE8,ACAN,HAPLN3* |
| rs7164649 | 15 | 90 | G | 0.98 | 3.37E-06 | 3 | | *MFGE8,ACAN,HAPLN3* |
| rs7163586 | 15 | 90 | G | -0.98 | 4.42E-06 | 6 | | *MFGE8,ACAN,HAPLN3* |
| rs12443377 | 15 | 84 | G | -0.97 | 1.37E-05 | 14 | | *MFGE8,ACAN,HAPLN3* |
| rs12103697 | 17 | 94 | A | 1.39 | 7.87E-06 | 11 | *SMTNL2* | *GGT6,MYBBP1A,ALOX15,SPNS2,PELP1* |
| rs894621 | 17 | 94 | G | 1.38 | 1.81E-06 | 1 | *MSI2* | |
| rs13306374 | 18 | 85 | G | 0.97 | 1.23E-05 | 13 | *GALR1* | |
| rs7276366 | 21 | 91 | A | 1.28 | 1.45E-05 | 16 | | |
| rs135195 | 22 | 75 | G | -1.02 | 3.85E-06 | 5 | | |

Only SNPs with MAF <10% and imputation accuracy > 0.9 were tested. Age, sex, BMI (body mass index) and altitude were used as covariates.

Table S10. 20 SNPs with lowest oxygen saturation association p-values within Oromo.

| SNP | Chr | N | A1 | β | P | Rank | Genes (within 10kb) | Genes (within 100kb) |
|---|---|---|---|---|---|---|---|---|
| rs7592938 | 2 | 85 | G | -3.74 | 1.25E-05 | 14 | *CTNNA2* | |
| rs4676890 | 3 | 98 | A | 2.48 | 8.91E-06 | 7 | | |
| rs164819 | 5 | 98 | A | -2.25 | 1.01E-05 | 10 | | *TBCA* |
| rs2569340 | 5 | 98 | A | -2.81 | 7.62E-06 | 6 | *PHF15* | *SAR1B* |
| rs2589403 | 5 | 85 | A | -2.99 | 1.23E-05 | 13 | *PHF15* | *SAR1B,SEC24A* |
| rs7822400 | 8 | 98 | A | -2.62 | 5.34E-06 | 2 | *XKR6* | |
| rs6985930 | 8 | 94 | G | -2.83 | 1.11E-05 | 11 | *OXR1* | |
| rs10109202 | 8 | 98 | A | -2.54 | 1.27E-05 | 18 | | |
| rs4876526 | 8 | 98 | A | -2.54 | 1.27E-05 | 18 | | |
| rs4876528 | 8 | 98 | A | -2.54 | 1.27E-05 | 18 | | |
| rs2123385 | 8 | 94 | A | -2.54 | 8.97E-06 | 8 | | |
| rs10106553 | 8 | 98 | G | -2.54 | 1.27E-05 | 18 | | |
| rs1452757 | 8 | 98 | G | -2.78 | 6.92E-06 | 4.5 | | |
| rs4876314 | 8 | 98 | A | -2.54 | 1.27E-05 | 18 | | |
| rs10086147 | 8 | 98 | A | -2.78 | 6.92E-06 | 4.5 | | |
| rs902991 | 10 | 98 | G | 2.51 | 4.36E-06 | 1 | *NEURL* | *SH3PXD2A* |
| rs3781366 | 10 | 95 | A | 2.32 | 1.18E-05 | 12 | *NEURL,SH3PXD2A* | |
| rs12413931 | 10 | 98 | A | -3.05 | 6.24E-06 | 3 | *SH3PXD2A* | *NEURL* |
| rs7102442 | 11 | 90 | A | 2.21 | 1.01E-05 | 9 | | |
| rs4784651 | 16 | 98 | A | -2.37 | 1.26E-05 | 15 | *GNAO1* | *DKFZP434H168,AMFR,LOC283856* |

Only SNPs with MAF <10% and imputation accuracy > 0.9 were tested. Age, sex, BMI (body mass index) and altitude were used as covariates.

Table S11. 20 SNPs with lowest hemoglobin association p-values within high altitude Oromo.

| SNP | Chr | N | A1 | β | P | Rank | Genes (within 10kb) | Genes (within 100kb) |
|---|---|---|---|---|---|---|---|---|
| rs12063638 | 1 | 59 | A | 1.80 | 1.16E-05 | 10 | | *PRDM2,PDPN* |
| rs323936 | 1 | 62 | G | -1.23 | 1.54E-05 | 14.5 | | |
| rs323937 | 1 | 62 | A | -1.23 | 1.54E-05 | 14.5 | | |
| rs6682943 | 1 | 62 | A | -1.24 | 1.74E-05 | 16 | | |
| rs6532200 | 4 | 52 | A | 1.48 | 2.25E-05 | 19 | *MMRN1* | *SNCA* |
| rs12189506 | 5 | 58 | G | 1.55 | 2.39E-05 | 20 | | |
| rs2108288 | 7 | 62 | A | 1.63 | 1.52E-05 | 13 | *RAPGEF5* | |
| rs11765705 | 7 | 62 | A | 1.81 | 5.87E-06 | 4 | *POU6F2* | |
| rs721123 | 7 | 62 | G | 1.82 | 1.11E-06 | 1 | *CNTNAP2* | |
| rs1997560 | 8 | 62 | A | 1.60 | 5.91E-06 | 5 | | *GDAP1,JPH1* |
| rs1350172 | 10 | 51 | G | 1.39 | 1.31E-06 | 2 | | *C10orf107* |
| rs780159 | 10 | 62 | A | 1.63 | 9.58E-06 | 9 | *ZMIZ1* | *LOC283050* |
| rs780151 | 10 | 60 | A | 1.60 | 2.10E-05 | 18 | *ZMIZ1* | |
| rs7120319 | 11 | 62 | A | 2.06 | 2.62E-06 | 3 | *OR51B6,OR51B5* | *HBG1,HBBP1,OR51Q1,OR51I1, HBG2,OR51M1,OR51B2,HBE1,OR51B4* |
| rs11038860 | 11 | 57 | G | 1.40 | 9.26E-06 | 7 | *CREB3L1* | *MDK,AMBRA1,CHRM4,DGKZ* |
| rs7476 | 11 | 62 | A | 1.30 | 1.36E-05 | 12 | *CREB3L1* | *MDK,AMBRA1,CHRM4,DGKZ* |
| rs10129651 | 14 | 51 | A | 1.36 | 1.29E-05 | 11 | | *SERPINA13,GSC,SERPINA3* |
| rs3744793 | 17 | 62 | A | -1.32 | 9.27E-06 | 8 | *USP36* | *TIMP2,PSCD1* |
| rs135195 | 22 | 48 | G | -1.33 | 8.23E-06 | 6 | | |
| rs138978 | 22 | 58 | G | -1.29 | 1.78E-05 | 17 | *SCUBE1* | *TSPO,TTLL12,BIK,MCAT* |

Only SNPs with MAF <10% and imputation accuracy > 0.9 were tested. Age, sex and BMI (body mass index) were used as covariates.

Table 12. 20 SNPs with lowest oxygen saturation p-values within high altitude Oromo.

| SNP | Chr | N | A1 | β | P | Rank | Genes (within 10kb) | Genes (within 100kb) |
|---|---|---|---|---|---|---|---|---|
| rs9557 | 1 | 48 | G | -4.491 | 3.73E-06 | 4 | *MAN1A2* | *FAM46C* |
| rs36235 | 3 | 58 | A | -3.698 | 3.97E-06 | 5 | *PRICKLE2* | *PSMD6* |
| rs7650033 | 3 | 58 | A | -4.654 | 2.79E-06 | 2 | | *SUCLG2* |
| rs1173228 | 5 | 63 | A | -3.468 | 9.35E-06 | 11 | | |
| rs1392411 | 5 | 63 | G | -3.987 | 1.04E-05 | 12 | | |
| rs164819 | 5 | 63 | A | -3.43 | 1.30E-05 | 17 | | *TBCA* |
| rs2569340 | 5 | 63 | A | -3.676 | 1.17E-05 | 13 | *PHF15* | *SAR1B* |
| rs4580847 | 6 | 55 | A | -4.749 | 2.26E-06 | 1 | | *LRFN2* |
| rs11786531 | 8 | 61 | G | -4.29 | 9.30E-06 | 10 | | *IMPAD1* |
| rs9297382 | 8 | 59 | G | -3.395 | 8.17E-06 | 8 | *OXR1* | |
| rs6985930 | 8 | 60 | G | -4.439 | 2.87E-06 | 3 | *OXR1* | |
| rs2123385 | 8 | 59 | A | -3.535 | 1.31E-05 | 18 | | |
| rs1452757 | 8 | 63 | G | -3.647 | 1.17E-05 | 14.5 | | |
| rs10086147 | 8 | 63 | A | -3.647 | 1.17E-05 | 14.5 | | |
| rs7067934 | 10 | 63 | A | -5.298 | 5.60E-06 | 6 | *SH2D4B* | *TSPAN14* |
| rs613587 | 11 | 63 | G | -4.361 | 8.90E-06 | 9 | *FLI1* | *KCNJ1* |
| rs4150282 | 13 | 62 | A | 3.598 | 1.33E-05 | 20 | *ERCC5* | *C13orf27,LOC121952,BIVM,KDELC1* |
| rs904864 | 15 | 63 | G | -4.063 | 6.55E-06 | 7 | | *NOX5,TMEM84,GLCE* |
| rs1345856 | 16 | 48 | A | -4.063 | 1.31E-05 | 19 | | |
| rs6075602 | 20 | 63 | G | -3.775 | 1.20E-05 | 16 | *RIN2* | *NAT5,CRNKL1,C20orf26* |

Only SNPs with MAF <10% and imputation accuracy > 0.9 were tested. Age, sex and BMI (body mass index) were used as covariates.

Table S13. 20 SNPs with lowest hemoglobin p-values within low altitude Oromo.

| SNP | Chr | N | A1 | β | P | Rank | Genes (within 10kb) | Genes (within 100kb) |
|---|---|---|---|---|---|---|---|---|
| rs4553221 | 1 | 32 | A | 1.591 | 2.63E-05 | 17 | *CAMTA1* | |
| rs696119 | 1 | 32 | A | -1.395 | 3.65E-06 | 5 | *C1orf173* | *CRYZ,TYW3* |
| rs13405506 | 2 | 32 | A | -1.158 | 1.19E-06 | 3 | *RAMP1* | *RBM44,UBE2F,LRRFIP1,SCLY* |
| rs2168494 | 4 | 31 | A | 1.748 | 1.45E-05 | 11.5 | | |
| rs959628 | 4 | 31 | A | 1.748 | 1.45E-05 | 11.5 | | |
| rs1378923 | 4 | 30 | A | 1.748 | 2.14E-05 | 16 | | |
| rs6842580 | 4 | 32 | G | 1.528 | 3.60E-06 | 4 | | |
| rs17343168 | 5 | 32 | G | 1.806 | 5.22E-06 | 6 | | |
| rs10484824 | 6 | 32 | G | 1.346 | 3.88E-07 | 1 | *KIF6* | |
| rs10456473 | 6 | 32 | A | 1.193 | 9.44E-07 | 2 | *KIF6* | |
| rs6458179 | 6 | 32 | G | 0.9942 | 9.63E-06 | 8 | | |
| rs4548215 | 8 | 32 | G | 1.288 | 8.68E-06 | 7 | | |
| rs3737147 | 9 | 32 | G | 1.764 | 1.46E-05 | 13.5 | *NOL8,CENPP* | *OGN,OMD,IARS,SNORA84* |
| rs7872423 | 9 | 32 | G | 1.764 | 1.46E-05 | 13.5 | *NOL8,CENPP* | *OGN,OMD,IARS,SNORA84* |
| rs10999916 | 10 | 28 | A | 1.477 | 2.64E-05 | 18 | *CDH23* | |
| rs1327314 | 13 | 29 | G | 1.334 | 2.80E-05 | 19 | | |
| rs1077918 | 15 | 32 | A | 1.826 | 1.69E-05 | 15 | *VPS18* | *RHOV,DLL4,CHAC1,INOC1, SPINT1,ZFYVE19,PPP1R14D* |
| rs1261084 | 18 | 31 | A | 1.074 | 1.28E-05 | 9 | *TCF4* | |
| rs5755469 | 22 | 25 | G | 1.377 | 1.33E-05 | 10 | | *ISX* |

Only SNPs with MAF <10% and imputation accuracy > 0.9 were tested. Age, sex and BMI (body mass index) were used as covariates.

Table S14. 20 SNPs with lowest oxygen saturation p-values within low altitude Oromo.

| SNP | Chr | N | A1 | β | P | Rank | Genes (within 10kb) | Genes (within 100kb) |
|---|---|---|---|---|---|---|---|---|
| rs12088462 | 1 | 35 | A | -1.93 | 6.56E-06 | 4 | SLC44A5 | |
| rs4598448 | 1 | 34 | A | -1.93 | 2.02E-05 | 19 | SLC44A5 | |
| rs1999493 | 1 | 34 | G | -1.91 | 1.10E-05 | 9 | SLC44A5 | |
| rs9819197 | 3 | 35 | A | -1.88 | 6.60E-06 | 5 | | |
| rs149473 | 5 | 35 | G | -1.12 | 4.39E-06 | 2 | | |
| rs12662109 | 6 | 35 | G | -1.67 | 5.82E-06 | 3 | | |
| rs17128582 | 8 | 34 | G | -1.89 | 9.01E-06 | 8 | ChGn | |
| rs16912021 | 9 | 35 | A | -1.36 | 1.26E-05 | 10 | FANCC | C9orf3 |
| rs12574036 | 11 | 33 | A | -1.56 | 1.92E-05 | 17 | LUZP2 | |
| rs2512637 | 11 | 35 | G | -1.65 | 1.74E-05 | 15 | | ODZ4 |
| rs10774174 | 12 | 35 | G | -1.41 | 1.27E-05 | 12.5 | PARP11 | EFCAB4B |
| rs12424633 | 12 | 35 | A | -1.41 | 1.27E-05 | 12.5 | PARP11 | EFCAB4B |
| rs7965615 | 12 | 35 | A | -1.41 | 1.27E-05 | 12.5 | PARP11 | EFCAB4B |
| rs11062846 | 12 | 35 | A | -1.41 | 1.27E-05 | 12.5 | PARP11 | EFCAB4B |
| rs3825374 | 12 | 34 | A | -1.51 | 1.24E-06 | 1 | PARP11 | EFCAB4B |
| rs11829730 | 12 | 33 | G | -1.70 | 8.90E-06 | 7 | SOX5 | |
| rs9520256 | 13 | 35 | G | 1.14 | 1.95E-05 | 18 | | |
| rs4414463 | 15 | 35 | A | 1.14 | 8.73E-06 | 6 | CSPG4 | SNUPN,ODF3L1,SH3PX3,IMP3 |
| rs16950701 | 18 | 35 | A | -1.38 | 2.14E-05 | 20 | LIPG | |
| rs6038686 | 20 | 33 | G | -1.96 | 1.88E-05 | 16 | | |

Only SNPs with MAF <10% and imputation accuracy > 0.9 were tested. Age, sex and BMI (body mass index) were used as covariates.

Table S15. 20 SNPs with lowest hemoglobin p-values within total Ethiopian sample.

| SNP | Chr | N | A1 | β | P | Rank | Genes (within 10kb) | Genes (within 100kb) |
|---|---|---|---|---|---|---|---|---|
| rs1574106 | 1 | 238 | A | -0.52 | 1.54E-05 | 15 | | TINAGL1,SERINC2,LOC284551 |
| rs11811630 | 1 | 241 | A | -0.58 | 1.66E-05 | 17 | NCF2 | APOBEC4,SMG7,RGL1,ARPC5 |
| rs17453871 | 2 | 220 | A | 0.65 | 9.72E-06 | 9 | | YWHAQ,ADAM17 |
| rs2742347 | 2 | 223 | A | -0.66 | 1.60E-05 | 16 | TTN | CCDC141 |
| rs2627037 | 2 | 222 | A | -0.61 | 7.30E-06 | 8 | TTN | CCDC141 |
| rs7590740 | 2 | 226 | G | -0.70 | 1.07E-05 | 11 | TTN | CCDC141 |
| rs6556187 | 5 | 241 | G | 0.85 | 2.49E-06 | 4 | | |
| rs6978495 | 7 | 241 | G | 0.50 | 1.72E-05 | 18 | PDE1C | |
| rs1062831 | 7 | 230 | G | 0.70 | 5.45E-06 | 7 | RELN | SLC26A5 |
| rs4147310 | 8 | 241 | G | 0.53 | 1.23E-05 | 13 | | |
| rs7068383 | 10 | 201 | G | -0.60 | 1.42E-05 | 14 | KIAA1217 | KIAA1217 |
| rs6480379 | 10 | 241 | G | 0.54 | 1.01E-05 | 10 | | DDX21,SRGN,VPS26A,KIAA1279 |
| rs4745975 | 10 | 232 | G | 0.62 | 1.38E-06 | 1 | | DDX21,SRGN,VPS26A,KIAA1279 |
| rs1524250 | 12 | 226 | G | -0.54 | 1.77E-05 | 19 | CNOT2 | KCNMB4 |
| rs2619133 | 12 | 196 | A | 0.56 | 1.92E-05 | 20 | | SPIC,MYBPC1 |
| rs4424941 | 17 | 236 | A | 0.65 | 3.75E-06 | 6 | | PRKCA,CCDC46,APOH |
| rs4305120 | 17 | 237 | A | 0.64 | 3.60E-06 | 5 | | PRKCA,CCDC46,APOH |
| rs8081614 | 17 | 241 | G | 0.66 | 1.63E-06 | 2 | | PRKCA,CCDC46,APOH |
| rs16959046 | 17 | 241 | G | 0.75 | 2.24E-06 | 3 | PRKCA | APOH |
| rs440624 | 21 | 226 | G | -0.75 | 1.21E-05 | 12 | CHODL | |

Only SNPs with MAF <10% and imputation accuracy > 0.9 were tested. Age, sex, BMI (body mass index), collection year, altitude and ethnicity were used as covariates.

Table S16. 20 SNPs with lowest oxygen saturation p-values within total Ethiopian sample.

| SNP | Chr | N | A1 | β | P | Rank | Genes (within 10kb) | Genes (within 100kb) |
|---|---|---|---|---|---|---|---|---|
| rs12733615 | 1 | 259 | A | 1.24 | 7.85E-06 | 7 | | UHMK1,UAP1,FLJ13137,SH2D1B |
| rs4675041 | 2 | 259 | A | -1.39 | 8.70E-06 | 9 | | |
| rs6802224 | 3 | 246 | G | 1.29 | 3.85E-06 | 3 | | |
| rs7703046 | 5 | 253 | G | -1.36 | 8.07E-06 | 8 | | |
| rs11738661 | 5 | 256 | A | -1.30 | 1.47E-05 | 19 | | |
| rs337715 | 5 | 259 | G | -1.43 | 1.16E-05 | 15 | KCNN2 | |
| rs1803989 | 6 | 258 | A | -1.73 | 1.38E-05 | 18 | LOC729603, IGF2R | SLC22A2,SLC22A1 |
| rs1035153 | 7 | 259 | A | -1.52 | 9.26E-06 | 10 | | |
| rs6996198 | 8 | 259 | A | -1.25 | 9.28E-06 | 11 | | BHLHB5,CYP7B1,LOC401463 |
| rs10104685 | 8 | 255 | G | -1.40 | 1.08E-05 | 14 | | |
| rs2123385 | 8 | 252 | A | -1.43 | 1.02E-05 | 12 | | |
| rs1452757 | 8 | 259 | G | -1.51 | 1.28E-05 | 16.5 | | |
| rs10086147 | 8 | 259 | A | -1.51 | 1.28E-05 | 16.5 | | |
| rs6420192 | 8 | 258 | A | -1.38 | 3.90E-06 | 4 | | PARP10,NRBP2,SCRIB,EPPK1, PLEC1,PUF60,GRINA |
| rs12380152 | 9 | 254 | A | -1.57 | 1.57E-05 | 20 | PTPRD | |
| rs10869434 | 9 | 246 | A | 1.40 | 2.39E-06 | 2 | PIP5K1B | FAM122A,PIP5K1B |
| rs17392931 | 9 | 251 | G | 1.29 | 4.60E-06 | 5 | PIP5K1B | FAM122A,PIP5K1B |
| rs11144066 | 9 | 245 | G | 1.35 | 2.12E-06 | 1 | PIP5K1B | FAM122A,PIP5K1B |
| rs11857947 | 15 | 234 | A | 1.34 | 5.70E-06 | 6 | CIB2 | IDH3A,ACSBG1,TBC1D2B,hCG_38941 |
| rs176096 | 17 | 253 | A | 1.34 | 1.04E-05 | 13 | FLJ45455 | |

Only SNPs with MAF <10% and imputation accuracy > 0.9 were tested. Age, sex, BMI (body mass index), collection year, altitude and ethnicity were used as covariates.

Table S17. 20 SNPs with lowest hemoglobin p-values within the total high altitude Ethiopian sample.

| SNP | Chr | N | A1 | β | P | Rank | Genes (within 10kb) | Genes (within 100kb) |
|---|---|---|---|---|---|---|---|---|
| rs1880418 | 1 | 147 | A | 0.71 | 1.49E-05 | 6 | | |
| rs7548781 | 1 | 154 | A | 0.87 | 3.19E-05 | 17.5 | HSD11B1 | LAMB3,TRAF3IP3,C1orf107, G0S2,C1orf74,IRF6 |
| rs13400823 | 2 | 140 | A | -0.72 | 2.29E-05 | 14 | MYT1L | |
| rs2624520 | 5 | 141 | A | 0.95 | 2.08E-05 | 13 | | |
| rs1586656 | 6 | 156 | G | 1.05 | 2.94E-05 | 15 | EGFL11 | |
| rs7801660 | 7 | 156 | A | 0.81 | 1.92E-05 | 11 | POU6F2 | |
| rs6952464 | 7 | 155 | A | 0.79 | 3.41E-05 | 19 | POU6F2 | |
| rs11765705 | 7 | 156 | A | 1.05 | 4.65E-06 | 2 | POU6F2 | |
| rs10951593 | 7 | 148 | A | 1.09 | 1.01E-05 | 3 | POU6F2 | |
| rs10260368 | 7 | 156 | G | -0.78 | 1.48E-05 | 5 | ST7OT2,ST7 | |
| rs4627213 | 13 | 155 | A | 1.10 | 3.61E-06 | 1 | | |
| rs2322233 | 13 | 150 | G | 1.05 | 1.15E-05 | 4 | | |
| rs10129651 | 14 | 135 | A | 0.74 | 1.83E-05 | 8 | | SERPINA3,GSC,SERPINA13 |
| rs915378 | 14 | 155 | A | 0.76 | 2.06E-05 | 12 | | |
| rs4842899 | 15 | 144 | A | 0.81 | 1.87E-05 | 10 | | KLHL25,AKAP13 |
| rs7164649 | 15 | 145 | A | -0.65 | 3.19E-05 | 17.5 | | ACAN,HAPLN3,MFGE8 |
| rs7163586 | 15 | 149 | G | -0.66 | 3.04E-05 | 16 | | HAPLN3,MFGE8,ACAN |
| rs16959046 | 17 | 156 | G | 0.89 | 1.72E-05 | 7 | PRKCA | APOH |
| rs681307 | 18 | 155 | A | -0.69 | 1.86E-05 | 9 | | |

Only SNPs with MAF <10% and imputation accuracy > 0.9 were tested. Age, sex, BMI (body mass index), collection year and ethnicity were used as covariates.

Table S18. 20 SNPs with lowest oxygen saturation p-values within the total high altitude Ethiopian sample.

| SNP | Chr | N | A1 | β | P | Rank | Genes (within 10kb) | Genes (within 100kb) |
|---|---|---|---|---|---|---|---|---|
| rs6751526 | 2 | 164 | G | -2.32 | 1.48E-05 | 18 | | |
| rs7438763 | 4 | 162 | G | 2.08 | 1.54E-05 | 20 | LRIT3,RRH | EGF,NOLA1,CFI |
| rs1990250 | 4 | 156 | G | 2.30 | 5.97E-06 | 10 | LRIT3,RRH | EGF,NOLA1,CFI |
| rs13160737 | 5 | 148 | G | -2.27 | 1.18E-05 | 15 | ZFR | MTMR12 |
| rs9464002 | 6 | 162 | A | -2.68 | 2.16E-06 | 1 | LRRC1 | C6orf142 |
| rs6937549 | 6 | 164 | A | -2.50 | 1.08E-05 | 14 | C6orf176 | |
| rs10269222 | 7 | 164 | A | -2.15 | 1.19E-05 | 16 | SDK1 | |
| rs10104685 | 8 | 160 | G | -1.99 | 4.15E-06 | 4 | | |
| rs2123385 | 8 | 158 | A | -2.08 | 2.63E-06 | 2 | | |
| rs1452757 | 8 | 164 | G | -2.09 | 5.82E-06 | 8.5 | | |
| rs10086147 | 8 | 164 | A | -2.09 | 5.82E-06 | 8.5 | | |
| rs6986020 | 8 | 164 | G | -1.83 | 1.24E-05 | 17 | | |
| rs11144066 | 9 | 154 | G | 1.72 | 1.50E-05 | 19 | PIP5K1B | FAM122A,PIP5K1B |
| rs602347 | 11 | 141 | A | 2.09 | 4.78E-06 | 5 | EHD1 | MEN1,ATG2A,MAP4K2,GPHA2, LOC283129,PPP2R5B,CDC42BPG |
| rs1647105 | 12 | 164 | A | 1.64 | 1.07E-05 | 13 | | |
| rs11857947 | 15 | 144 | A | 1.83 | 5.41E-06 | 7 | CIB2 | IDH3A,ACSBG1,TBC1D2B,hCG_38941 |
| rs10775410 | 17 | 164 | A | -1.64 | 7.45E-06 | 12 | MYO1D | CDK5R1,PSMD11 |
| rs7263002 | 20 | 164 | G | -2.62 | 6.28E-06 | 11 | SNRPB,SNORD119 | TMC2,ZNF343,TGM6 |
| rs6114384 | 20 | 164 | A | -2.63 | 5.22E-06 | 6 | SNRPB,SNORD119 | TMC2,ZNF343,TGM6 |
| rs8114620 | 20 | 159 | G | -2.63 | 2.65E-06 | 3 | SNRPB,SNORD119 | TMC2,ZNF343,TGM6 |

Only SNPs with MAF <10% and imputation accuracy > 0.9 were tested. Age, sex, BMI (body mass index), collection year and ethnicity were used as covariates.

Table S19. 20 SNPs with lowest hemoglobin p-values within the total low altitude Ethiopian sample.

| SNP | Chr | N | A1 | β | P | Rank | Genes (within 10kb) | Genes (within 100kb) |
|---|---|---|---|---|---|---|---|---|
| rs12119834 | 1 | 85 | A | 0.98 | 1.39E-05 | 11 | *PCTK3* | *ELK4,MFSD4,LOC284578,LEMD1* |
| rs10510829 | 3 | 85 | G | 1.22 | 1.78E-05 | 14 | *FHIT* | |
| rs11711620 | 3 | 85 | G | 1.31 | 1.09E-06 | 1 | *ZBTB20* | |
| rs1506792 | 4 | 82 | A | 1.06 | 1.11E-05 | 9 | | |
| rs1506791 | 4 | 73 | A | 1.21 | 1.01E-05 | 8 | | |
| rs4146409 | 4 | 85 | A | -0.92 | 2.54E-06 | 2 | | |
| rs11944671 | 4 | 85 | A | -0.92 | 8.30E-06 | 5 | | |
| rs1363073 | 5 | 85 | A | 1.12 | 1.54E-05 | 12 | | |
| rs13195572 | 6 | 85 | G | 1.16 | 2.45E-05 | 19.5 | | *C6orf138* |
| rs13198330 | 6 | 85 | G | 1.16 | 2.45E-05 | 19.5 | | *C6orf138* |
| rs11992396 | 8 | 80 | A | 1.09 | 1.79E-05 | 16 | *PSD3* | |
| rs1447293 | 8 | 85 | A | 0.90 | 8.77E-06 | 6 | *LOC727677* | *POU5F1P1* |
| rs11142705 | 9 | 85 | A | 0.89 | 1.25E-05 | 10 | *TRPM3* | |
| rs12351692 | 9 | 77 | A | 1.35 | 9.03E-06 | 7 | *CORO2A,TRIM14* | *TBC1D2,NANS* |
| rs10788462 | 10 | 85 | A | 1.25 | 1.91E-05 | 17 | *GRID1* | |
| rs2928011 | 10 | 84 | A | 1.17 | 1.78E-05 | 15 | *GRID1* | |
| rs7919376 | 10 | 84 | A | 1.24 | 2.38E-05 | 18 | *GRID1* | |
| rs4757615 | 11 | 84 | A | -0.84 | 7.40E-06 | 3 | *SAAL1* | *MRGPRX4,TPH1,SERGEF,MRGPRX3,SAA3P* |
| rs734309 | 16 | 84 | A | -0.85 | 1.60E-05 | 13 | *GPT2* | *LOC388272,DNAJA2* |
| rs6089316 | 20 | 80 | A | 0.82 | 8.30E-06 | 4 | *TAF4* | *PSMA7,SS18L1,LSM14B* |

Only SNPs with MAF <10% and imputation accuracy > 0.9 were tested. Age, sex, BMI (body mass index), collection year and ethnicity were used as covariates.

Table S20. 20 SNPs with lowest oxygen saturation p-values within the total low altitude Ethiopian sample.

| SNP | Chr | N | A1 | β | P | Rank | Genes (within 10kb) | Genes (within 100kb) |
|---|---|---|---|---|---|---|---|---|
| rs6690390 | 1 | 95 | A | 1.16 | 8.52E-06 | 8 | | |
| rs11582141 | 1 | 95 | G | 0.97 | 8.61E-06 | 9 | | |
| rs2792251 | 1 | 95 | G | 0.84 | 1.42E-05 | 14 | *PBX1* | |
| rs4551629 | 1 | 74 | G | -1.20 | 7.85E-06 | 7 | | *LOC441931,ZNF496,ZNF124, VN1R5,FLJ45717* |
| rs7581685 | 2 | 90 | G | -1.00 | 1.82E-05 | 19 | | *SPAG16* |
| rs10458166 | 6 | 95 | G | -0.93 | 1.17E-05 | 11 | *PRPF4B* | *C6orf146,C6orf201* |
| rs980962 | 6 | 91 | G | -1.15 | 8.51E-07 | 2 | *ZNF391* | *FKSG83,ZNF184,ZNF204* |
| rs9373984 | 6 | 76 | G | 0.88 | 5.88E-06 | 5 | *SCML4* | *SEC63* |
| rs7784712 | 7 | 90 | A | -0.98 | 1.78E-05 | 17 | | *HDAC9,TWIST1* |
| rs2392591 | 7 | 90 | A | -0.82 | 6.03E-06 | 6 | | *AMPH,LOC340286,VPS41* |
| rs16890858 | 8 | 95 | A | -1.04 | 8.80E-06 | 10 | *SAMD12* | |
| rs10988467 | 9 | 95 | A | -0.80 | 4.83E-06 | 4 | *PRRX2* | *C9orf32,ASB6,C9orf50,PTGES* |
| rs10786940 | 10 | 85 | G | 0.95 | 1.89E-05 | 20 | | |
| rs7970151 | 12 | 95 | A | -1.21 | 1.78E-05 | 18 | | |
| rs8038032 | 15 | 90 | G | 0.82 | 1.50E-05 | 15 | | *ZNF770,AQR,hCG_1787519* |
| rs11083340 | 18 | 95 | A | -0.86 | 1.61E-05 | 16 | | |
| rs4369752 | 18 | 88 | A | -0.88 | 2.16E-06 | 3 | | |
| rs183049 | 19 | 92 | A | -0.88 | 1.41E-05 | 13 | *SLC6A16* | *FLJ32658,DKKL1,TEAD2,CD37* |
| rs659555 | 19 | 91 | A | -0.90 | 1.25E-05 | 12 | *SLC6A16,CD37* | *FLJ32658,DKKL1,TEAD2,PTH2* |
| rs35270176 | 19 | 89 | A | -1.27 | 1.73E-07 | 1 | *LOC199800,CPT1C,PRMT1* | *AP2A1,SCAF1,IRF3,PRR12, RRAS,BCL2L12,TSKS,PRRG2* |

Only SNPs with MAF <10% and imputation accuracy > 0.9 were tested. Age, sex, BMI (body mass index), collection year and ethnicity were used as covariates.

Table S21. Power calculations within the Ethiopian samples.

| | | Tibetans[1] | | Amhara | | | Oromo | | | Ethiopia | | |
|---|---|---|---|---|---|---|---|---|---|---|---|---|
| SNP | Gene | $\beta$[2] | half $\beta$[3] | N | MAF | Power[3] | N | MAF | Power[3] | N | MAF | Power[3] |
| rs961154 | EGLN1 | 1.70 | 0.85 | 141 | 0.39 | 100 | 92 | 0.41 | 100 | 233 | 0.40 | 100 |
| rs2790859 | EGLN1 | 1.70 | 0.85 | 141 | 0.39 | 100 | 92 | 0.41 | 100 | 233 | 0.40 | 100 |
| rs1992846 | EPAS1 | 0.84 | 0.42 | 124 | 0.36 | 90 | 78 | 0.42 | 74 | 202 | 0.38 | 99 |
| rs7594278 | EPAS1 | 0.52 | 0.26 | 131 | 0.31 | 49 | 77 | 0.24 | 27 | 208 | 0.28 | 66 |
| rs6544887 | EPAS1 | 0.79 | 0.40 | 130 | 0.34 | 86 | 86 | 0.34 | 70 | 216 | 0.34 | 98 |
| rs17035010 | EPAS1 | 0.84 | 0.42 | 136 | 0.39 | 93 | 85 | 0.30 | 72 | 221 | 0.36 | 99 |
| rs3768729 | EPAS1 | 0.80 | 0.40 | 133 | 0.44 | 91 | 86 | 0.44 | 75 | 219 | 0.44 | 99 |
| rs7583554 | EPAS1 | 0.94 | 0.47 | 138 | 0.46 | 98 | 90 | 0.41 | 89 | 228 | 0.44 | 100 |
| rs7583088 | EPAS1 | 0.92 | 0.46 | 144 | 0.30 | 96 | 94 | 0.22 | 75 | 238 | 0.27 | 100 |
| rs11678465 | EPAS1 | 0.85 | 0.43 | 144 | 0.30 | 92 | 93 | 0.22 | 67 | 237 | 0.27 | 99 |
| rs6712143 | EPAS1 | 0.94 | 0.47 | 145 | 0.40 | 98 | 92 | 0.32 | 86 | 237 | 0.37 | 100 |
| rs4953342 | EPAS1 | 0.90 | 0.45 | 147 | 0.21 | 89 | 94 | 0.12 | 52 | 241 | 0.18 | 97 |
| rs2121266 | EPAS1 | 1.02 | 0.51 | 147 | 0.38 | 99 | 94 | 0.30 | 91 | 241 | 0.35 | 100 |
| rs9973653 | EPAS1 | 0.52 | 0.26 | 147 | 0.35 | 57 | 94 | 0.30 | 37 | 241 | 0.33 | 77 |
| rs1374749 | EPAS1 | 0.88 | 0.44 | 147 | 0.47 | 97 | 94 | 0.49 | 87 | 241 | 0.48 | 100 |
| rs4953353 | EPAS1 | 0.97 | 0.49 | 147 | 0.31 | 98 | 94 | 0.31 | 88 | 241 | 0.31 | 100 |
| rs6756667 | EPAS1 | 0.93 | 0.47 | 147 | 0.37 | 98 | 94 | 0.32 | 86 | 241 | 0.35 | 100 |
| rs7571218 | EPAS1 | 0.71 | 0.36 | 147 | 0.50 | 87 | 94 | 0.49 | 69 | 241 | 0.50 | 98 |

[1] Genotype-phenotype association beta coefficients for EGLN1 were obtained from Simonson *et al* [16] while those for EPAS1 were obtained from Beall *et al* [17].
[2] $\beta$ indicates the observed linear coefficient for the relationship between SNP genotype and Hb levels.
[3] half $\beta$ indicates half of the observed linear coefficient for the relationship between SNP genotype and Hb levels.
[3] Power refers to the probability of detecting a significant association (p<0.05) between SNP genotype and Hb level given the MAF and the sample size in the Ethiopian populations assuming that the $\beta$ coefficient is higher than half of the observed in Tibetan

Table S22. Proportions of Response to Hypoxia genic SNPs relative to the proportion of all other genic SNPs in the tails of the PBS distributions for each population trio tested. * and ** denote support from ≥ 95% and 99% of bootstrap replicates, respectively.

| | Populations | | Hypoxia genes : all other genes PBS tail cut-off | | |
|---|---|---|---|---|---|
| A | B | C | 0.005 | 0.01 | 0.05 |
| **Amhara** | Maasai | Luhya | 3.40* | 3.41** | 1.84* |
| **Amhara** | Maasai | Yoruba | 3.432 | 3.058* | 1.929 |
| **Amhara** | LA Oromo | HA Oromo | 2.628 | 1.951 | 2.203* |
| **Amhara** | Luhya | Europe | 1.899 | 2.074 | 1.468 |
| **Amhara** | Luhya | Yoruba | 1.151 | 2.502 | 1.586 |
| **Amhara** | Maasai | Europe | 1.862 | 2.124 | 1.406 |
| **HA Amhara** | LA Amhara | LA Oromo | 3.31 | 2.90 | 1.92* |
| **HA Amhara** | LA Amhara | Oromo | 2.215 | 2.003 | 1.557 |
| **HA Amhara** | LA Amhara | Yoruba | 1.029 | 1.843 | 2.03** |
| **HA Amhara** | LA Amhara | Maasai | 1.675 | 2.147 | 1.774 |
| **HA Amhara** | LA Amhara | Europe | 2.297 | 1.619 | 1.121 |
| **HA Amhara** | LA Amhara | Luhya | 0.585 | 1.167 | 2.088* |
| **LA Amhara** | HA Oromo | LA Oromo | 1.174 | 1.528 | 1.149 |
| **LA Amhara** | HA Amhara | Oromo | 0.897 | 0.81 | 1.131 |
| **LA Amhara** | HA Amhara | Maasai | 1.172 | 0.995 | 0.944 |
| **LA Amhara** | HA Amhara | Europe | 1.112 | 0.678 | 1.026 |
| **LA Amhara** | HA Amhara | Luhya | 1.215 | 0.608 | 0.939 |
| **LA Amhara** | HA Amhara | Yoruba | 1.134 | 0.567 | 0.778 |
| **LA Amhara** | HA Amhara | LA Oromo | 0.69 | 0.704 | 0.895 |
| **Ethiopia** | Maasai | Yoruba | 3.991 | 3.162* | 1.578 |
| **Ethiopia** | Luhya | Europe | 3.458 | 2.334 | 1.226 |
| **Ethiopia** | Luhya | Yoruba | 1.358 | 1.675 | 1.581 |
| **Ethiopia** | Maasai | Europe | 1.184 | 1.169 | 1.077 |
| **HA Ethiopia** | LA Ethiopia | Yoruba | 2.105 | 2.367* | 1.442* |
| **HA Ethiopia** | LA Ethiopia | Luhya | 0.551 | 1.378 | 1.747* |
| **HA Ethiopia** | LA Ethiopia | Maasai | 1.128 | 0.824 | 1.323 |
| **HA Ethiopia** | LA Ethiopia | Europe | 0.584 | 0.82 | 0.971 |
| **LA Ethiopia** | HA Ethiopia | Europe | 1.104 | 1.077 | 1.247 |
| **LA Ethiopia** | HA Ethiopia | Yoruba | 1.756 | 1.164 | 0.653 |
| **LA Ethiopia** | HA Ethiopia | Maasai | 1.132 | 0.891 | 0.9 |
| **LA Ethiopia** | HA Ethiopia | Luhya | 1.18 | 1.176 | 0.687 |
| **Europe** | Luhya | Oromo | 1.74 | 1.917 | 1.542 |
| **Europe** | LA Oromo | HA Oromo | 2.062 | 1.463 | 1.769* |
| **Europe** | Maasai | Oromo | 1.071 | 2.039 | 1.737 |
| **Europe** | LA Ethiopia | HA Ethiopia | 1.695 | 1.364 | 1.332 |
| **Europe** | Luhya | Ethiopia | 0.001 | 1.558 | 1.411 |
| **Europe** | Luhya | Amhara | 0.044 | 1.149 | 1.448 |
| **Europe** | LA Amhara | HA Amhara | 0.199 | 1.303 | 1.213 |
| **Europe** | Maasai | Ethiopia | 0.008 | 1.173 | 1.375 |
| **Europe** | Maasai | Amhara | 0 | 1.11 | 1.39 |
| **Luhya** | LA Amhara | HA Amhara | 2.274 | 1.675 | 1.55 |
| **Luhya** | LA Ethiopia | HA Ethiopia | 2.317 | 1.686 | 1.357 |
| **Luhya** | Ethiopia | Europe | 1.202 | 1.923 | 1.741 |
| **Luhya** | Amhara | Europe | 1.174 | 1.693 | 1.818* |
| **Luhya** | Oromo | Europe | 1.722 | 1.431 | 1.286 |
| **Luhya** | Amhara | Yoruba | 1.667 | 2.068 | 1.494 |
| **Luhya** | LA Oromo | HA Oromo | 1.952 | 1.513 | 1.222 |
| **Luhya** | Ethiopia | Yoruba | 1.13 | 1.4 | 1.363 |
| **Luhya** | Oromo | Yoruba | 0.881 | 1.388 | 1.38 |
| **Luhya** | Maasai | Amhara | 0.457 | 1.151 | 0.953 |
| **Luhya** | Maasai | Oromo | 0.516 | 0.584 | 0.888 |
| **Maasai** | Amhara | Europe | 3.022 | 2.468 | 1.704 |
| **Maasai** | LA Amhara | HA Amhara | 4.125 | 3.497* | 1.706 |
| **Maasai** | Ethiopia | Europe | 1.851 | 2.408 | 1.59 |
| **Maasai** | LA Ethiopia | HA Ethiopia | 2.343 | 2.645 | 1.542 |
| **Maasai** | LA Oromo | HA Oromo | 4.41 | 2.683 | 0.964 |
| **Maasai** | Oromo | Europe | 1.541 | 1.408 | 1.377 |
| **Maasai** | Amhara | Yoruba | 1.675 | 1.122 | 1.202 |
| **Maasai** | Ethiopia | Yoruba | 1.746 | 0.873 | 1.048 |
| **Maasai** | Amhara | Luhya | 1.753 | 1.144 | 0.821 |
| **Maasai** | Oromo | Yoruba | 1.689 | 1.127 | 0.824 |
| **Maasai** | Oromo | Luhya | 1.194 | 0.785 | 0.972 |
| **Oromo** | LA Amhara | HA Amhara | 2.69 | 2.19 | 2.114** |
| **Oromo** | Maasai | Luhya | 2.988 | 1.756 | 1.445 |
| **Oromo** | Maasai | Yoruba | 3.464 | 2.002 | 1.038 |
| **Oromo** | Luhya | Europe | 1.261 | 1.719 | 1.165 |
| **Oromo** | Luhya | Yoruba | 0.916 | 1.391 | 1.116 |
| **Oromo** | Maasai | Europe | 1.175 | 1.261 | 0.936 |
| **HA Oromo** | LA Amhara | LA Oromo | 1.858 | 1.567 | 1.155 |
| **HA Oromo** | LA Oromo | Amhara | 1.324 | 1.857 | 0.835 |
| **HA Oromo** | LA Oromo | Europe | 0.502 | 1.622 | 1.161 |
| **HA Oromo** | LA Oromo | Yoruba | 0.001 | 1.353 | 1.089 |
| **HA Oromo** | LA Oromo | Maasai | 0.606 | 0.913 | 0.993 |
| **HA Oromo** | LA Oromo | Luhya | 0.53 | 0.266 | 1.159 |
| **LA Oromo** | HA Oromo | Europe | 2.869 | 1.972 | 1.355 |
| **LA Oromo** | HA Oromo | Yoruba | 3.754 | 1.903 | 1.162 |
| **LA Oromo** | HA Oromo | Amhara | 1.999 | 1.558 | 1.166 |
| **LA Oromo** | HA Oromo | Maasai | 1.671 | 1.478 | 1.064 |
| **LA Oromo** | LA Amhara | HA Amhara | 0.219 | 1.189 | 1.393 |
| **LA Oromo** | LA Amhara | HA Oromo | 0.889 | 0.898 | 0.837 |
| **LA Oromo** | HA Oromo | Luhya | 0.53 | 0.266 | 1.159 |
| **Yoruba** | LA Ethiopia | HA Ethiopia | 1.588 | 1.309 | 1.587 |
| **Yoruba** | LA Amhara | HA Amhara | 1.057 | 2.15 | 1.617 |
| **Yoruba** | LA Oromo | HA Oromo | 2 | 1.259 | 1.32 |
| **Yoruba** | Maasai | Amhara | 1.71 | 1.092 | 1.312 |
| **Yoruba** | French | Oromo | 0.532 | 1.703 | 1.808 |
| **Yoruba** | Maasai | Ethiopia | 1.597 | 1.077 | 1.227 |
| **Yoruba** | Maasai | Oromo | 1.597 | 0.799 | 1.213 |
| **Yoruba** | Luhya | Ethiopia | 2.149 | 1.074 | 0.773 |
| **Yoruba** | Luhya | Oromo | 2.2 | 1.1 | 0.737 |
| **Yoruba** | Luhya | Amhara | 2.181 | 1.091 | 0.739 |

Table S23. List of the 20 SNPs with the largest Amhara PBS (*vs*. Maasai and Luya).

| SNP | Chr | Nt. pos. | Rank | Gene (within 10kb) | Gene (within 100kb) |
|---|---|---|---|---|---|
| rs619660 | 3 | 43002953 | 2 | LOC729085 | *C3orf39,ZNF662,CCBP2* |
| rs9853065 | 3 | 174475785 | 5 | NA | *NA* |
| rs6994475 | 8 | 1260833 | 3 | NA | *NA* |
| rs6558447 | 8 | 1261463 | 9 | NA | *NA* |
| rs2357735 | 8 | 115943237 | 13 | NA | *NA* |
| rs10112994 | 8 | 116016470 | 14 | NA | *NA* |
| rs1888076 | 9 | 72945254 | 17 | TRPM3 | *TRPM3* |
| rs7860423 | 9 | 140075368 | 4 | CACNA1B | *NA* |
| rs11137360 | 9 | 140077378 | 12 | CACNA1B | *NA* |
| rs7940567 | 11 | 7078885 | 1 | RBMXL2 | *ZNF214,NLRP14* |
| rs4375446 | 11 | 66667402 | 10 | FBXL11 | *RHOD,SYT12* |
| rs674499 | 11 | 66720001 | 11 | FBXL11 | *ANKRD13D,ADRBK1* |
| rs4542419 | 11 | 66727173 | 6 | FBXL11 | *ANKRD13D,ADRBK1* |
| rs3927807 | 11 | 66755431 | 8 | FBXL11 | *SSH3,ADRBK1,ANKRD13D* |
| rs1638566 | 11 | 66880909 | 15.5 | POLD4,CLCF1 | *PTPRCAP,CORO1B,GPR152,TBC1D10C,KIAA1394,RPS6KB2 CABP4,PPP1CA,ADRBK1,ANKRD13D,SSH3,RAD9A,FBXL11* |
| rs1638567 | 11 | 66881799 | 15.5 | POLD4,CLCF1 | *PTPRCAP,CORO1B,GPR152,TBC1D10C,KIAA1394,RPS6KB2 CABP4,PPP1CA,ADRBK1,ANKRD13D,SSH3,RAD9A,FBXL11* |
| rs10842162 | 12 | 23444386 | 7 | NA | *NA* |
| rs735480 | 15 | 42939663 | 19 | NA | *TRIM69,C15orf43* |
| rs1011489 | 16 | 5607394 | 18 | NA | *NA* |
| rs2306192 | 19 | 14004016 | 20 | IL27RA,RLN3 | *LOC113230,PRKACA,ASF1B,SAMD1,PODNL1,LOC90379,RFX1* |

Only SNPs with imputation accuracy > 0.9 were tested.

Table S24. 20 SNPs with largest high altitude Amhara PBS (*versus* low altitude Amhara and low altitude Oromo).

| SNP | Chr | Nt. pos. | Rank | Genes (within 10kb) | Genes (within 100kb) |
|---|---|---|---|---|---|
| rs2814778 | 1 | 157441307 | 2 | *DARC,CADM3* | *FCER1A* |
| rs1487529 | 1 | 192399145 | 12 | | |
| rs10863766 | 1 | 204892228 | 17 | *DYRK3* | *LGTN,RASSF5,MAPKAPK2* |
| rs7539843 | 1 | 242957549 | 5 | | *FAM152A,C1orf101* |
| rs10180234 | 2 | 153095625 | 11 | *FMNL2* | |
| rs349003 | 2 | 222660565 | 10 | | |
| rs349010 | 2 | 222663592 | 14 | | |
| rs3768889 | 2 | 225073300 | 3 | *CUL3* | *FAM124B* |
| rs13181143 | 5 | 140186928 | 9 | *PCDHA1, PCDHA2, PCDHA3, PCDHA4, PCDHA5, PCDHA6, PCDHA7* | *PCDHAC1,PCDHA8,PCDHA9, PCDHA10,PCDHA11, PCDHA12,PCDHA13* |
| rs4377754 | 5 | 153096868 | 16 | *GRIA1* | |
| rs7764351 | 6 | 32529609 | 1 | *HLA-DRA* | *HLA-DRB5,BTNL2,HLA-DRB1,C6orf10,HLA-DRB6* |
| rs7753021 | 6 | 145310595 | 19 | | *UTRN* |
| rs10094452 | 8 | 117810022 | 15 | *EIF3H* | *C8orf53* |
| rs2048803 | 12 | 57841104 | 18 | | |
| rs1983521 | 14 | 21688320 | 20 | | |
| rs16940947 | 18 | 20533441 | 13 | | |
| rs10416242 | 19 | 13611618 | 4 | | *CCDC130* |
| rs1422837 | 19 | 13615667 | 6 | | *CCDC130* |
| rs398717 | 21 | 42874233 | 7 | *SLC37A1* | *PDE9A,SLC37A1,RSPH1* |
| rs133352 | 22 | 40763968 | 8 | *WBP2NL* | *NDUFA6,CYP2D6,NAGA,FAM109B, C22orf32,LOC339674,CENPM,SEPT3* |

Only SNPs with imputation accuracy > 0.9 were tested.

Table S25. SNPs with oxygen saturation ($O_2$ Sat) association p-value < 0.05 among the top 20 Amhara PBS SNPs.

| Test1 | Test1 | SNP | Chr | Nt. pos. | Hb P | $O_2$ Sat P | PBS P | Genes (within 10kb) | Genes (within 100kb) |
|---|---|---|---|---|---|---|---|---|---|
| PBS | $O_2$ Sat | rs619660 | 3 | 43002953 | 7.00E-02 | 1.70E-02 | 3.60E-06 | *LOC72985* | *C3orf39,ZNF662,CCBP2* |
| PBS | $O_2$ Sat | rs9853065 | 3 | 1.74E+08 | 3.80E-01 | 2.70E-02 | 9.10E-06 | | |

Table S26. 20 SNPs with largest high altitude Amhara PBS or MR and Hb or $O_2$ sat association pvalue < 0.05 within HA Amhara.

| Test 1 | Test 2 | SNP | Chr | Nt. pos. | Hb P | $O_2$ Sat P | Genes (within 10kb) | Genes (within 100kb) |
|---|---|---|---|---|---|---|---|---|
| PBS | $O_2$ Sat | rs7753021 | 6 | 145310595 | 9.50E-01 | 5.40E-03 | | *UTRN* |
| PBS | $O_2$ Sat | rs1983521 | 14 | 21688320 | 6.70E-01 | 3.50E-02 | | |
| PBS | Hb | rs16940947 | 18 | 20533441 | 2.70E-02 | 5.30E-01 | | |
| MR | Hb | rs926130 | 21 | 15136788 | 8.66E-03 | 5.67E-01 | | |

Table S27. 20 SNPs with largest high altitude Amhara MR score.

| SNP | Chr | Nt. pos. | Rank | Genes (within 10kb) | Genes (within 100kb) |
|---|---|---|---|---|---|
| rs682600 | 1 | 199730648 | 14 | *CSRP1* | *PHLDA3,LOC376693,TNNI1,LAD1* |
| rs6810492 | 4 | 11833486 | 18 | | |
| rs2660343 | 4 | 41672523 | 15 | *WDR21B* | *TMEM33,SLC30A9* |
| rs2660342 | 4 | 41672561 | 8 | *WDR21B* | *TMEM33,SLC30A9* |
| rs12510722 | 4 | 100366124 | 4 | *ADH6* | *ADH1A,ADH1B,ADH4* |
| rs2173199 | 4 | 100390402 | 5 | | *ADH1A,ADH1B,ADH1C,ADH6* |
| rs6532814 | 4 | 100392991 | 6 | | *ADH1A,ADH1B,ADH1C,ADH6* |
| rs1826909 | 4 | 100436766 | 19 | *ADH1A,ADH1B* | *ADH1C,ADH6* |
| rs13103321 | 4 | 100439263 | 12 | *ADH1A,ADH1B* | *ADH1C,ADH6* |
| rs1353621 | 4 | 100460598 | 9 | *ADH1B* | *ADH1C,ADH1A,ADH7* |
| rs7661978 | 4 | 100503222 | 10 | | *ADH1B,ADH1C,ADH7,ADH1A* |
| rs729147 | 4 | 100552290 | 2 | *ADH7* | *ADH1C,C4orf17,ADH1B* |
| rs325502 | 5 | 104036032 | 7 | | |
| rs2662891 | 7 | 85327629 | 1 | | |
| rs1994056 | 8 | 3610422 | 11 | *CSMD1* | |
| rs1412060 | 9 | 121329170 | 20 | | |
| rs4842631 | 12 | 87450918 | 13 | *KITLG* | |
| rs9323043 | 14 | 40618377 | 3 | | |
| rs8091228 | 18 | 69134448 | 16 | | |
| rs926130 | 21 | 15136788 | 17 | | |

Only SNPs with imputation accuracy > 0.9 were tested.

Table S28. 40 CpG sites with highest high versus low methylation difference within Oromo. TSS denotes transcription start site.

| CpG | Chr | Nt. pos. | P | Rank | Distance to TSS | Genes |
|---|---|---|---|---|---|---|
| cg04286933 | 22 | 37802689 | 2.69E-07 | 1 | 393 | APOBEC3G |
| cg22198623 | 16 | 55259369 | 7.18E-07 | 2 | 109 | MT1G |
| cg14056644 | 4 | 111778554 | 1.27E-06 | 3 | 597 | PITX2 |
| cg09931793 | 9 | 113130082 | 1.54E-06 | 4 | 452 | OR2K2 |
| cg15089487 | 15 | 68972306 | 1.95E-06 | 5 | 499 | THAP10 |
| cg14785479 | 22 | 19122535 | 6.52E-06 | 6 | 389 | SCARF2 |
| cg20988728 | 22 | 30018502 | 8.99E-06 | 7 | 28 | MGC17330 |
| cg07200897 | 3 | 123995511 | 9.24E-06 | 8 | 171 | HSPBAP1 |
| cg07426848 | 1 | 151788336 | 9.63E-06 | 9 | 22 | S100A3 |
| cg11237738 | 4 | 5578196 | 1.06E-05 | 10 | 412 | C4orf6 |
| cg24237576 | 16 | 2225776 | 1.08E-05 | 11 | 693 | DNASE1L2 |
| cg13006591 | 4 | 38508060 | 1.20E-05 | 12 | 505 | TLR6 |
| cg21053015 | 22 | 38076000 | 1.25E-05 | 13 | 100 | SYNGR1 |
| cg03533858 | 1 | 2314256 | 1.46E-05 | 14 | 1403 | MORN1 |
| cg02592124 | 2 | 202811724 | 1.96E-05 | 15 | 157 | SUMO1 |
| cg10313633 | 11 | 44929466 | 2.55E-05 | 16 | 456 | TP53I11 |
| cg10574499 | 16 | 66476255 | 2.65E-05 | 17 | 27 | UNQ2446 |
| cg18410627 | 9 | 128717703 | 2.88E-05 | 18 | 829 | RALGPS1 |
| cg15983520 | 8 | 145699914 | 2.93E-05 | 19 | 317 | GPT |
| cg26856388 | 7 | 64975330 | 3.11E-05 | 20 | 362 | VKORC1L1 |
| cg13210534 | 11 | 111289538 | 3.52E-05 | 21 | 829 | HSPB2 |
| cg08611714 | 17 | 36877384 | 3.55E-05 | 22 | 220 | KRTHA2 |
| cg04081402 | 22 | 29333587 | 4.07E-05 | 23 | 426 | TCN2 |
| cg24585690 | 5 | 135259413 | 4.10E-05 | 24 | 2 | IL9 |
| cg17043155 | 2 | 75727427 | 4.15E-05 | 25 | 7 | MRPL19 |
| cg22036988 | 3 | 142253035 | 4.33E-05 | 26 | 91 | SPSB4 |
| cg10525372 | 6 | 112482026 | 4.40E-05 | 27 | 55 | WISP3 |
| cg19464252 | 16 | 30582734 | 4.43E-05 | 28 | 1155 | FBS1 |
| cg21830413 | 10 | 70554257 | 4.47E-05 | 29 | 295 | VPS26 |
| cg05670596 | 3 | 46423500 | 4.84E-05 | 30 | 225 | CCRL2 |
| cg00795812 | 2 | 242450682 | 5.78E-05 | 31 | 951 | PDCD1 |
| cg18168989 | 9 | 76834070 | 5.90E-05 | 32 | 940 | C9orf41 |
| cg17964955 | 10 | 48010942 | 5.93E-05 | 33 | 55 | RBP3 |
| cg10150813 | 4 | 25474276 | 5.94E-05 | 34 | 734 | KIAA0746 |
| cg08859675 | 19 | 10424771 | 6.09E-05 | 35 | 134 | PDE4A |
| cg22396755 | 1 | 21867684 | 6.16E-05 | 36 | 702 | RAP1GA1 |
| cg22858308 | 6 | 143137306 | 6.69E-05 | 37 | 369 | HIVEP2 |
| cg06852744 | 7 | 44109498 | 6.73E-05 | 38 | 987 | AEBP1 |
| cg13439730 | 16 | 31054500 | 7.23E-05 | 39 | 152 | PRSS8 |
| cg15901783 | 13 | 76359427 | 7.51E-05 | 40 | 901 | KCTD12 |

Table S29. 40 CpG sites with highest high versus low methylation difference within Amhara. TSS denotes transcription start site.

| CpG | Chr | Nt. pos. | P | Rank | Distance to TSS | Genes |
|---|---|---|---|---|---|---|
| cg16351002 | 10 | 124600091 | 8.13E-06 | 1 | 208 | *CUZD1* |
| cg11378686 | 6 | 43444256 | 4.32E-05 | 2 | 903 | *ZNF318* |
| cg05244766 | 11 | 67107075 | 8.03E-05 | 3 | 787 | *GSTP1* |
| cg22445920 | 5 | 150663371 | 1.38E-04 | 4 | 149 | *SLC36A3* |
| cg12026956 | 5 | 72147764 | 2.03E-04 | 5 | 490 | *TNPO1* |
| cg08385610 | 8 | 20156963 | 2.23E-04 | 6 | 120 | *LZTS1* |
| cg10453758 | 3 | 133862880 | 2.50E-04 | 7 | 1264 | *ACAD11* |
| cg19764399 | 13 | 47567117 | 2.81E-04 | 8 | 124 | *MED4* |
| cg00922727 | 6 | 109521816 | 3.18E-04 | 9 | 154 | *SESN1* |
| cg23364287 | 3 | 48729780 | 3.99E-04 | 10 | 70 | *IHPK2* |
| cg04609640 | 1 | 45026679 | 4.21E-04 | 11 | 715 | *VMD2L2* |
| cg23743114 | 17 | 31352509 | 4.83E-04 | 12 | 616 | *CCL15* |
| cg27626899 | 15 | 89276929 | 5.27E-04 | 13 | 149 | *HDDC3* |
| cg23323879 | 15 | 63291587 | 5.85E-04 | 14 | 723 | *CILP* |
| cg16746462 | 5 | 96296194 | 6.36E-04 | 15 | 908 | *LNPEP* |
| cg07246225 | 2 | 144994129 | 6.50E-04 | 16 | 257 | *ZFHX1B* |
| cg05449607 | 6 | 111386305 | 6.78E-04 | 17 | 151 | *C6orf51* |
| cg10167296 | 1 | 204747251 | 7.16E-04 | 18 | 251 | *RASSF5* |
| cg10051054 | 1 | 3659032 | 7.66E-04 | 19 | 210 | *CCDC27* |
| cg23487586 | 16 | 83873424 | 8.29E-04 | 20 | 641 | *MGC22001* |
| cg09450020 | 7 | 89679371 | 8.31E-04 | 21 | 435 | *STEAP2* |
| cg21850254 | 19 | 42651693 | 8.43E-04 | 22 | 129 | *ZNF570* |
| cg08866753 | 1 | 39814648 | 8.66E-04 | 23 | 355 | *PABPC4* |
| cg00600684 | 10 | 122600148 | 8.71E-04 | 24 | 537 | *BRWD2* |
| cg05028306 | 4 | 17187761 | 8.95E-04 | 25 | 264 | *LAP3* |
| cg09180926 | 15 | 89338658 | 9.11E-04 | 26 | 150 | *PRC1* |
| cg04868764 | 15 | 38924311 | 9.17E-04 | 27 | 773 | *SPINT1* |
| cg13726191 | 4 | 15549131 | 9.20E-04 | 28 | 62 | *FGFBP1* |
| cg08303146 | 11 | 2423231 | 9.71E-04 | 29 | NA | *KCNQ1* |
| cg18493238 | 10 | 37454267 | 1.00E-03 | 30 | 524 | *ANKRD30A* |
| cg02449978 | 3 | 48205171 | 1.01E-03 | 31 | 366 | *CDC25A* |
| cg03077492 | 5 | 43448852 | 1.02E-03 | 32 | 623 | *CCL28* |
| cg24134767 | 11 | 113350848 | 1.03E-03 | 33 | 272 | *HTR3A* |
| cg23863670 | 5 | 43447686 | 1.04E-03 | 34 | 543 | *CCL28* |
| cg12820481 | 1 | 209915383 | 1.05E-03 | 35 | 207 | *NEK2* |
| cg07123548 | 19 | 45588003 | 1.09E-03 | 36 | 69 | *HIPK4* |
| cg16791508 | 12 | 51001958 | 1.13E-03 | 37 | 520 | *KRTHB3* |
| cg09126273 | 12 | 15367034 | 1.17E-03 | 38 | NA | *PTPRO* |
| cg07710481 | 13 | 87123385 | 1.21E-03 | 39 | 514 | *SLITRK5* |
| cg16961218 | 18 | 74930131 | 1.24E-03 | 40 | 254 | *ATP9B* |